\documentclass[longauth]{aa}

\usepackage{graphicx}
\usepackage{txfonts}
\usepackage{graphicx}	
\usepackage{amsmath}	
\usepackage{textcomp, gensymb}    
\usepackage{xspace}
\usepackage{url}
\usepackage{subcaption}
\usepackage{booktabs}
\usepackage{placeins}
\usepackage{float}

\usepackage[colorlinks=true,linkcolor=blue,citecolor=blue,urlcolor=blue]{hyperref}
\usepackage{svg}
\newcommand{\orcid}[1]{\href{https://orcid.org/#1}{\includegraphics[width=10pt]{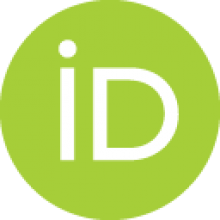}}}

\usepackage[colorinlistoftodos]{todonotes}
\usepackage{marginnote}

\begin{document}

\title{The ALPINE-CRISTAL-JWST Survey: Investigating the role of interstellar dust, gas and stars in high-$z$ galaxies at kpc-scales}
\subtitle{}

\author{
Fiona~Lopez\inst{1}\orcid{0009-0005-5605-4686}
\and 
M\'onica~Rela\~no\inst{2,3}\orcid{000-0003-1682-1148}
\and
Ilse~De~Looze\inst{4}\orcid{000-0001-9419-6355}
\and
Andreas L. Faisst\inst{5}\orcid{0000-0002-9382-9832}
\and
Rodrigo Herrera-Camus\inst{6,7}\orcid{0000-0002-2775-0595}
\and
Regina~A.~Jorgenson\inst{8,9}\orcid{0000-0003-2973-0472}
\and
Enrique Pérez-Montero\inst{10}\orcid{0000-0003-3985-4882} 
\and
Marco~Palla\inst{11,12}\orcid{0000-0002-3574-9578}
\and
C. Accard \inst{13} \orcid{0009-0005-9982-7239}
\and
Ricardo~O.~Amor\'{i}n\inst{10}\orcid{0000-0001-5758-1000} 
\and
Manuel~Aravena\inst{14,7}\orcid{0000-0002-6290-3198}
\and 
Roberto~J.~Assef\inst{14}\orcid{0000-0002-9508-3667}
\and 
Andrew J. Battisti\inst{15,16}\orcid{0000-0003-4569-2285}
\and
Médéric~Boquien\inst{17}\orcid{0000-0003-0946-6176}
\and
Elisabete~da~Cunha\inst{15}\orcid{0000-0001-9759-4797}
\and
Poulomi~Dam\inst{1}\orcid{0009-0007-7842-9930}
\and
Rebecca L. Davies\inst{19}\orcid{0000-0002-3324-4824} 
\and
Miroslava Dessauges-Zavadsky\inst{20}\orcid{0000-0003-0348-2917} 
\and 
Andrea~Ferrara\inst{21}\orcid{0000-0002-9400-7312}
\and
Seiji Fujimoto\inst{22,23}\orcid{0000-0001-7201-5066}
\and 
Michele Ginolfi\inst{24,25}\orcid{0000-0002-9122-1700}
\and 
Nicol Gutiérrez-Vera\inst{6,7}\orcid{0009-0005-8932-7783} 
\and
Ali Hadi\inst{26}\orcid{0009-0003-3097-6733}
\and 
Edo~Ibar\inst{7,27}\orcid{0009-0008-9801-2224} 
\and
Hanae~Inami\inst{28}\orcid{0000-0003-4268-0393}  
\and
Anton~M.~Koekemoer\inst{29}\orcid{0000-0002-6610-2048}  
\and
Lilian L. Lee\inst{30}\orcid{0000-0001-7457-4371} 
\and
Juno Li\inst{15}\orcid{0000-0002-8184-5229}
\and 
Juan Molina\inst{7,27}\orcid{0000-0002-8136-8127} 
\and
Ambra Nanni\inst{31,32}\orcid{0000-0001-6652-1069}
\and 
Desika Narayanan\inst{33,34}\orcid{0000-0002-7064-4309}
\and 
Francesca Pozzi\inst{11,12}\orcid{0000-0002-7412-647X}  
\and
Francesca~Rizzo\inst{35}\orcid{0000-0001-9705-2461}
\and
Michael Romano\inst{36,37}\orcid{0000-0002-9948-3916}
\and
David B. Sanders\inst{38}\orcid{0000-0002-1233-9998}
\and 
Prasad Sawant\inst{39}\orcid{0000-0002-0498-8074}
\and 
Livia~Vallini\inst{12}\orcid{0000-0002-3258-3672} 
\and
Stefan A. van der Giessen\inst{40}\orcid{0009-0001-4144-9635}
\and
Vicente~Villanueva\inst{7,14}\orcid{0000-0002-5877-379X}
\and 
Giovanni Zamorani\inst{12}\orcid{0000-0002-2318-301X}
}
\institute{
Department of Astronomy, New Mexico State University, Las Cruces, NM 88003, USA 
\and 
Dept. Física Teórica y del Cosmos, E-18071 Granada, Spain 
\and 
Instituto Universitario Carlos I de Física Te\'{o}rica y Computacional, Universidad de Granada, 18071, Granada, Spain 
\and 
Department of Physics and Astronomy, Ghent University, Proeftuinstraat 86, N3, B-9000 Ghent, Belgium 
\and
IPAC, California Institute of Technology, 1200 E. California Blvd. Pasadena, CA 91125, USA 
\and
Departamento de Astronomía, Universidad de Concepción, Barrio Universitario, Concepción, Chile 
\and
Millenium Nucleus for Galaxies (MINGAL), Av. Ej\'ercito 441, Santiago 8370191, Chile 
\and 
Maria Mitchell Observatory, Nantucket, MA 02554, USA 
\and 
Department of Physics \& Astronomy, California State Polytechnic University, Humboldt, Arcata, California 95521, USA 
\and
Instituto de Astrof\'{i}sica de Andaluc\'{i}a (CSIC), Apartado 3004, 18080 Granada, Spain 
\and
Dipartimento di Fisica e Astronomia “Augusto Righi”, Alma Mater Studiorum, Università di Bologna, Via Gobetti 93/2, 40129 Bologna, Italy 
\and
INAF – Osservatorio di Astrofisica e Scienza dello Spazio di Bologna, Via Gobetti 93/3, 40129 Bologna, Italy 
\and
Université de Strasbourg, CNRS, Observatoire Astronomique de Strasbourg, UMR 7550, 67000 Strasbourg, France
\and
Instituto de Estudios Astrof\'{i}sicos, Facultad de Ingenier\'{i}a y Ciencias, Universidad Diego Portales, Av. Ej\'{e}rcito 441, Santiago 8370191, Chile 
\and 
International Centre for Radio Astronomy Research (ICRAR), The University of Western Australia, M468, 35 Stirling Highway, Crawley, WA 6009, Australia 
\and 
Research School of Astronomy and Astrophysics, Australian National University, Cotter Road, Weston Creek, ACT 2611, Australia 
\and
Université Côte d'Azur, Observatoire de la Côte d'Azur, CNRS, Laboratoire Lagrange, 06000, Nice, France 
\and
Dipartimento di Fisica e Astronomia Galileo Galilei Universit{\`a} degli Studi di Padova, Vicolo dell’Osservatorio 3, 35122 Padova, Italy 
\and 
Centre for Astrophysics and Supercomputing, Swinburne University of Technology, Hawthorn, Victoria 3122, Australia 
\and 
D\'epartement d’Astronomie, Universit\'e de Gen\`eve, Chemin Pegasi 51, 1290 Versoix, Switzerland
\and
Scuola Normale Superiore, Piazza dei Cavalieri 7, 50126 Pisa, Italy 
\and 
David A. Dunlap Department of Astronomy and Astrophysics, \\ University of Toronto, 50 St. George Street, Toronto, Ontario, M5S 3H4, Canada
\and
Dunlap Institute for Astronomy and Astrophysics, 50 St. George Street, Toronto, Ontario, M5S 3H4, Canada
\and 
Universit\`a di Firenze, Dipartimento di Fisica e Astronomia, via G. Sansone 1, 50019 Sesto Fiorentino, Florence, Italy 
\and 
INAF -- Arcetri Astrophysical Observatory, Largo E. Fermi 5, I-50125, Florence, Italy 
\and 
Department of Physics and Astronomy, University of California, Riverside, 900 University Ave, Riverside, CA 92521, USA 
\and
Instituto de F\'{i}sica y Astronom\'{i}a, Universidad de Valpara\'{i}so, Avda. Gran Breta\~{n}a 1111, Valpara\'{i}so, Chile 
\and
Hiroshima Astrophysical Science Center, Hiroshima University, 1-3-1 Kagamiyama, Higashi-Hiroshima, Hiroshima 739-8526, Japan 
\and
Space Telescope Science Institute, 3700 San Martin Drive, Baltimore, MD 21218, USA 
\and 
Max-Planck-Institute f\"ur extratarrestrische Physik, Giessenbachstrasse 1, 85748 Garching, Germany
\and
National Centre for Nuclear Research, ul. Pasteura 7, 02-093 Warsaw, Poland 
\and 
INAF - Osservatorio astronomico d'Abruzzo, Via Maggini SNC, 64100, Teramo, Italy 
\and 
Department of Astronomy, University of Florida, 211 Bryant Space Sciences Center, Gainesville, FL 32611 USA 
\and 
Cosmic Dawn Center at the Niels Bohr Institute, University of Copenhagen and DTU-Space, Technical University of Denmark 
\and 
Kapteyn Astronomical Institute, University of Groningen, Landleven 12, 9747 AD, Groningen, The Netherlands 
\and 
Max-Planck-Institut für Radioastronomie, Auf dem Hügel 69, 53121 Bonn, Germany 
\and 
INAF - Osservatorio Astronomico di Padova, Vicolo dell’Osservatorio 5, I-35122 Padova, Italy 
\and
Institute for Astronomy, University of Hawaii, 2680 Woodlawn Drive, Honolulu, HI 96822, USA 
\and 
Institute of Astronomy, Faculty of Physics, Astronomy and Informatics, Nicolaus Copernicus University, Grudzi\k{a}dzka 5, 87-100 Toru\'n, Poland
\and
Université Paris-Saclay, Université Paris Cité, CEA, CNRS, AIM, 91191, Gif-sur-Yvette, France 
}

   \date{Received ; accepted }

\abstract{A comprehensive understanding of galaxy evolution requires a detailed analysis of gas and dust, their distribution, and evolution within galaxies. In this work, we analyse spatially resolved ($\sim\,1-2$\,kpc) rest-frame [C{\sc{ii}}] and dust continuum ALMA observations together with JWST/NIRSpec IFU data of a sample of six massive galaxies at $z\,\sim$\,4\,-\,5. We focus on deriving and comparing the spatial distributions of atomic and molecular gas masses, dust and stellar masses, star formation rates (SFR), and metallicity. The oxygen abundance maps have median values in the range of 12+log(O/H)\,$\sim8.1-8.3$ with variations of the order of 1\,dex within the same galaxy. In four galaxies of our sample the oxygen abundance shows a minimum at the peak of the H$\alpha$ emission and an enhancement in the outer parts. The calibration used in this study to estimate the atomic gas mass in our galaxies seems to overestimate the atomic gas mass surface densities up to 1-2 orders of magnitude. However, a calibration based on a spatially resolved [C{\sc{ii}}] conversion factor gives a more robust estimation of the total gas mass. The dust-to-gas surface mass density ratio varies between $\log \Sigma_{\rm dust}/\Sigma_{\rm gas}=-2.2{ 0}^{+0.24}_{-0.19}$ and $\log\Sigma_{\rm dust}/\Sigma_{\rm gas}=-2.69^{+0.21}_{-0.16}$, while the dust-to-stellar surface mass density ratio ranges from $\log\Sigma_{\rm dust}/\Sigma_{\rm star}=-2.01^{+0.13}_{-0.12}$ to $\log\Sigma_{\rm dust}/\Sigma_{\rm star}=-2.91^{+0.13}_{-0.13}$. 
A detailed analysis of the dust-to-stellar surface mass density ratio and the gas mass fraction versus metallicity reveals that galaxies exhibiting signatures of strong outflows have lower dust-to-stellar surface mass density ratio and lower gas mass fractions than the galaxies with no outflow signatures. While the latter is consistent with these galaxies ejecting part of their gas reservoir to the circumgalactic medium, the low dust-to-stellar surface mass density ratios also suggest that part of the dust content is expelled outside the galaxy, which reinforces previous theoretical suggestions of dust removal by stellar feedback or radiation pressure to explain the high UV luminosities in galaxies at very high-redshifts. Some galaxies of our sample have almost an order of magnitude higher dust-to-gas and dust-to-stellar surface mass density ratio than galaxies in the local Universe, suggesting that a large dust reservoir has been built up already at time scales of $\sim1$\,Gyr in these galaxies.}

\titlerunning{ }
   \maketitle

\section{Introduction}

Relations between galaxy properties are key to understand galaxy evolution in the Universe. Fundamental relations such as the star-forming main sequence (SFMS) or the mass-metallicity relations \citep[e.g.,][]{Speagle2014,Maiolino2019A&AR} help us to clarify how galaxies consume their gas reservoirs to form stars, how star formation proceeds, and how the chemical enrichment occurs throughout the lifetime of a galaxy. Although these scaling relations represent the equilibrium among the different processes influencing galaxy evolution (gas accretion, star formation, feedback processes, and chemical enrichment) \citep[e.g.,][]{Lilly2013,Feldmann2015}, departures from them showcase galaxies that might be or have been in the past strongly affected by intermittent star formation, intense inflows, high outflow metal loading, etc., that cause deviation from the expected evolution of the bulk of galaxies \citep[e.g.,][]{Dayal2013,Ma2016,Caplar2019,DeLucia2020,Korhonen2025,Pallottini2025}. 

Interstellar dust is an important component in galaxy evolution as it regulates molecular hydrogen formation and gas cooling, both processes crucial for star formation. It contributes to the chemical enrichment balance in galaxies by providing the reservoir of the solid-phase metal content. Interstellar dust also participates in feedback phenomena, as dust being ejected in feedback-driven outflows has recently been witnessed with the James Webb Space Telescope (JWST) \citep{Chastenet2024,Bolatto2024,Fisher2025} and archival Herschel observations  \citep{McCormick2018,Romano2024,Lesniewska2025}. 

Incorporating interstellar dust in the group of scaling relations represents therefore a crucial addition. Relations between the ratio of the dust and gas mass (dust-to-gas mass ratio), the dust mass per unit of stellar mass (dust-to-stellar mass ratio) and the metallicity in the gas phase have been extensively studied in the literature \citep[e.g.][]{daCunha2010,DeVis2017,DeLooze2020}. They help understand how dust is built up in galaxies with different physical properties (e.g. morphological classification, metal content, history of star formation, etc.), and to explore the main dust formation channels in the local Universe \citep[e.g,][]{Remy,Galliano2018,Salvestrini2025}, at intermediate redshifts \citep[$z\sim\,1-2.5$, e.g.,][]{Shapley2020,Popping2023} and up to $z\sim\,6$ \citep[e.g.][]{DeCia2016,Konstantopoulou2024,Sawant2025}. In particular, in the local Universe these relations allowed to identify galaxies that depart from the bulk of the local galaxy behaviour on spatially resolved \citep[e.g. Sextans A,][]{Park2024} and integrated scales \citep[e.g. SBS0335$-$052 and IZw18][respectively]{Hunt2014,Schneider2016}, which brings information on how dust build-up proceeds at different spatial scales and throughout the cosmic time. 

With the advent of JWST we are now able to explore the SFMS and mass-metallicity relations in the very early ages (up to z$\sim$\,10) of the Universe \cite[e.g.,][]{Chemerynska2024,Nakajima2023,Curti2024}. NIRSpec observations allow us to derive accurate estimates of the gas phase metallicity, showing evidence that the early Universe was quickly chemically enriched \cite[e.g.][]{Arellano2025,Scholte2025,Curti2023,Faisst2025_massmetal}. The high metallicities derived from observations are consistent with the large dust reservoirs inferred from single- and multi-band ALMA observations of high-redshift galaxies \citep{Pozzi2021,Inami2022,Bakx2020,Hashimoto2019,Algera2026,Palla2024,Villanueva2024}. 
Accurate estimates of dust masses in high-$z$ galaxies suffer from uncertainties due to upper limits in multi-band observations \citep[e.g.][]{Lower2024,Sommovigo2025MNRAS}, the assumption of a dust temperature to describe the dust spectral energy distribution \citep[e.g.][]{Sommovigo2020,Sommovigo2025MNRAS}, and the uncertainty in the dust optical thickness \citep[e.g.][]{Boquien2022}. However, despite these uncertainties and the challenge that this large dust reservoir implies for chemical and dust evolution models \citep[see][]{Ferrara2022,Palla2024,Sawant2025,Algera2026}, a dust and metal rich scenario is becoming a robust scenario to observationally describe galaxies in the early ($t\,\lesssim\,1-2$\,Gyr) Universe. A description of how this rapid dust build-up occurs and which are the main dust formation channels in galaxies at these short timescales is still an open question.

 Dust grains originate in the cool envelopes of the Asymptotic Giant Branch stars (AGBs) \citep[e.g.,][]{Ferrarotti2006,Nanni2013} and the expanding ejecta of core-collapse supernovae (SN) \citep[e.g.,][]{Matsuura2015,DeLooze2017}. Once in the ISM, dust grains can then grow further by accreting metals from the surrounding gas \citep[e.g.][]{Asano2013,Zhukovska2008}. Explaining the large amount of dust at high-redshifts poses some challenges in these dust formation mechanisms: the short time scales make AGB stars inefficient dust producers \citep[][but see also \citealt{Valiante2009}]{Leniewska2019}, and a certain, still unknown, fraction of the dust mass produced by SN is expected to be destroyed by the passage of the reverse shock \citep{Bianchi2007}. Dust growth in the ISM seems to be, therefore, the most relevant mechanism to account for the large amount of dust in the Universe at high-redshift \citep{Algera2024,Burgarella2025,Witstok2023,Narayanan2025}. However, dust growth itself is not exempt from difficulties \citep{Ferrara2016,Ceccarelli2018}. It is efficient in high density environments where significant chemical gas enrichment has occurred, but a detailed physical description of the sticking process by which the gaseous metals are finally incorporated into the dust grains is not available yet \citep[e.g.][]{Choban2022}. Moreover, the final dust mass budget in a galaxy represents a balance between the efficiency of the dust formation mechanisms and the destruction processes due to the strong shocks happening in the ISM \citep{Slavin2015,Kirchschlager2022,Kirchschlager2024}, dust depletion due to star formation, and dust leaving the galaxy into the circumgalactic medium (CGM) \citep{Chastenet2024,Bolatto2024}. In high-redshift galaxies one should expect a large fraction of the dust grains to be destroyed due to the strong stellar feedback (through mergers and starburst phenomena) happening in the early Universe, unless the shocks are efficiently stalled by a dense wind-driven shell \citep{Martinez2019}. 

Tracing the interplay between the different mechanisms that account for the dust budget in galaxies is challenging. Dust and chemical gas enrichment, as well as dust destruction and accretion, occur at the scales of the star-forming regions. In recent years there has been an increased interest in applying chemical and dust evolution models on spatially resolved scales \citep[e.g.][]{Calura2023,Giessen2024} and in the derivation of scaling relations at different locations within galaxy disks \citep[e.g.][]{Casasola,Park2024,Viaene2014}. These studies show that scaling relations hold typically on scales of $1-3$\,kpc, providing evidence that the main physical processes regulating the properties and evolution of galaxies tend to occur locally. Moreover, variations on how the dust-to-gas and dust-to-stellar mass ratios behave with metallicity at different galactocentric radii \citep[e.g.,][]{Chiang2021,Vilchez2019,Casasola2017} provide key information on how the dust formation and destruction channels are balanced at different locations within the galaxy.

The CRISTAL ([C\,{\sc{ii}}] Resolved Ism in STar-forming galaxies with ALma) program has been designed to study early galaxies in [C\,{\sc{ii}}] emission on highly resolved scales \citep{Herrera-Camus2025}. CRISTAL builds upon the ALPINE large program \citep{Faisst2020_Alpine}, which obtained 1\arcsec\ resolution imaging of a sample of 75 star-forming  galaxies at $z=4-6$. CRISTAL targeted a mass-selected representative sample of 19 massive star forming galaxies from ALPINE, aiming to spatially resolve the [C\,{\sc{ii}}] line and dust continuum emission at matched resolution with HST and JWST. By combining deep surveys of stellar ultraviolet (UV) light with HST, ALMA observations of cold dust and gas, CRISTAL  created the first systematic census of the gas, dust, and stars on approximately 1 kpc-scales in typical star-forming galaxies, when the Universe was $\sim$\,1\,Gyr old. Follow-up JWST NIRSpec/IFU observations of most of the 19 CRISTAL sample have been recently obtained. This ALPINE/CRISTAL-JWST survey \citep{Faisst2026_survey} is designed to extract physical properties of the ionized gas, including metallicities, at spatially resolved scales. The survey provides an unprecedented observational data set to fully explore the spatially resolved physical properties of star-forming galaxies at $z\sim4-6$. 

This paper capitalises on the exceptionally comprehensive data set and focuses on deriving spatially resolved $\sim$\,1\,kpc relations between dust-to-gas and dust-to-stellar surface mass density ratios with metallicity and gas mass fractions of a subsample of ALPINE/CRISTAL galaxies with a two-fold purpose: 1) to explore if the spatially resolved observed scaling relations shown for local galaxies hold at early time scales of the Universe after $\sim$1\,Gyr of evolution, and 2) to identify departures of the main relations that might show evidence of strong mechanisms (e.g. mergers, strong feedback processes) affecting the evolution of galaxies at these early epochs. Identifying those objects deviating from the more general evolution of local disk galaxies serves as an observational test for dust and chemical evolution models trying to explain the main processes characterizing galaxy formation and evolution in the early ages of the Universe. 

The paper is structured as follows: in Section\,\ref{sec:Observations_and_galaxysample} we present the galaxy sample and the data analysis to derive maps of the most intense rest-frame optical emission lines. In Section\,\ref{sec:metal} we derive metallicity maps, as well as estimate gas masses from different calibrations proposed in the literature. Spatially resolved scaling relations will be shown in Section\,\ref{sec:resolved_scaling_relations}, and a comparison with a set of chemical and dust evolution models from the literature will be presented in Section\,\ref{sec:models}. Conclusions are found in Section\,\ref{sec:conclusions}. Throughout this work, we assume a $\Lambda$CDM cosmology with $H_0 = 70$\,km\,s$^{-1}$\,Mpc$^{-1}$, $\Omega_{\Lambda}=0.7$, and  $\Omega_{\rm m}=0.3$. 

\section{Observations and galaxy sample}\label{sec:Observations_and_galaxysample}
\subsection{Sample selection}\label{sec:sample}
Our sample is obtained from a subsample of 13 CRISTAL galaxies which have been observed by ALMA at a spatial resolution ($\sim$ 0.1 $-$ 0.3\arcsec) that have \textit{HST} and \textit{JWST}/NIRCam observations.  Applying models to the observed UV-far-infrared SED of these galaxies, \citet{Li_2024} derived stellar mass and SFR maps, among other physical properties.
We chose 11 objects from \citet{Li_2024} (C01b, DC-848185 (C02), DC-536534 (C03), VC-5100822662 (C04a), VC-5100541407 (C06), DC-873321 (C07ab), DC-630594 (C11), VC-5100994794 (C13), HZ7 (C21), DC-873756 (C24), and VC-5101218326 (C25)) having observations at rest-frame UV wavelengths (1600$\AA$) from the COSMOS HST/ACS F814W imaging \citep{Koekemoer2007} -- the rest-frame UV image was used to assist in the derivation of spatially resolved dust mass maps (Rela\~no et al in prep). Furthermore, we discarded one of the galaxies, C01b, as the ALMA [C\,{\sc{ii}}] map 
exceeded the resolution of the stellar mass and SFR maps obtained in \citet{Li_2024}. This leaves us with a sample of 10 CRISTAL galaxies. 

The ALPINE/CRISTAL-JWST program \citep[JWST Cycle 2, GO PID:3045,][]{Faisst2026_survey} observed 19 representative main-sequence galaxies at 4.4\,<\,$z$\,<\,5.7 with the NIRSpec-IFU instrument on board the \textit{JWST}. Nine of these galaxies are contained in the subsample of  CRISTAL galaxies we have selected for this study. The observations made with NIRSpec allow us to obtain emission lines at each location of the field of view and derive spatially resolved physical properties of the ionised gas, key to understand the chemical enrichment and history of star formation of these galaxies. As we will see in Section\,\ref{sec:EM}, our emission line procedure forced us to exclude some other galaxies from our original sample, finally remaining with 6 galaxies in total.  The objects included in our final sample are shown in Table \ref{tab:gal}, together with the DEIMOS-COSMOS and VUDS-COSMOS names and some other physical galaxy properties derived from the SED modelling performed in \citet{Li_2024} and \citet{Mitsuhashi}. In this work, we will use the SFR, stellar mass, [C\,{\sc{ii}}] and dust continuum maps from \citet{Li_2024} with the same pixel grid as the one used in that work \citep[see section\,3.2.1 and figure\,1 in][for a detailed explanation on the grid selection]{Li_2024}. The grid size in  \citet{Li_2024} was adapted to ensure that the ALMA resolution is consistently resolved with $\simeq$\,4 spatial bins, each one corresponding to physical sizes from $\simeq 0.5$ to $\lesssim 1$\,kpc.

\begin{table*}
\caption{Properties of our final sub-sample of CRISTAL galaxies.}
\setlength{\tabcolsep}{1.5mm}

\label{tab:gal}      
\centering          
\begin{tabular}{lcccccccc}     
\hline\hline       
CRISTAL & Alternative  & Redshift & RA & Dec & $\log\rm M_{*}$ & $\log\rm SFR$ &  $\log\rm  M_{gas}$ & $\log\rm M_{\rm dust}$  \\
ID & name & $z_{\rm [CII]}$ &h:m:s & d:m:s & [$\rm M_{\odot}$] &  [$\rm M_{\odot}yr^{-1}$] & [$\rm M_{\odot}$] & [$\rm M_{\odot}$] \\
\hline 
        C02  & DC-848185 &  5.294 &  10:00:21.50 & +02:35:11.08 & 10.30 & 2.25  &  10.14$^{+0.07}_{-0.07}$ & 7.67$\pm 0.08$\\
        C03  & DC-536534 &  5.689 &  09:59:53.26 & +02:07:05.42 & 10.40 & 1.79 & 9.49$^{+0.08}_{-0.08}$ & 7.89$\pm 0.14$\\
        C04a  & VC-5100822662 &  4.520 & 09:58:57.91 & +02:04:51.48 & 10.15 & 1.89 & 9.57$^{+0.07}_{-0.08}$ & 6.94$\pm 0.11$\\
        C07ab  & DC-873321 &   5.154 & 10:00:04.06 & +02:37:35.76 & 10.00 & 1.89 & 9.78$^{+0.09}_{-0.09}$ & 7.10$\pm 0.20$\\
        C11  & DC-630594 &  4.439 & 10:00:32.62 &  +02:15:28.40  & 9.68 & 1.57 & 9.58$^{+0.07}_{-0.08}$ & 6.85$\pm 0.17$\\
        C13  & VC-5100994794 &  4.579 &  10:00:41.16 & +02:17:14.13 & 9.65 & 1.51 & 9.56$^{+0.14}_{-0.14}$ & 6.99$\pm 0.12$\\
\hline                  
\end{tabular}
\tablefoot{Stellar masses and SFRs shown here are integrated values and are obtained using UV-far-infrared SED modelling \citep[see][]{Mitsuhashi}. The spatially resolved stellar masses and SFRs are presented in \citet{Li_2024}. Integrated gas masses have been obtained using the $\alpha_{\rm C{II}}$ conversion factor given in equation\,1 in \citet{Vallini_25} together with the metallicity estimates presented in \citet{Faisst2026_survey} and the [C{\sc ii}] luminosities reported in \citet{Ikeda2025}. Dust masses are obtained using the calibration given in Relaño et al. (in prep).}
\end{table*}

\subsection{JWST data processing}\label{sec:jwst}

The \textit{JWST} NIRSpec integral field unit (IFU) \citep{JWST-IFU} is optimised for studies of astronomical targets that are extended over a few arcseconds. It provides spatially resolved imaging spectroscopy over a 3\arcsec\,$\times$\,3\arcsec\ square region with each spatial element in the resulting IFU data cube being 0.1\arcsec\,$\times$\,0.1\arcsec\ in size. 
The ALPINE/CRISTAL-JWST program \citep[JWST Cycle 2, GO PID:3045,][]{Faisst2026_survey} used both the G235M/F170LP and G395M/F290LP disperser-filter combination to cover an observed range of 1.6 to 5.2\,$\mu$m at a spectral resolution of $R\sim$\,1000. Details of the data reduction can be found in \citet{Fujimoto2025arXiv} and the observations of the full ALPINE/CRISTAL galaxy sample is presented in \citet{Faisst2026_survey}. 

Since we are interested in comparing dust masses, SFR and stellar masses with metallicity at spatially resolved scales we convolve our NIRSpec data set to the coarse resolution ($\sim$\,0.5\,\arcsec) of the data products obtained in \citet{Li_2024}.  The convolution was done by applying a kernel
that weights each pixel based on its neighbouring pixels, effectively smoothing each plane of the data cube to the ALMA resolution of each galaxy. The convolved data cube was then registered in the same reference frame as the derived maps in \citet{Li_2024}, creating a data set for each galaxy whose data products can be directly compared. 

\begin{figure*}
    \centering
    \includegraphics[width=\linewidth]{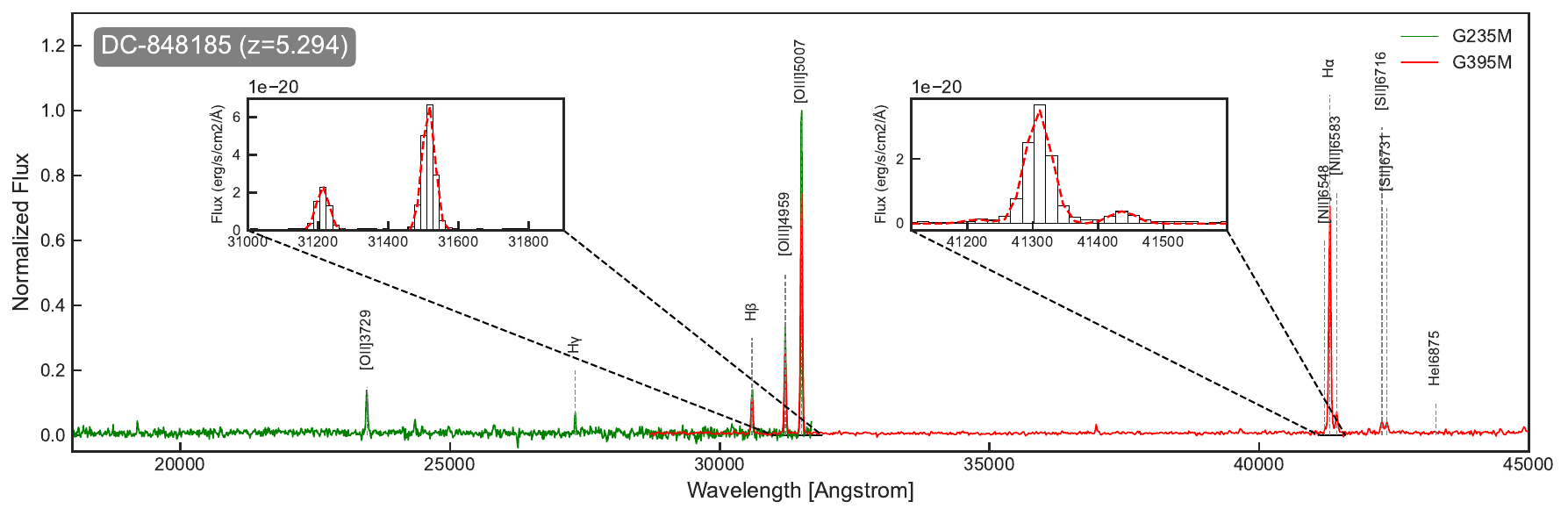}
    \caption {Spectrum for a single spaxel in the brightest H$\alpha$ area of C02 (DC-848185) obtained as a combination of G235M/F170LP (green line) and G395M/F290LP (red line). The most intense emission lines have been identified. The insets show the Gaussian fit to the H$\alpha$+[N\,{\sc{ii}}]$\lambda\lambda$6548,6583 and to the doublet [O{\sc{iii}}]$\lambda\lambda$4959,5007. The spectrum is normalised to the maximum of the most intense emission line ([O{\sc{iii}}]$\lambda$5007).
    }
    \label{fig:spec_spaxel}
\end{figure*}

The \textit{JWST}/NIRSpec convolved data cubes were processed using {\texttt{MPDAF}}, the MUSE Python Data Analysis Framework \citep{mpdaf}. {\texttt{MPDAF}}\footnote{\url{https://mpdaf.readthedocs.io/en/latest/}} is an open source Python package which provides tools to work with MUSE-specific IFU data cubes. For each spaxel, the spectrum was extracted and a Gaussian fitting was performed with the $\texttt{MPDAF gauss\_fit}$ package. 
The initial parameter set for the fitting process includes the minimum and maximum wavelength defining the emission line, the central wavelength and the width (FWHM) of the emission line. We modified the relevant scripts within $\texttt{MPDAF gauss\_fit}$ to fit simultaneously H$\alpha$, [N{\sc{ii}}]$\lambda$6583, and [N{\sc{ii}}]$\lambda$6548 while keeping the flux ratio of the two nitrogen lines fix to a certain value. For the case of [O{\sc{iii}}]$\lambda\lambda$4959,5007 and [N{\sc{ii}}]$\lambda\lambda$6548,6583 we keep the flux ratio fixed to a value of 2.98 and 3.07, following \citet{Storey2000} and \citet{Osterbrock2006}, respectively. In Figure\,\ref{fig:spec_spaxel} we show the spectrum of a spaxel in DC-848185 (C02), with the most intense emission lines identified. As an illustration, the results of the fitting procedure for the  [O{\sc{iii}}]$\lambda\lambda$4959,5007 and  [N{\sc{ii}}]$\lambda\lambda$6548,6583 are shown in the insets of Figure\,\ref{fig:spec_spaxel}\footnote{The [N{\sc{ii}}]$\lambda$6548 is not detected in spaxels with low S/N, which gave non-physical values for the nitrogen fluxes when fitting the doublet [N{\sc{ii}}]$\lambda\lambda$6548,6583 simultaneously, together with the H$\alpha$ emission line. Since we want to obtain maps as extended as possible we did not include [N{\sc{ii}}]$\lambda$6548 in the fit. Differences in the H$\alpha$ flux when excluding [N{\sc{ii}}]$\lambda$6548 are less than 2\%.}. 

In our analysis, we focus on fitting several key emission lines that provide information on the oxygen abundances. Specifically, we target [O{\sc{ii}}]$\lambda\lambda$3727,3729, [O{\sc{iii}}]$\lambda\lambda$4959,5007, H$\beta$, H$\alpha$, and [N{\sc{ii}}]$\lambda\lambda$6548,6583. The observed data cubes together cover a wavelength range from 1.6 to 5.2\,$\mu$m, providing the optical rest-frame wavelength range where all these emission lines are observed.

\subsection{Emission line maps}\label{sec:EM}
For each galaxy in the sample an emission line map (2D map) was produced for the relevant emission lines. 
This was created by iterating the procedure described in Section~\ref{sec:jwst} over each spaxel within the field of view. In order to avoid contamination of pixels where the emission line fitting is not robust, we identified which spaxels in the data cube have low $\textit{S/N}$ by creating a continuum noise map for each emission line. The noise for the continuum was obtained using an average of the root-mean-square (rms) on each side of the emission line (making sure that the chosen wavelength range did not include contamination from other adjacent emission lines).

We eliminate the spaxels with a $\textit{S/N}$ below 2 to ensure that only regions with sufficient $\textit{S/N}$ were included in the final analysis. Another mask was applied to the emission line maps using stellar mass maps that were derived from the SED fitting in \citet{Li_2024}. This mask ensures that we are only considering spaxels where the $\textit{S/N}$ is sufficient for robust fitting, as well as spaxels where stellar mass and SFR derivations are available. This dual masking approach allows us to focus on the most reliable regions of the data for analysis.

\subsection{Final galaxy sample}
\begin{figure*}
    \centering
    \includegraphics[width=\linewidth]{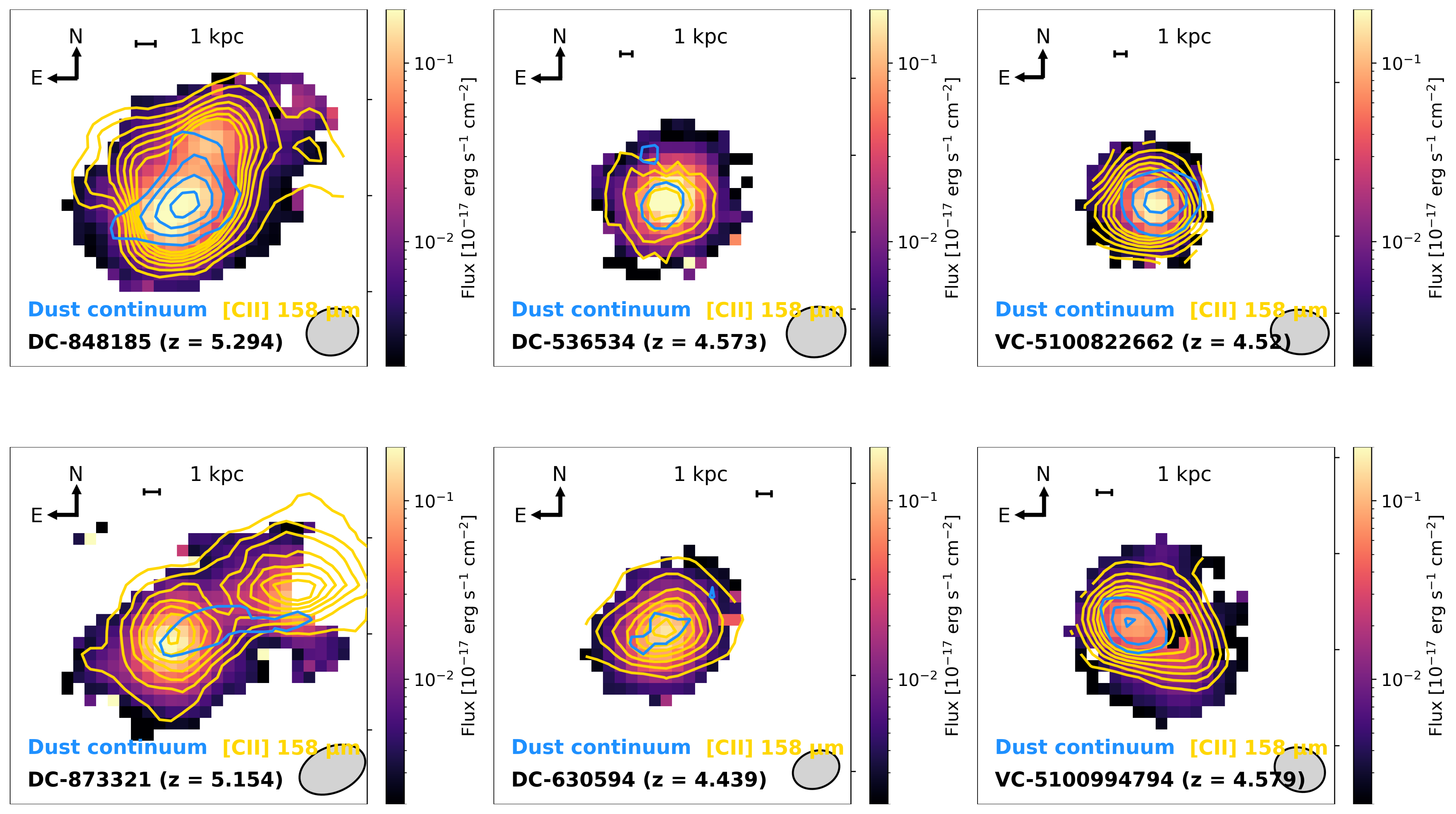}
    \caption {Observed H$\alpha$ emission line maps obtained with our Gaussian fitting procedure (see Section~\ref{sec:jwst} and~\ref{sec:EM}) for the sample of CRISTAL galaxies given in Table\,\ref{tab:gal}. Each map has contours of the rest-frame 158\,$\mu$m dust continuum (blue) and [C\,{\sc{ii}}] (yellow) ALMA emission. Contours start at 3$\sigma$ and increase in steps of 2$\sigma$. The synthesized beam (FWHM) size for ALMA observations is shown in grey in the bottom right corner of each panel.
    }
    \label{halpha maps}
\end{figure*}

Our emission line procedure forced us to exclude some other galaxies from our original sample. These were VC-5100541407 (C06), DC-873756 (C24), and VC-5101218326 (C25). The $\textit{S/N}$ of the observations of VC-5101218326 (C25) was too low to even fit the H$\alpha$ emission line in a significant number of spaxels. For DC-873756 (C24), although H$\alpha$ and [N{\sc{ii}}] are detected, the emission lines are blended together, likely due to the presence of high-velocity gas flows. In the case of VC-5100541407 (C06), in most of the pixels [O{\sc{ii}}], [O{\sc{iii}}], and H$\beta$ are not detected with a $\textit{S/N}$ above 2. With these considerations our final galaxy sample where we can robustly compare metallicity, dust and stellar masses, and SFR consists of 6 objects presented in Table\,\ref{tab:gal}. Figure\,\ref{halpha maps} shows the H$\alpha$ emission line maps obtained with our procedure together with contours of the dust continuum and [C\,{\sc{ii}}] ALMA emission. We see that we are able to fit the H$\alpha$ emission line in a significant fraction of the area of each individual object of our galaxy sample.

\section{Results}\label{sec:metal}

We use the emission line maps with the $\textit{S/N}$ threshold applied to produce metallicity maps for our galaxy sample. Beforehand, we corrected the observed emission line fluxes from extinction. An estimation of the extinction for each galaxy was obtained using the total H$\alpha$ and H$\beta$ fluxes integrated over the region where both emission lines are detected and assuming a theoretical H$\alpha$/H$\beta$ ratio of 2.89\footnote{For an electron temperature and density of T$_{\rm e}=9000$\,K and $\rm n_{\rm e}=100\,cm^{-3}$, respectively \citep{Luridiana2015}.}. We have considered other physical conditions that could be more representative of high redshift galaxies \citep{Isobe2023} and derive oxygen abundances assuming gas at T$_{\rm e}=15000$\,K and $\rm n_{\rm e}=200\,cm^{-3}$, and T$_{\rm e}=20000$\,K and $\rm n_{\rm e}=300\,cm^{-3}$. We find differences in oxygen abundances of less than 0.01\,dex. The Cardelli extinction law \citep{Cardelli1989} was then applied to correct the flux of each emission line\footnote{Differences with the Calzetti extinction law \citep{Calzetti2000ApJ} are less than 0.003\,mag in the fluxes of the optical emission lines analysed in this paper.}. We obtain $\rm A_{V}$ values between 0.06\,mag for DC-848185 (C02) and 0.2\,mag for DC-536534 (C03).  
Since the [O{\sc{iii}}]$\lambda$4363 auroral emission line is not detected at spatially resolved scales\footnote{The [O{\sc{iii}}]$\lambda$4363 emission line was not detected for any of our targets in the integrated spectrum obtained by adding all the pixels contained in the emission line maps shown in Figure\,\ref{halpha maps}.}, we use two different methods to derive the oxygen abundance: i) strong-line calibration methods derived for high-redshift galaxies \citep{Sanders2025,Cataldi2025}, and ii) photoionisation models that make use of different observed emission lines to robustly quantify oxygen abundances \citep[\textsc{HII-CHI-Mistry}, ][]{PerezMontero2014}.

\subsection{Strong-line metallicity calibrators}\label{sec:metal:strong}
With the advent of JWST NIRSpec observations, there has been an increasing number of works providing metallicity calibrations using strong emission line ratios \citep[e.g.,][]{Sanders,Sanders2025,Cataldi2025,Scholte2025,Backhaus2025}. In this paper we focus on the strong-line ratios involving the oxygen emission lines. Another widely used metallicity tracer for local studies and also applied to high-redshift galaxies is the [N{\sc{ii}}]/H$\alpha$ ratio \citep{Pettini2004}. However, this ratio is sensitive to the nitrogen-to-oxygen abundance ratio \citep{PerezMontero2009} and recent studies have shown a systematic enhancement of N/O at a fixed metallicity that requires further explanation \citep{Cataldi_NO2025,Arellano2025,Bhattacharya2025}. 
Due to this possible bias in the use of the [N{\sc{ii}}]/H$\alpha$ ratio \citep[see also][]{Scholte2025,Sanders2025}, we therefore decided to avoid it and focus on oxygen-based strong-line calibrations. In particular, we use the strong-line calibrations O2, O32, and R23
derived in \cite{Sanders2025} and \cite{Cataldi2025} where: 
\begin{equation}
\text{O2} = [\text{O}\textsc{ii}]\lambda\lambda3727,3729/\text{H}\beta,
\label{O2 relation}
\end{equation}
\begin{equation}
    \text{O32}=[\text{O}\textsc{iii}]\lambda5008/[\text{O}\textsc{ii}]\lambda\lambda3727,3729
\end{equation}  
\begin{equation}
    \text{R23}=([\text{O}\textsc{iii}]\lambda\lambda4959,5008+[\text{O}\textsc{ii}]\lambda\lambda3727,3729)/\text{H}\beta,
\end{equation}
We produced emission line ratio maps for O2, O32 and R23 using the corresponding extinction corrected fluxes. Only spaxels with fluxes above our $\textit{S/N}$ threshold (see Section~\ref{sec:EM}) were taken into account.  We then applied the corresponding metallicity calibrations presented in \cite{Sanders2025} and \cite{Cataldi2025}.\footnote{Several of the calibrations have multiple solutions, with some giving unrealistic oxygen abundances that were discarded. We checked among all the outcomes from the different calibrations and chose the corresponding root that gave consistent oxygen abundances across the calibrations.} Using the uncertainties from the emission line fitting and those reported for the calibrations, and propagating the uncertainties accordingly, we estimate average oxygen abundance uncertainties of 0.09-0.13, 0.04-0.06, and 0.08-0.43\,dex for O32, O2, and R23, respectively. The median values, average uncertainties, and standard deviations of the oxygen abundances for each galaxy are shown in Table\,\ref{tab:met_results} and Table\,\ref{ap:tab:met_results} for \cite{Sanders2025} and \cite{Cataldi2025} calibrations, respectively. The median oxygen abundances of each galaxy obtained with \cite{Sanders2025} and \cite{Cataldi2025} are in agreement taking into account the average uncertainties. This is reassuring when using one or the other calibration. \cite{Cataldi2025} calibrations are based on 16 star forming galaxies within $z\sim2-3$ and a compilation of galaxies with data in the literature, making a total of 128 galaxies. The sample from \cite{Sanders2025} are 33 galaxies spanning a wider range in redshifts ($z\sim1.4-7.2$, with a median of $z\sim$4.24), with a combination of data from literature that makes altogether a sample of 131 galaxies. Since the redshift from the galaxy sample in \cite{Sanders2025} is closer to the redshift of our galaxy sample, we decided to use \cite{Sanders2025} strong-line calibrations when comparing with the oxygen abundances derived from photoionisation models (see next section).

\subsection{Metallicity estimates with \textsc{HII-CHI-Mistry}}\label{sec:metal:HIICh}

\textsc{HII-CHI-Mistry} (hereinafter, HCM) is a suite of python scripts\footnote{\textsc{HII-CHI-Mistry} involves emission line ratios in different wavelength ranges. Full description of the code is in {\url {https://home.iaa.csic.es/~epm/HII-CHI-mistry.html}} and the git repository can be found in {\url {https://github.com/Estallidos/HII-CHI-Mistry}}.} designed to analyse spectroscopic observations to derive chemical abundances in the gas phase by using predictions from photoionisation models within extensive grids of different ionisation scenarios. HCM makes use of all the different observed emission lines and derives oxygen abundances with a robust statistical methodology. The validity of the results provided by HCM has been confirmed in several works \citep{PerezMontero2017,Ontiveros2021,PerezDiaz2022} and, in general, HCM gives consistent oxygen abundances\footnote{Uncorrelated differences with the physical conditions of the ionised gas, T$_{\rm e}$ and ionization parameter, are up to \,0.5\,dex.} as those predicted from auroral T$_{\rm e}$ direct abundance determinations \citep[see figures\,6 and\,7 in][]{Zurita2021}.
We run HCM for each spaxel of each galaxy presenting [O{\sc{ii}}]$\lambda\lambda$3727,3729, [O{\sc{iii}}]$\lambda\lambda$4959,5007, H$\beta$, H$\alpha$, and [N{\sc{ii}}]$\lambda\lambda$6583 emission line fluxes above $S/N=2$. The photoionisation models were built using a stellar population typical of a star-forming galaxy with a burst of 1\,Myr \citep[POPSTAR,][]{Molla2009} and the IMF from \citet{Chabrier2003}. Making use of a different stellar population (similar to extreme emission line galaxies) did not significantly change the resulting oxygen abundances. Average uncertainties in the oxygen abundances estimates are 0.07-0.11, with standard deviations in the metallicity distribution for each galaxy being larger than the average errors (see the last two columns of Table\,\ref{tab:met_results}).    
\begin{table*}
\caption{Oxygen abundances (12 + log (O/H)) using strong-line calibrations from \citet{Sanders2025} and \textsc{HII-CHI-Mistry}.}
\label{tab:met_results}      
\centering 
\begin{tabular}{ccccccccccccc}
\toprule
 & \multicolumn{9}{c}{\citet{Sanders2025} } & \multicolumn{3}{c}{\textsc{HII-CHI-Mistry}} \\
\cmidrule(lr){2-10}
Diagnostic & \multicolumn{3}{c}{O2} & \multicolumn{3}{c}{O32} & \multicolumn{3}{c}{R23} & \multicolumn{3}{c}{} \\
\midrule
Galaxy &  median  & av. err. & std &  median & av. err. & std &  median & av. err. & std &  median & av. err. & std  \\
\midrule
C02 & 8.11 & 0.12 & 0.28 & 8.25 & 0.05 & 0.16 & 8.28 & 0.32 & 0.47 & 8.27 & 0.07 & 0.15\\
C03 & 8.03 & 0.09 & 0.28 & 8.25 & 0.06 & 0.27 & 8.20 & 0.25 & 0.35 & 8.14 & 0.09 & 0.20 \\
C04a & 8.21 & 0.12 & 0.24 & 8.29 & 0.06 & 0.18 & 8.21 & 0.43 & 0.52 & 8.28 & 0.09 & 0.17 \\
C07ab & 8.10 & 0.11 & 0.29 & 8.17 & 0.06 & 0.21 & 8.29 & 0.29 & 0.37 &
8.21 & 0.07 & 0.16 \\
C11 & 8.22 & 0.12 & 0.25 & 8.30 & 0.06 & 0.14 & 8.20 & 0.31 & 0.48 & 8.21 & 0.09 & 0.17 \\
C13 & 8.11 & 0.13 & 0.25 & 8.32 & 0.06 & 0.19 & 8.56 & 0.25 & 0.36 & 8.25 & 0.11 & 0.18 \\
\bottomrule
\end{tabular}      
\tablefoot{For each calibration we show the median values, average uncertainties, and standard deviations of the oxygen abundances for the spaxel distribution in each galaxy.}           
\end{table*}

\subsection{Strong-line metallicity callibrations versus \textsc{HII-CHI-Mistry}}

In Table\,\ref{tab:met_results} we show the median values for each galaxy obtained using the strong-line calibrations from \citet{Sanders2025} and those derived from HCM. Except C03 and C13, for which the abundances derived from O2 and O32 do not overlap taking into account the average uncertainties, the rest of the galaxies show consistent abundances within the errors for the three calibrations. In general, O2 tends to give slightly lower values than O32 and R23, and R23 gives metallicities with larger average uncertainties. The median abundances derived using HCM exhibit lower uncertainties and agree with those derived using \citet{Sanders2025} calibrations.

\begin{figure*}
    \
    \centering
    \includegraphics[width=0.32\linewidth]{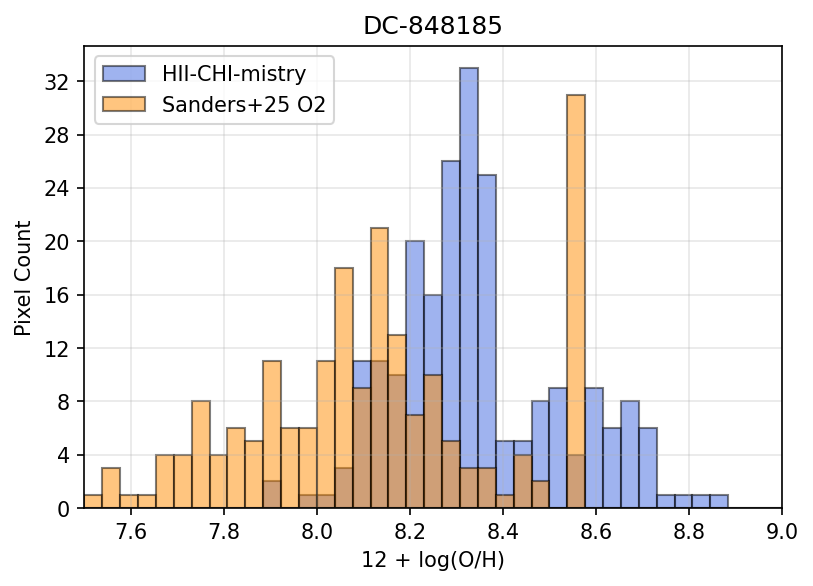}
    \includegraphics[width=0.32\linewidth]{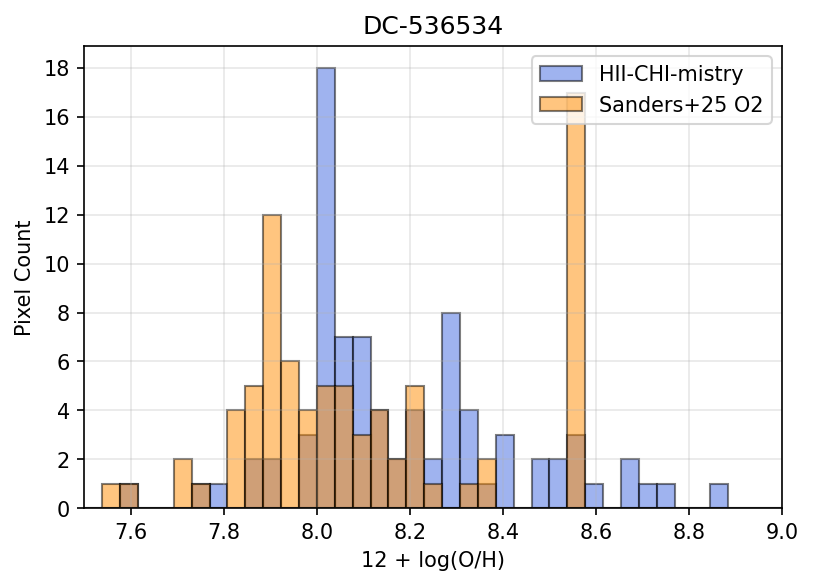}
    \includegraphics[width=0.32\linewidth]{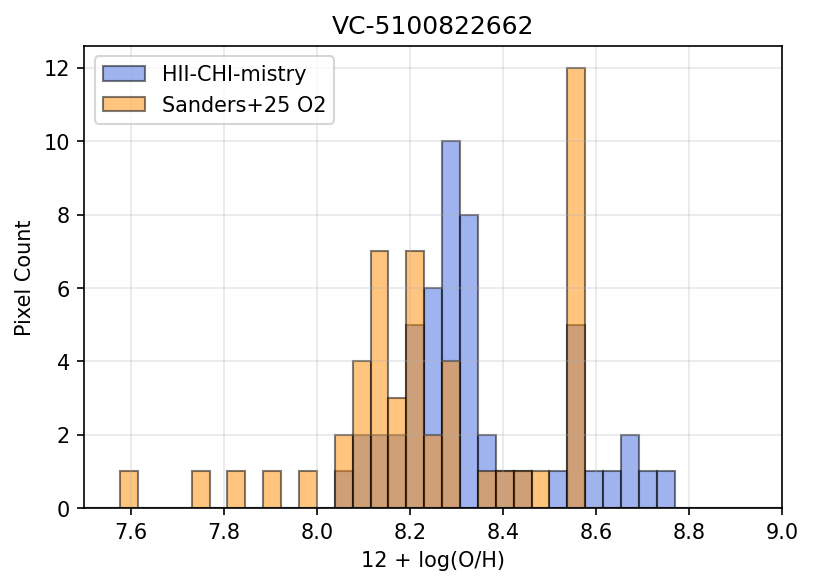}
\\[2mm]
    \includegraphics[width=0.32\linewidth]{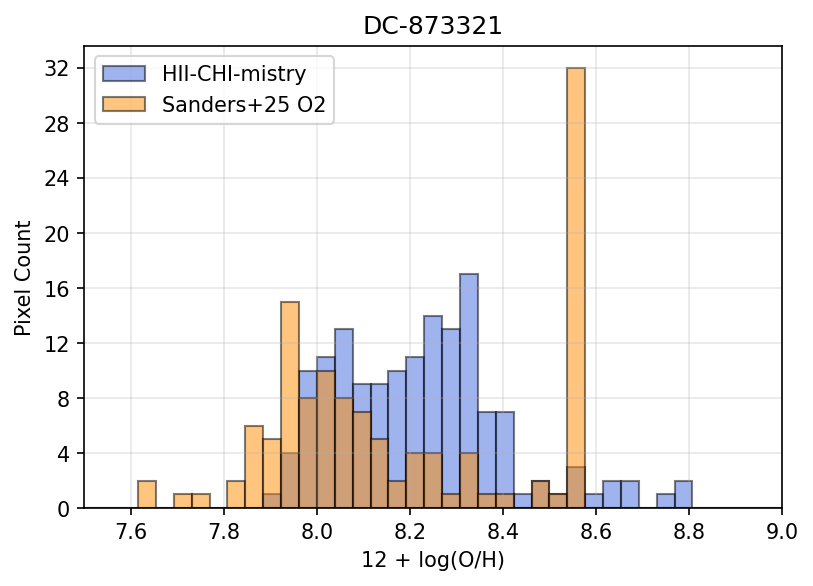}
    \includegraphics[width=0.32\linewidth]{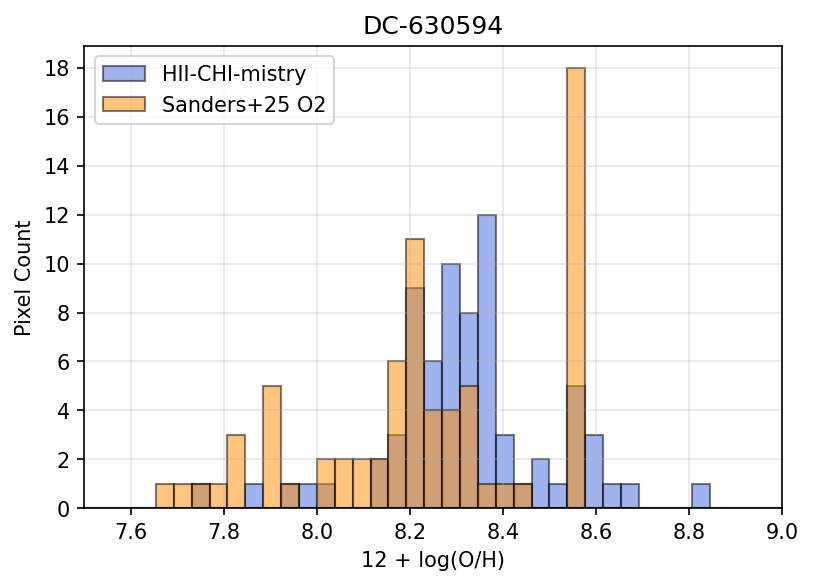}
    \includegraphics[width=0.32
    \linewidth]{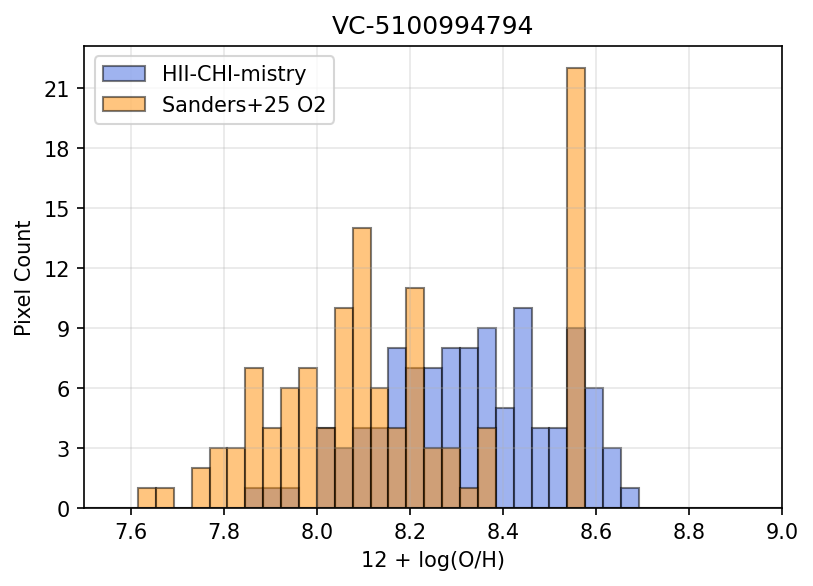}
    \caption {Comparison of the oxygen abundance distribution obtained using \textsc{HII-CHI-Mistry} (blue) \citep{PerezMontero2014} and that derived using the O2 strong-line calibrator from \citet{Sanders2025} (orange).}
    \label{fig:met_strongline_h2chemistry}
\end{figure*}

\begin{figure*}
    \
    \centering
    \includegraphics[width=0.32\linewidth]{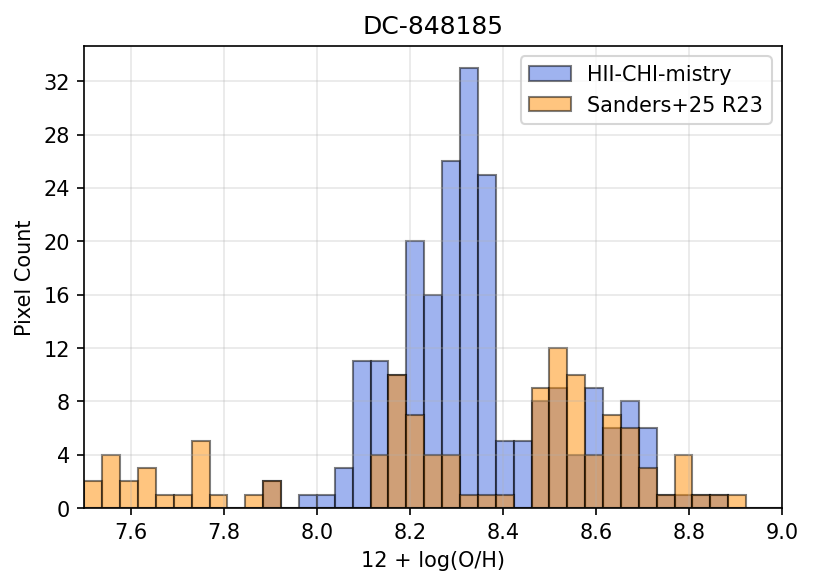}
    \includegraphics[width=0.32\linewidth]{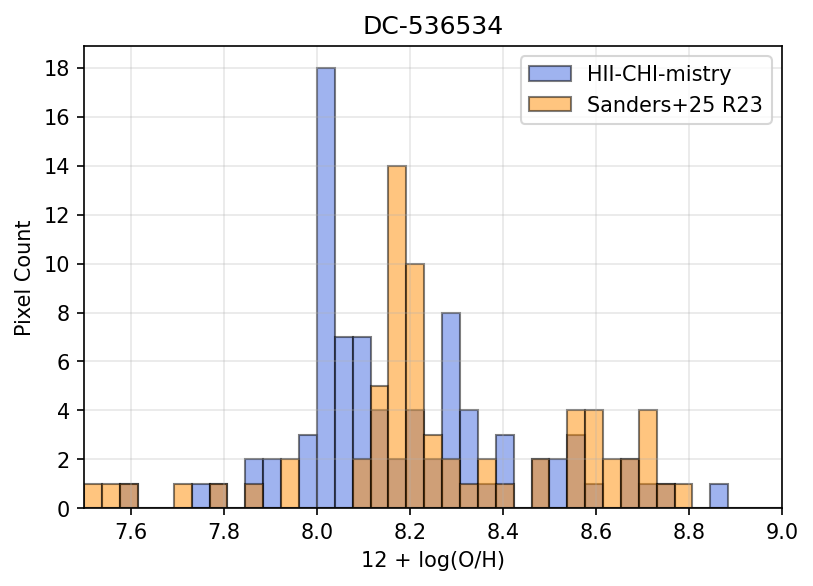}
    \includegraphics[width=0.32\linewidth]{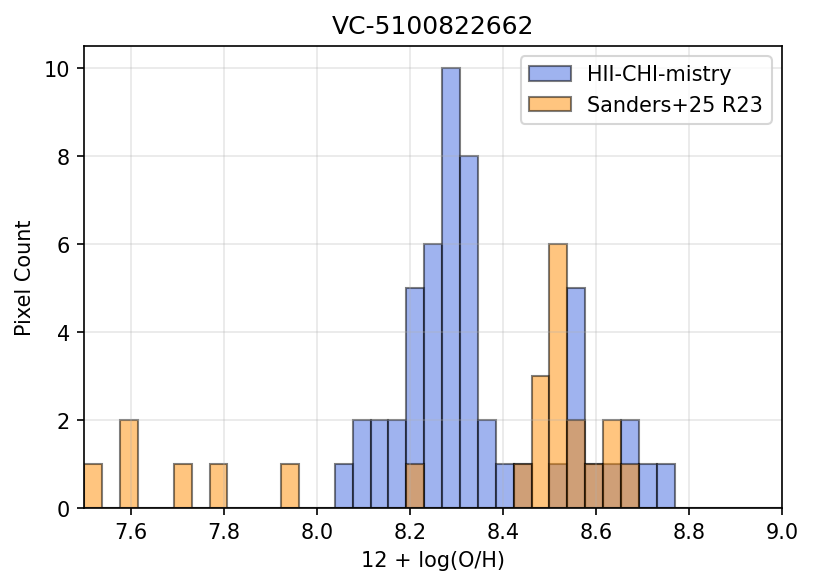}
\\[2mm]
    \includegraphics[width=0.32\linewidth]{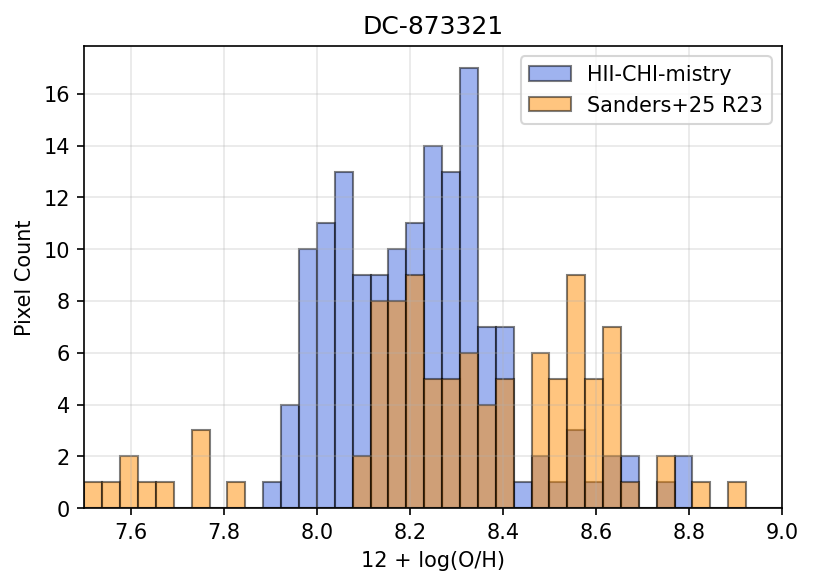}
    \includegraphics[width=0.32\linewidth]{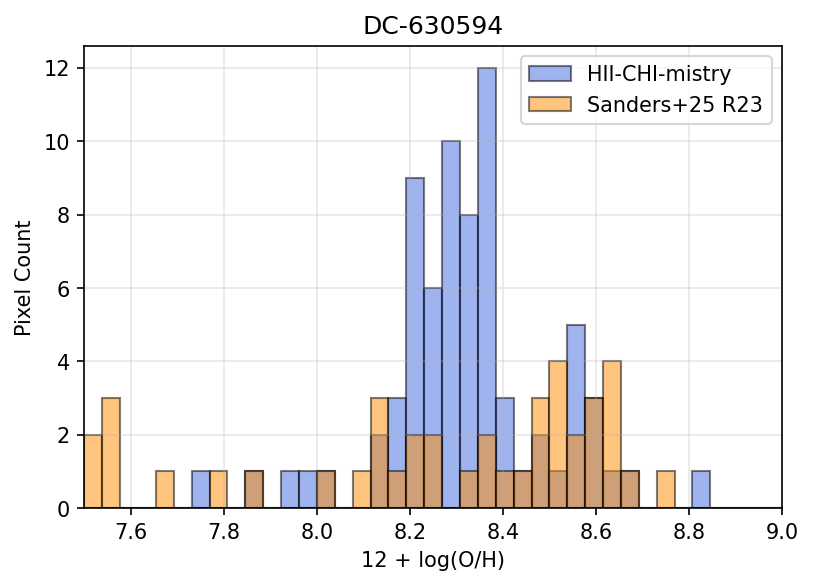}
    \includegraphics[width=0.32
    \linewidth]{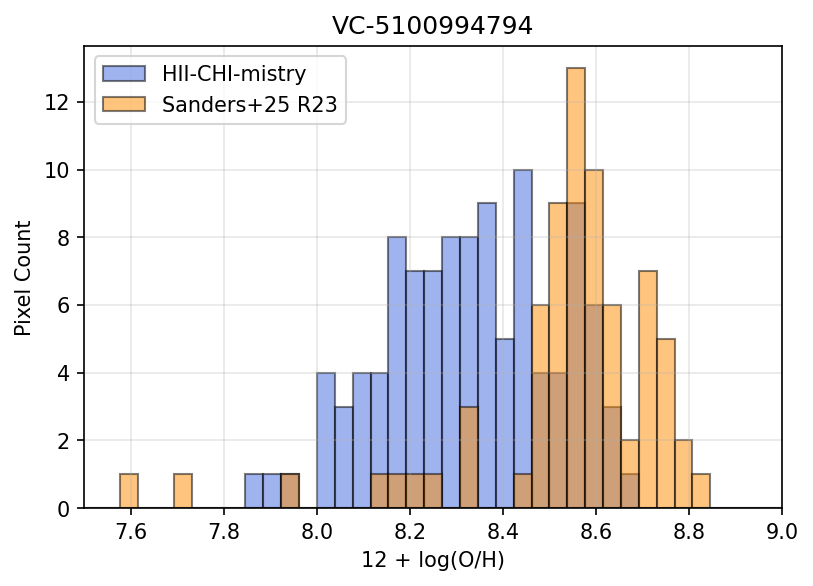}
    \caption {The same as in Figure\,\ref{fig:met_strongline_h2chemistry} but using the R23 calibrator from  \citet{Sanders2025}.}
    \label{fig:met_strongline_h2chemistry_R23}
\end{figure*}

\begin{figure*}
    \
    \centering
    \includegraphics[width=0.32\linewidth]{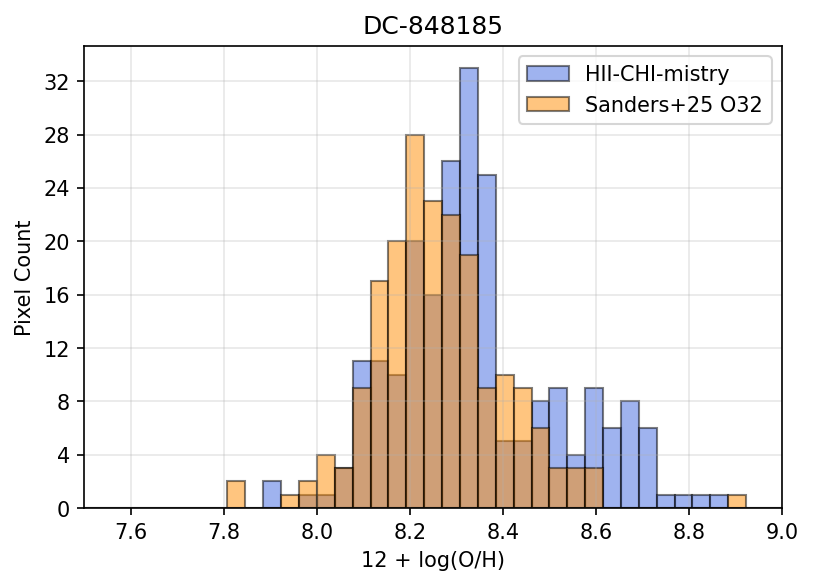}
    \includegraphics[width=0.32\linewidth]{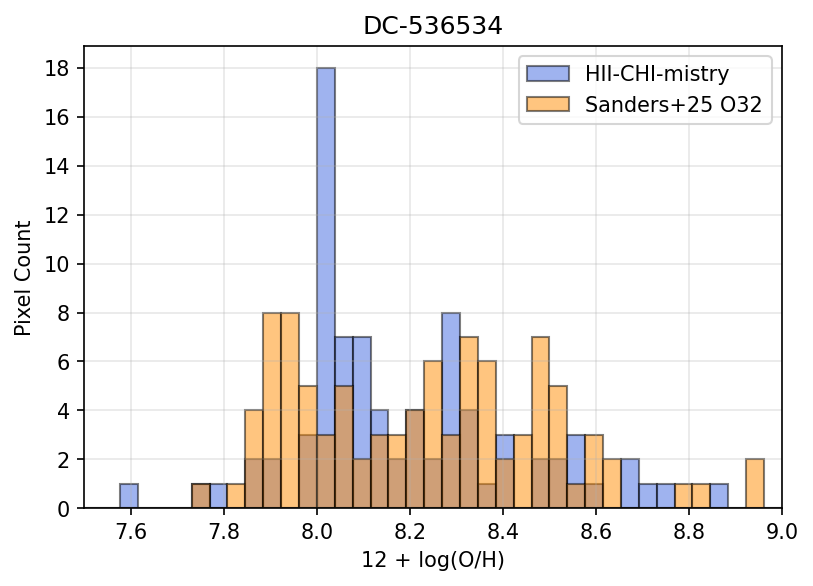}
    \includegraphics[width=0.32\linewidth]{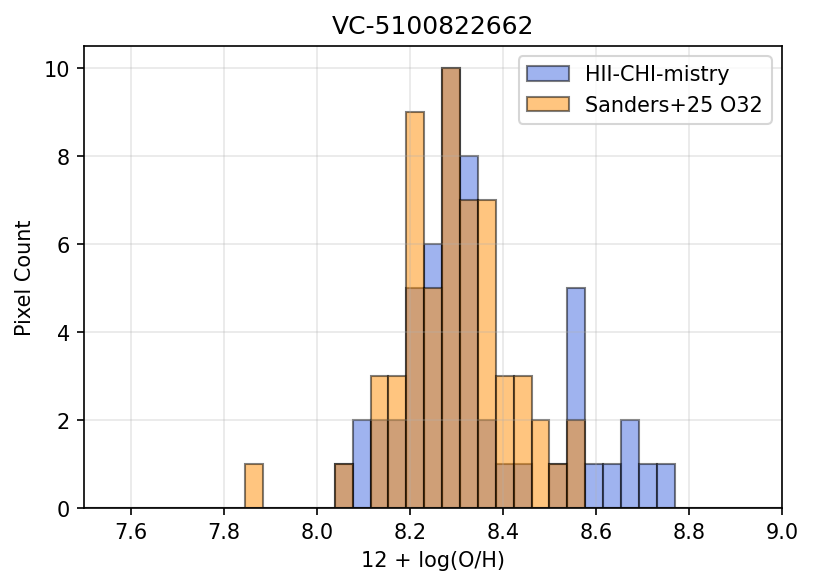}
\\[2mm]
    \includegraphics[width=0.32\linewidth]{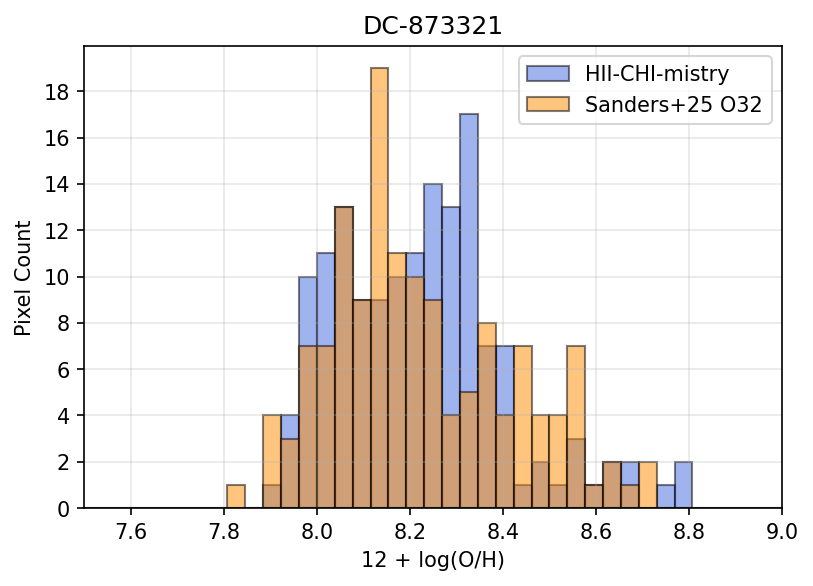}
    \includegraphics[width=0.32\linewidth]{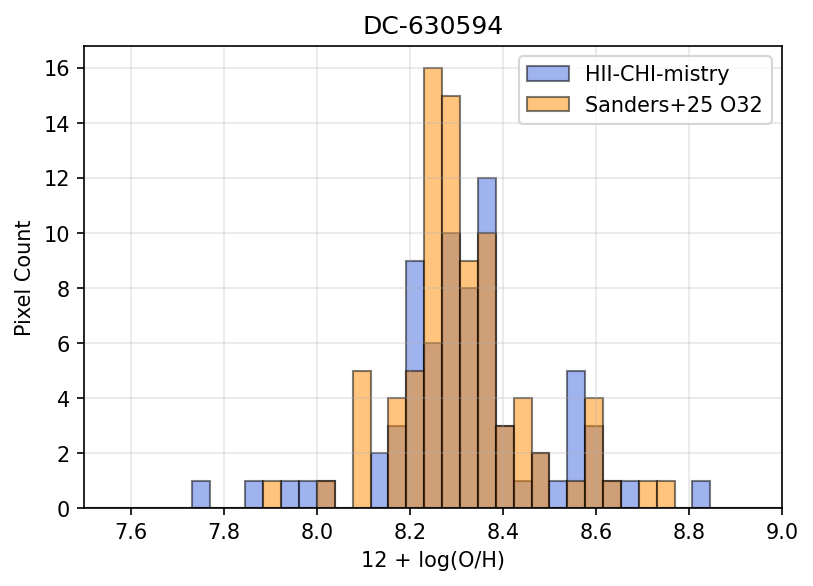}
    \includegraphics[width=0.32
    \linewidth]{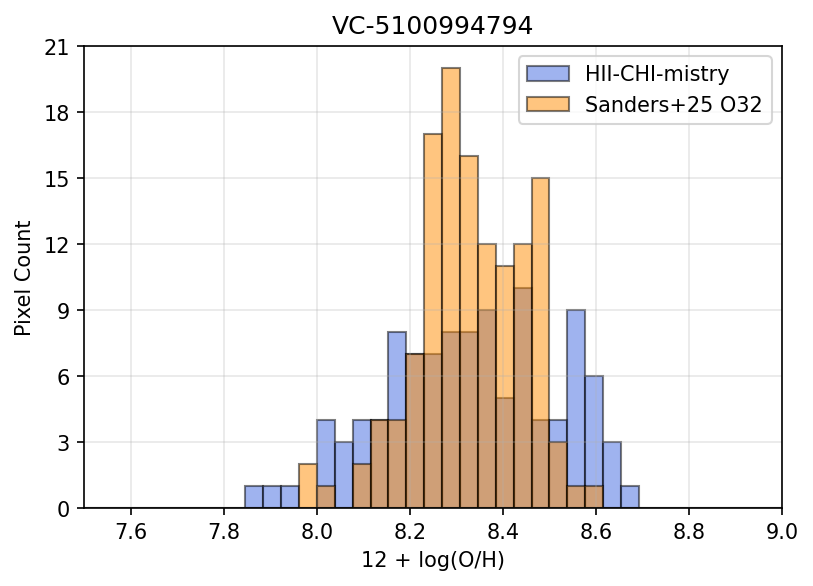}
    \caption {The same as in Figure\,\ref{fig:met_strongline_h2chemistry} but using the O32 calibrator from  \citet{Sanders2025}.}
    \label{fig:met_strongline_h2chemistry_O32}
\end{figure*}

In Figures\,\ref{fig:met_strongline_h2chemistry}, \,\ref{fig:met_strongline_h2chemistry_R23}, and\,\ref{fig:met_strongline_h2chemistry_O32} we compare the oxygen abundance distribution for each galaxy obtained from HCM with the abundances derived using O2, R23, and O32 calibrations from \citet{Sanders2025}, respectively. This comparison gives us further insight into the use of the different calibrations. The metallicity distribution for O2 tends to give lower values than those derived from HCM. In particular, the peak in the histogram at 12 + log(O/H)\,$\sim$\,8.55 shows the limitation in using O2 for ratios above log(O2)\,$\gtrsim$\,0.5 as the O2 calibration flattens for abundances above 8.4 \citep[see top-right panel of figure\,9 in][]{Sanders2025}. The R23 calibration in \citet{Sanders2025} shows a wide range of oxygen abundances. We note here that this calibrator is not sensitive to oxygen abundances between 12 + log(O/H)\,$\sim$\,7.6-8.4 due to the flat dependence of log(R23) with 12 + log(O/H) in this metallicity range \citep[see bottom-left panel of figure\,9 in][]{Sanders2025}. As we can see in Figure\,\ref{fig:met_strongline_h2chemistry_R23}, the metallicity distribution derived using R23 at 12 + log/O/H)\,$\gtrsim$\,8.4 agrees relatively well with that  obtained from HCM. The O23 calibration gives results with a similar distribution as those derived from HCM for the whole metallicity range we are probing. This reinforces the usage of the O23 calibration for our galaxy sample among these three classical oxygen strong-line calibrations that are typically used in the literature.  

Considering the limitations of the strong-line metallicity calibrations we have shown above, for this work we take the oxygen abundances derived from HCM as our fiducial ones. 
Moreover, previous studies in the literature \citep[e.g.][]{Zurita2021} show the consistency between the metallicity derived from HCM and those obtained using the direct method, therefore we consider that applying HCM represents a more reliable methodology than the strong-line calibrations.   

The median metallicities derived using HCM (column 11 in Table\,\ref{tab:met_results}) agree within the errors with the metallicities obtained from strong-line calibrations applied to integrated emission line fluxes \citep[see Table\,A.2 in][]{Faisst2026_survey}. Moreover, for the three galaxies with $\rm T_{\rm e}$ determination (DC-848185 (C02), DC-630594 (C011) and VC-5100994794 (C13)), the median metallicities reported here agree within the errors with those derived from the $\rm T_{\rm e}$-direct method \citep{Faisst2025_massmetal}. 

\subsection{Metallicity maps}

\begin{figure*}
    \
    \centering
    
    \includegraphics[width=\linewidth]{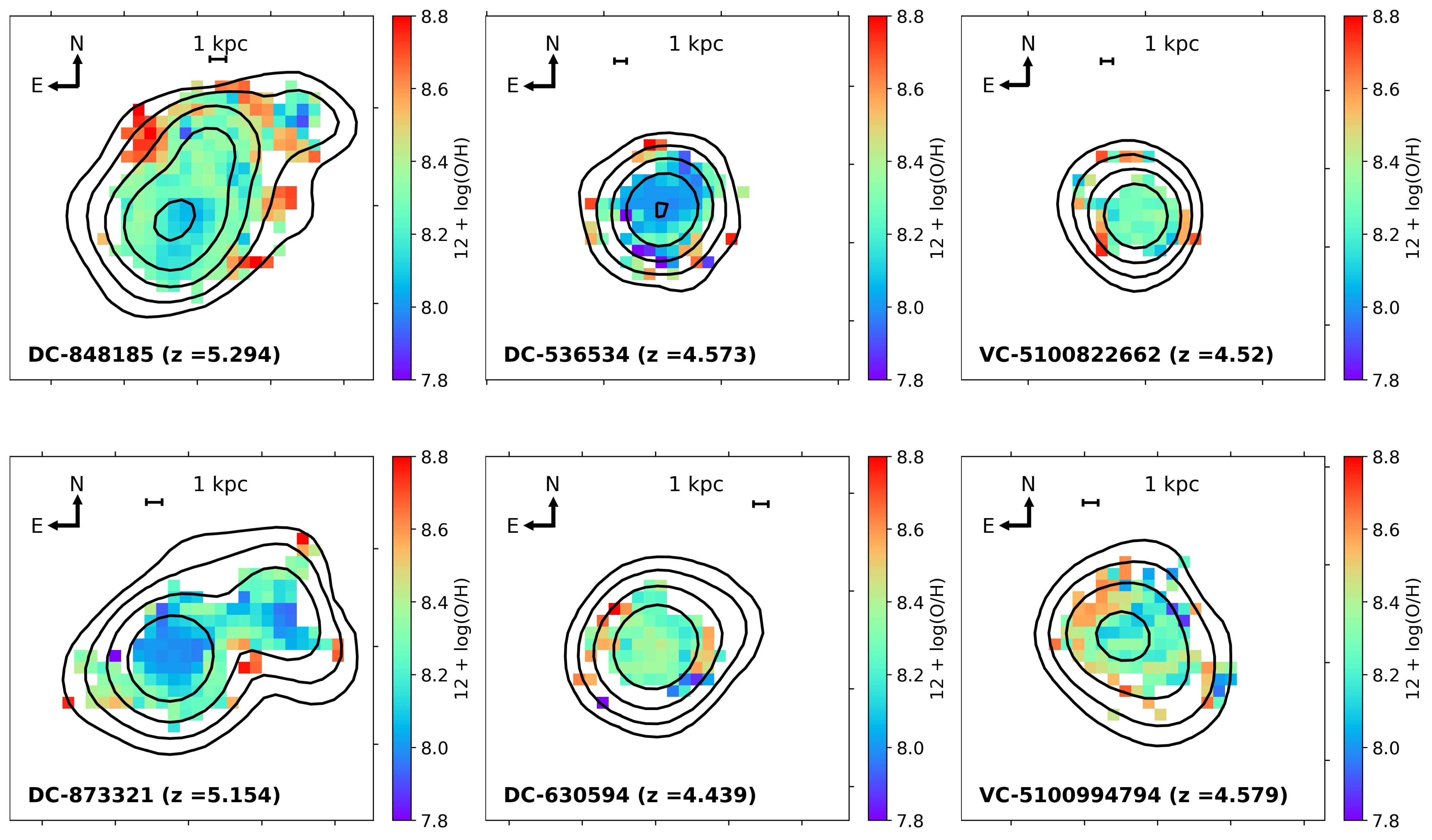}
    
    \caption {Oxygen abundance maps for our sample of CRISTAL galaxies using HCM (see Section\,\ref{sec:metal:HIICh}). The contours trace the H$\alpha$ emission derived from the emission-line fitting, starting at 3$\sigma$ and increasing in powers of three. A 1\,kpc bar is depicted to show the linear scales at which we see differences in the oxygen abundance.}
    \label{fig:metHCMplusHa}
\end{figure*}

In Figure\,\ref{fig:metHCMplusHa} we show the oxygen abundance maps obtained using HCM overlaid with H$\alpha$ contours. The values of 12 + log(O/H) for our galaxies range from 12 + log(O/H)\,$\sim$\,7.8 to 12 + log(O/H)\,$\sim$\,8.8, with local variations of $\sim$\,0.3-0.4\,dex, larger than the uncertainties in the metallicity derivations (see Table\,\ref{tab:met_results}). In general, the oxygen abundances shown in this figure are lower in the centre of the galaxies where star formation occurs and the H$\alpha$ emission peaks. In the outer parts of the galaxy the oxygen abundance tends to have larger values. This trend is particularly clearly seen for DC-848185 (C02), DC-536534 (C03) and DC-873321 (C07ab), where differences in the metallicity at the peak of  H$\alpha$ emission and the outer parts are more pronounced than for the other three galaxies. Interestingly, these galaxies have observed signatures of outflows \citep[][]{Ren2025} and interactions \citep{Romano2021}. The enhancement of oxygen abundances detected in our galaxy sample is in line with the positive metallicity gradients found in these galaxies by \citet{Lee2026}. Indeed, these authors show positive metallicity gradients for all our galaxies, with DC-536534 (C03) and DC-873321 (C07ab) showing  the steepest and DC-873321 (C11) and VC-5100994794 (C13) presenting the flattest, which agrees with the observed trend seen in the metallicity maps shown in Figure\,\ref{fig:metHCMplusHa}. 

A similar anticorrelation between metallicity and H$\alpha$  has been obtained from metallicity maps in galaxies at lower redshifts \citep[$z\sim3.4$,][]{Troncoso2014}.
Although the decrease of metallicity at the peak of H$\alpha$ emission is more pronounced for galaxies presenting broad emission lines indicative of strong outflows, we cannot rule out the possibility that inflow of pristine gas supporting star-formation could cause metal dilution in the centres of these galaxies \citep[][]{SanchezAlmeida2014,Troncoso2014,PerezDiaz2024NatAs}.

\begin{figure*}
    \centering
    \includegraphics[width=\linewidth]{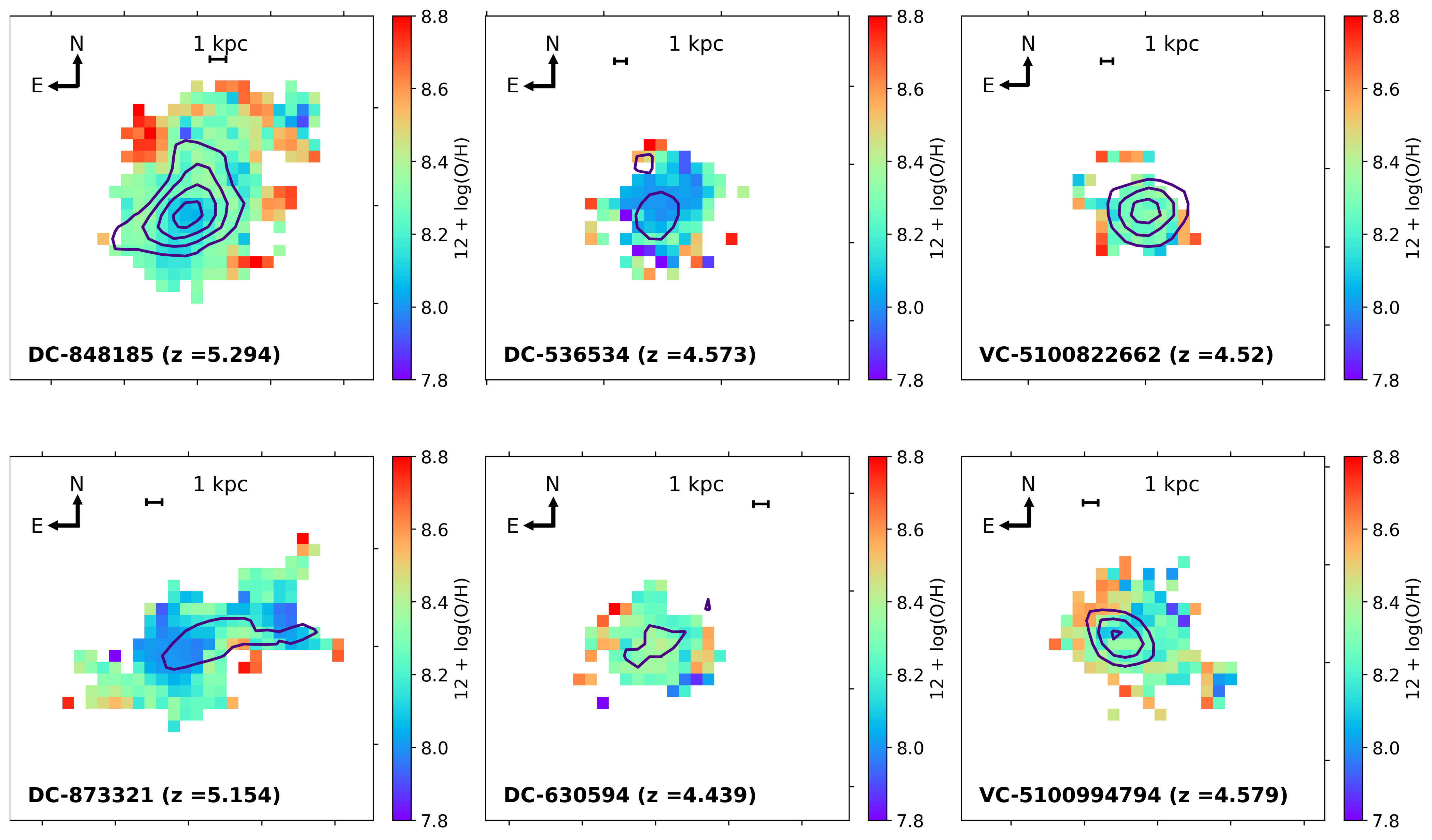}
    \caption {Oxygen abundance maps for our sample of CRISTAL galaxies obtained using HCM (see Section\,\ref{sec:metal:HIICh}). Contours correspond to the dust continuum obtained from the emission line fitting. Contours start at 3$\sigma$ and increase in steps of 2$\sigma$. A 1\,kpc bar is depicted to show the linear scales at which we see differences in the oxygen abundance.}
    \label{fig:metHCMplusdust}
\end{figure*}

Surprisingly, the minima in metallicity correspond to maxima in the dust continuum emission (see Figure\,\ref{fig:metHCMplusdust}) for DC-848185 (C02), DC-536534 (C03) and DC-873321 (C07ab). For C04 (VC-5100822662), DC-630594 (C11), and VC-5100994794 (C13) a similar trend could be slightly observed, but the depth in metallicity at the centres of these galaxies is not so pronounced, thus making it difficult to draw any conclusion in this regard.  

For DC-848185 (C02) we find a trend of increasing metallicity towards the northeastern side of the main body of the galaxy. This area of enhanced metallicity could be spatially related to the biconical strong outflow detected in this galaxy \citep{Davies2026}. DC-536534 (C03) is the galaxy showing the lowest values of oxygen abundances, with a general uniform value below 12+log(O/H)$\sim$8.0. We will comment further on these galaxies in Section\,\ref{sec:scalingdust}. 

\subsection{Gas and dust mass map estimates}\label{sec:gascal}
In this section, we derive gas and dust mass maps that we will use in combination with the previously obtained metallicity maps to produce spatially resolved scaling relations. 

For the gas mass estimates, we will use the 158\,$\mu$m emission line coming from ionized carbon, which is the brightest line in the FIR wavelength range. The [C\,{\sc{ii}}]158\,$\mu$m emission line traces all the gas phases in the ISM, but predominantly the cold neutral medium and photodissociation regions (PDRs) \citep{Wolfire2022}. We will use this emission line to estimate the amount of atomic and molecular gas content in our galaxies. 

For the case of the H{\sc{i}} gas content we adopt the calibration given in  \cite{Heintz_2021} derived from measurements in the line of sight through the ISM of galaxies at $z\gtrsim2$ hosting $\gamma$-ray bursts within a range of metallicity of $Z\sim$\,0.01-0.5\,$Z_{\odot}$\footnote{12 + log (O/H) = 8.69 for log (Z/Z$_{\odot}$) = 0 \citep{Asplund_2009}.}:
\begin{equation}\label{eq:heintz21}
    \log \left(\frac{{L}_{\text{CII}}}{\text{L}_{\odot}}\right) = (0.87 \pm 0.09) \log \left(\frac{\text{Z}}{\text{Z}_{\odot}}\right) - (1.48 \pm 0.12) + \log \left(\frac{M_{\rm HI}}{\text{M}_{\odot}}\right),
\end{equation}
where M$_{\rm H\textsc{i}}$ and $L_{\text{CII}}$ are in units of M$_{\odot}$ and L$_{\odot}$, respectively.

We use the metallicity maps derived in the previous section and the [C\,{\sc{ii}}] emission line maps from \citet{Li_2024} to compute the atomic gas mass maps for each galaxy. 

For estimating the molecular H$_{2}$ gas mass, we use the average [C\,{\sc{ii}}]-to-H$_{2}$ ratio found in simulations by \citet{Vizgan_2022} for galaxies at $z\sim$\,6 with mean values for the stellar mass, SFR and metallicity of 10$^{8.07}$  M$_{\odot}$, 1.9\,M$_{\odot}$\,yr$^{-1}$, and 0.18\,$Z_{\odot}$, respectively. The calibration given in \citet{Vizgan_2022} is:
\begin{equation}\label{eq:vizgan22}
    \frac{M_{\rm H_{2}}}{\text{M}_{\odot}} = (18 \pm 10) \frac{L_{\text{CII}}}{\text{L}_{\odot}}.
\end{equation}

There is evidence that these [C\,{\sc{ii}}]-to-H$_{2}$ and [C\,{\sc{ii}}]-to-H{\sc{i}} conversion factors could overestimate the total gas mass \citep{Palla2024,Algera2026} (see Section\,\ref{sec:bestgasmass}). We, therefore, also apply a metallicity-dependent spatially resolved [C\,{\sc{ii}}] conversion factor derived in \citet{Vallini_25}. This conversion factor was derived using galaxy simulations within $z\sim4-9$ with stellar masses M$_{\star}$>10$^{8.07}$\,M$_{\odot}$ and SFR$\sim0.01-70$\,M$_{\odot}$\,yr$^{-1}$. The calibration to obtain the total gas mass from the [C\,{\sc{ii}}] line luminosity in \citet{Vallini_25} is:
\begin{equation}\label{eq:vallini25}
    \text{log}W_{\text{[CII]}}=-0.324\text{log}\left(\frac{Z}{Z_{\odot}}\right)-0.355\text{log}\left(\frac{\Sigma_{\text{[CII]}}}{\rm L_{\odot}\rm kpc^{-2}}\right)+3.37,
\end{equation}
where $\log W_{[\rm CII]}$ is the spatially resolved [C\,{\sc{ii}}]-to-gas conversion factor in units of $\rm M_{\odot}\,kpc^{-2}/(L_{\odot}\,kpc^{-2})$. $W_{[\rm CII]}$ is not a constant value for the entire galaxy, but it changes with the metallicity and the [C\,{\sc{ii}}] surface density at each location within the galaxy. 

We will explore in Section\,\ref{sec:bestgasmass} which of these calibrations more robustly estimates the gas mass surface densities for our galaxy sample. We note here that \citet{Lee2025} derived gas masses using the ALMA Band-7 dust continuum emission for the same galaxy sample. In Figure\,\ref{ap:fig:compfgas} of the appendix we compare the gas mass fractions derived using \citet{Vallini_25} calibrations with those derived in \citet{Lee2025}, finding agreement in all the galaxies except two that deviate slightly from the one-to-one relation.

As for the dust mass maps, they were obtained making use of the dust continuum observations from ALMA provided in \citet{Li_2024}. We rely on a novel methodology presented in Rela\~no et al. (in prep) that exploits the relation between the SFR surface density and the dust temperature observed in local galaxies \citep[see][]{Chiang}. We summarize briefly here the methodology to estimate the dust masses presented in that work. Firstly, we recalibrated the relation between the SFR surface density and the dust temperature for high redshift galaxies using dust temperature estimates from the literature derived with multi-band FIR observations. Then, we use the SFR surface density maps from \citet{Li_2024} to estimate the dust temperature applying the recalibrated $\Sigma_{\rm SFR}$-T$_{\rm dust}$ relation. Finally, using the dust continuum observations, dust mass surface density maps are derived by assuming that the dust continuum is described by a modified black body in the optically thin approximation. We refer the reader to Relaño et al. (in prep) for a detailed description of the methodology and a comparison with other methods used in the literature to estimate dust masses from single-band ALMA observations. For this work, we derive dust temperature maps for all the pixels with S/N above 3 in the dust continuum maps. We find dust temperatures between $\sim$\,30\,K and $\sim$60\,K, similar to what has been obtained for the ALPINE galaxies in \citet{Sommovigo2022b}. We will make use of the dust mass surface density maps derived for our objects with this methodology to analyse the spatially resolved scaling relations that will be presented in the next section.

\begin{table*}
\caption{Average values of the per-pixel logarithmic dust-to-gas, dust-to-star and dust-to-metals ratio for our galaxy sample.}
\label{tab:properties}      
\centering          
\begin{tabular}{ccccccccc}     
\hline\hline
CRISTAL & Alternative & $\log\Sigma_{\rm dust}$ & $\log\Sigma_{\rm gas}$ &  $\langle\log(\Sigma_{\rm dust}/\Sigma_{\rm gas})\rangle$ & $\langle\log(\Sigma_{\rm dust}/\Sigma_{\rm star})\rangle$ & $\langle\log(\Sigma_{\rm dust}/\Sigma_{\rm Z})\rangle$ & pixels & Area \\
ID & Name & [$\rm M_{\odot}/pc^{2}$] & [$\rm M_{\odot}/pc^{2}$] &  &  & & & kpc$^{-2}$  \\

\hline \\
        C02  & DC-848185 & 1.63$^{+0.01}_{-0.01}$ & 4.37$^{+0.02}_{-0.02}$ & -2.69 $^{+0.21}_{-0.16}$ & -2.91$^{+0.13}_{-0.13}$  &  -0.53$^{+0.17}_{-0.14}$ & 66 & 35.40 \\\\
        C03  & DC-536534 & 0.90$^{+0.04}_{-0.04}$ & 3.29$^{+0.07}_{-0.05}$ &  -2.52$^{+0.23}_{-0.17}$ & -2.65$^{+0.17}_{-0.14}$ & -0.30$^{+0.16}_{-0.14}$ & 12 & 9.32 \\\\
        C04a  & VC-5100822662 & 1.46$^{+0.02}_{-0.02}$ & 3.67$^{+0.04}_{-0.03}$ &  -2.29$^{+0.21}_{-0.15}$ & -2.01$^{+0.13}_{-0.12}$& -0.31$^{+0.15}_{-0.13}$ & 32 & 31.27 \\\\
        C07ab  & DC-873321  & 1.39$^{+0.02}_{-0.02}$ &  3.61$^{+0.04}_{-0.03}$ & -2.20$^{+0.24}_{-0.19}$ & -2.53$^{+0.15}_{-0.15}$ & -0.22$^{+0.17}_{-0.16}$ & 30 & 16.53 \\\\
        C11  & DC-630594   & 0.84$^{+0.04}_{-0.04}$ & 3.45$^{+0.06}_{-0.04}$ &  -2.61$^{+0.23}_{-0.18}$  & -2. 58$^{+0.16}_{-0.16}$ & -0.53$^{+0.21}_{-0.18}$ & 11 & 6.99 \\\\
        C13  & VC-5100994794 & 1.{26}$^{+0.02}_{-0.02}$ & 3.67$^{+0.05}_{-0.03}$ & -2.41$^{+0.22}_{-0.16}$ & -2.21$^{+0.10}_{-0.12}$ & -0.36$^{+0.16}_{-0.14}$ & 21 & 12.97 \\
\hline\hline                 
\end{tabular}
\tablefoot{Note that the area considered for each galaxy (and presented in column\,8 and\,9 in pixels and kpc$^{2}$, respectively) is determined by the area for which dust mass can be derived, corresponding to the pixels with S/N higher than 3 in the dust continuum observations. Gas surface mass densities were derived using the spatially resolved [C{\sc ii}] conversion factor given in \citet{Vallini_25} (see Section\,\ref{sec:bestgasmass} for more details).
}
\end{table*}

\section{Spatially resolved scaling relations for high-z galaxies}\label{sec:resolved_scaling_relations} 

 We study in this section the relation between dust-to-gas, dust-to-stellar, and dust-to-metals surface mass density ratios with the metallicity and gas mass fraction at local spatially resolved scales. The main goal is to explore how dusty our high-redshift objects are compared to galaxies in the local Universe and how their dust and gas content relate to the metal abundance within the galaxy. Since these galaxies are expected to undergo non-secular processes such as mergers and/or strong outflows, we would like to relate the resolved metal and dust scaling relations with their evolutionary stage and possibly with their metal and gas retention and redistribution. 

For this purpose, we compare our spatially resolved relations with those obtained for a set of star-forming disk galaxies in the local Universe \citep{Casasola}. Using a sample of 18 large spiral disk galaxies from the DustPedia sample, \cite{Casasola} derived spatially resolved dust mass, stellar mass and SFR surface density maps to investigate scaling relations in a range of scales between 0.3 and 3.4\,kpc. They used metallicity radial gradients to compute azimuthally averaged metallicity maps and analysed how the dust and gas masses relate with the metal content at local scales. These scaling relations, which involve dust, gas-phase metallicity, stellar mass, and SFR, offer insights into the physical processes that govern galaxy evolution on subgalactic scales. Given the established correlations observed in local galaxies, it is essential to determine whether these scaling relations hold in high-redshift galaxies as well.

\subsection{Atomic, molecular and total gas mass from [C\,{\sc{ii}}] observations}\label{sec:bestgasmass}

In Section\,\ref{sec:gascal} we present two different calibrations from the literature that allow gas mass estimations from the [C\,{\sc{ii}}] emission line. We explore here which would be the most suitable one for our galaxy sample. 

\begin{figure*}
\centering
    {\includegraphics[width=0.45\linewidth]{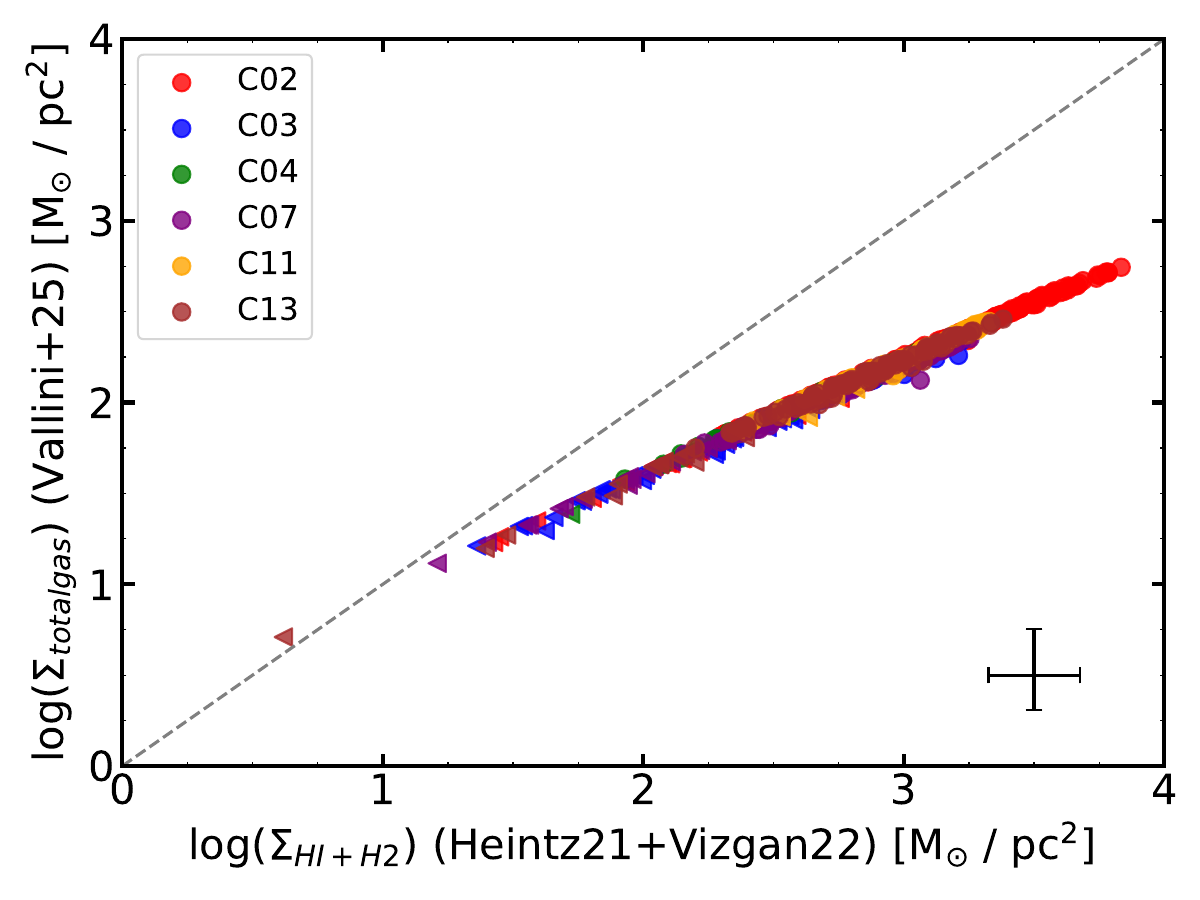}}
    {\includegraphics[width=0.45\linewidth]{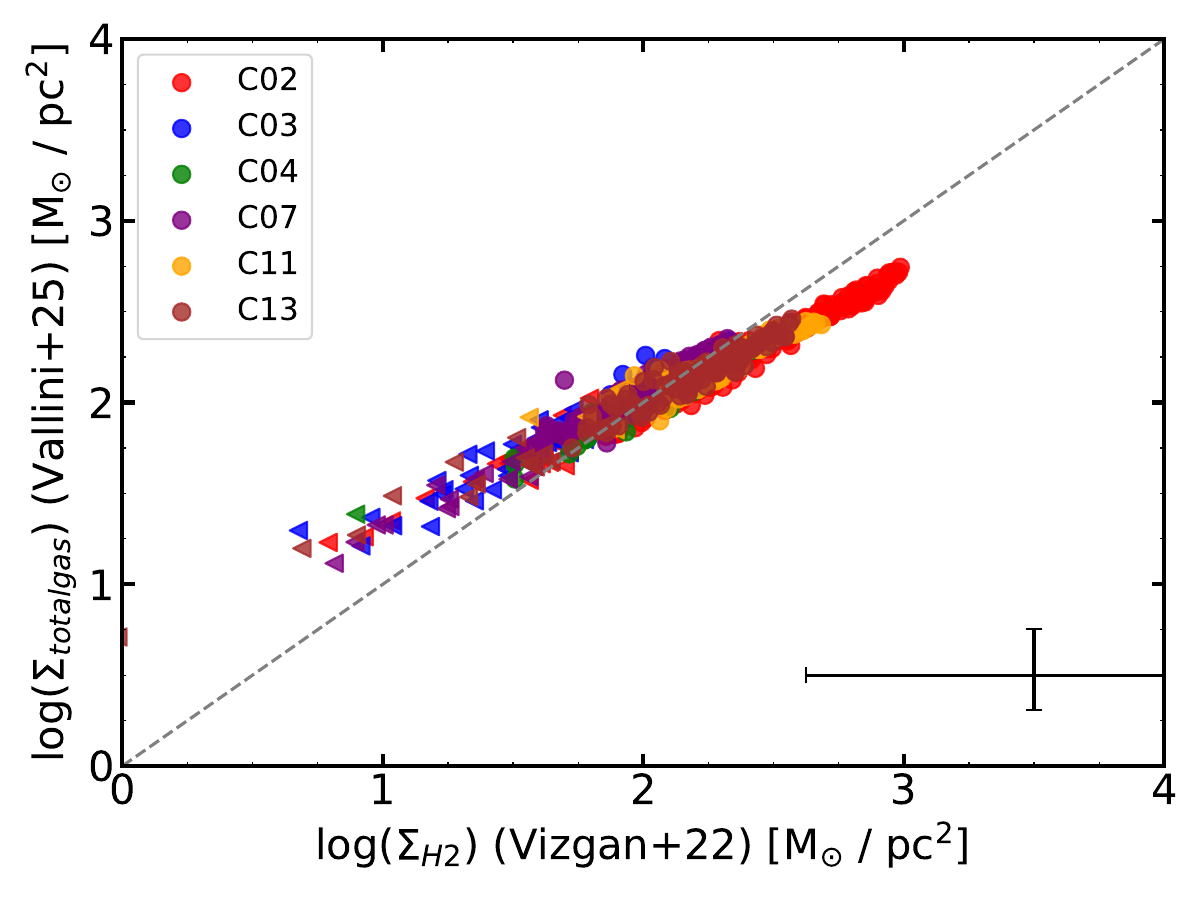}}
 \caption{Left: Comparison of the total gas mass surface density obtained using \citet{Vallini_25}  and the total gas mass surface density derived as a combination of the atomic gas mass given by \citet{Heintz_2021} and the molecular gas mass using \citet{Vizgan_2022}. Each colour represents a different galaxy in our sample. Right:  The total gas mass surface density obtained using \citet{Vallini_25} versus the total gas mass surface density obtained using \citet{Vizgan_2022} calibration. In both figures, left triangles correspond to the pixels in the [C{\sc ii}] luminosity map with values below 3$\sigma$.}
  \label{fig:gascomparison}
\end{figure*}

In the left-hand panel of Figure\,\ref{fig:gascomparison} we present the comparison of the total gas mass obtained as a sum of the atomic and molecular gas masses estimated from the calibrations by \citet{Heintz_2021} and \citet{Vizgan_2022}, respectively, and the total gas mass derived using the spatially resolved [C\,{\sc{ii}}] conversion factor derived from galaxy simulations in \citet{Vallini_25}. We use the [C{\sc ii}] luminosity maps given in \citet{Li_2024} and the metallicity maps presented in Figure\,\ref{fig:metHCMplusdust} to derive the gas surface mass density maps presented here. We clearly see that the two  calibrations do not produce the same total amount of gas. The total gas mass estimated using \citet{Heintz_2021} and \citet{Vizgan_2022} is approximately one order of magnitude larger than that predicted from the simulations in \citet{Vallini_25}. This has already been reported in \citet{Palla2024} and other studies \citep[e.g.][]{Heintz2023}, where they found that the total gas masses obtained using these calibrations turn out to produce highly anomalous values.

In the right-hand panel of Figure\,\ref{fig:gascomparison} we can see that the total gas mass estimated from the simulations agrees better with the molecular gas mass derived from \citet{Vizgan_2022} calibration than with the combination of atomic and molecular gas mass presented in the left-hand panel. This is not surprising considering that the fraction of [C\,{\sc{ii}}] emission coming from the atomic phase in S\'IGAME simulations \cite[those used by ][]{Vizgan_2022} is very low   \citep[$<$2-3\%, see figure\,2 in][]{Vizgan_2022}, and that the ISM mass fraction representing the atomic gas mass in these simulations is less than $\sim$10\%. Besides, \citet{Vallini_25} trace the [C\,{\sc{ii}}] emission coming from the cold gas\footnote{\citet{Vallini_25} did not make distinction between atomic and molecular gas phase, thus they trace the cold molecular and atomic gas mass.} (T\,$\lesssim$\,100\,K) in photodissociation regions, and therefore not including the diffuse gaseous  phase. Therefore, the good agreement between \citet{Vizgan_2022} and \citet{Vallini_25} is somehow expected. However, this raises the question whether the atomic gas masses are indeed very low in these systems and \citet{Heintz_2021} calibration significantly overestimates the total gas mass, or if the total gas masses derived using the calibrations from \citet{Vizgan_2022} or \citet{Vallini_25} would underestimate the total gas mass. 

\begin{figure*}[h]
\centering
    {\includegraphics[width=0.45\linewidth]{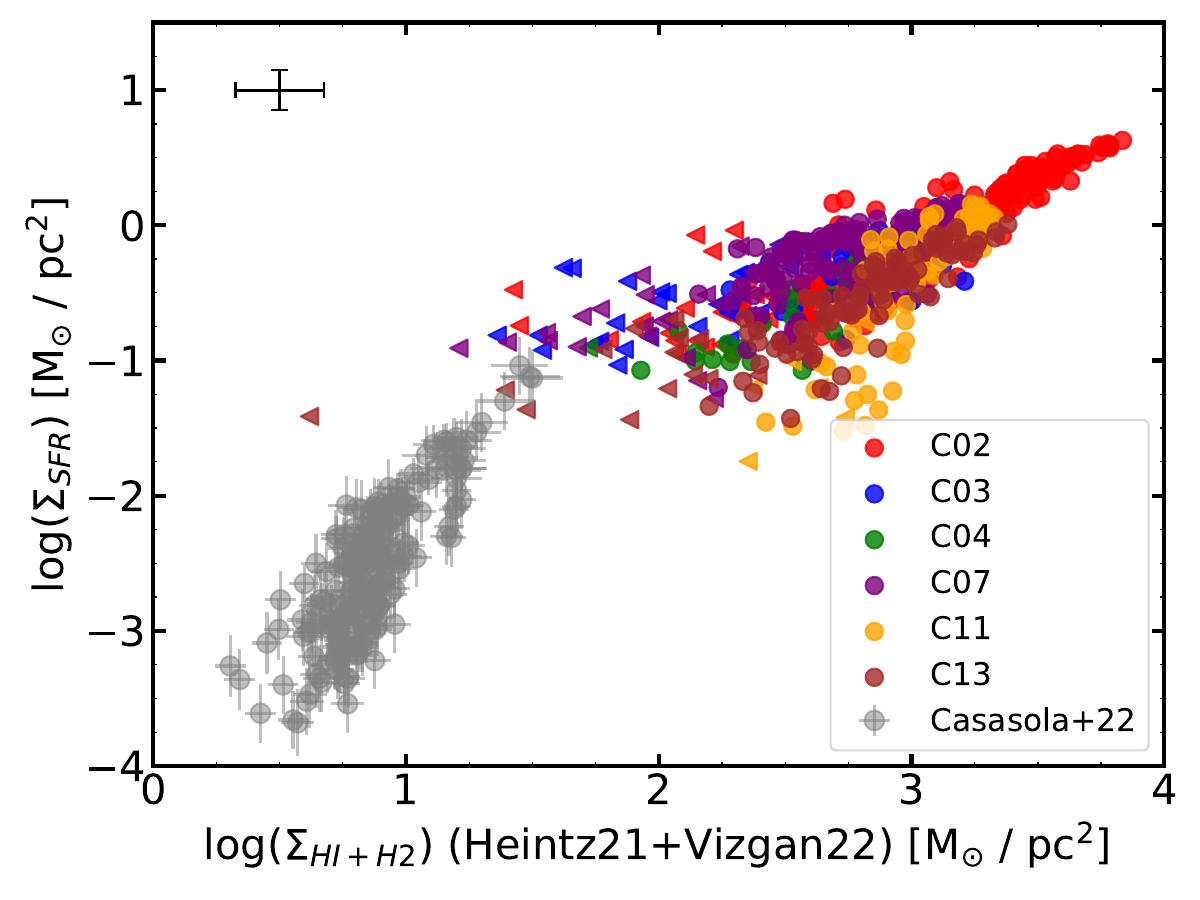}}
    {\includegraphics[width=0.45\linewidth]{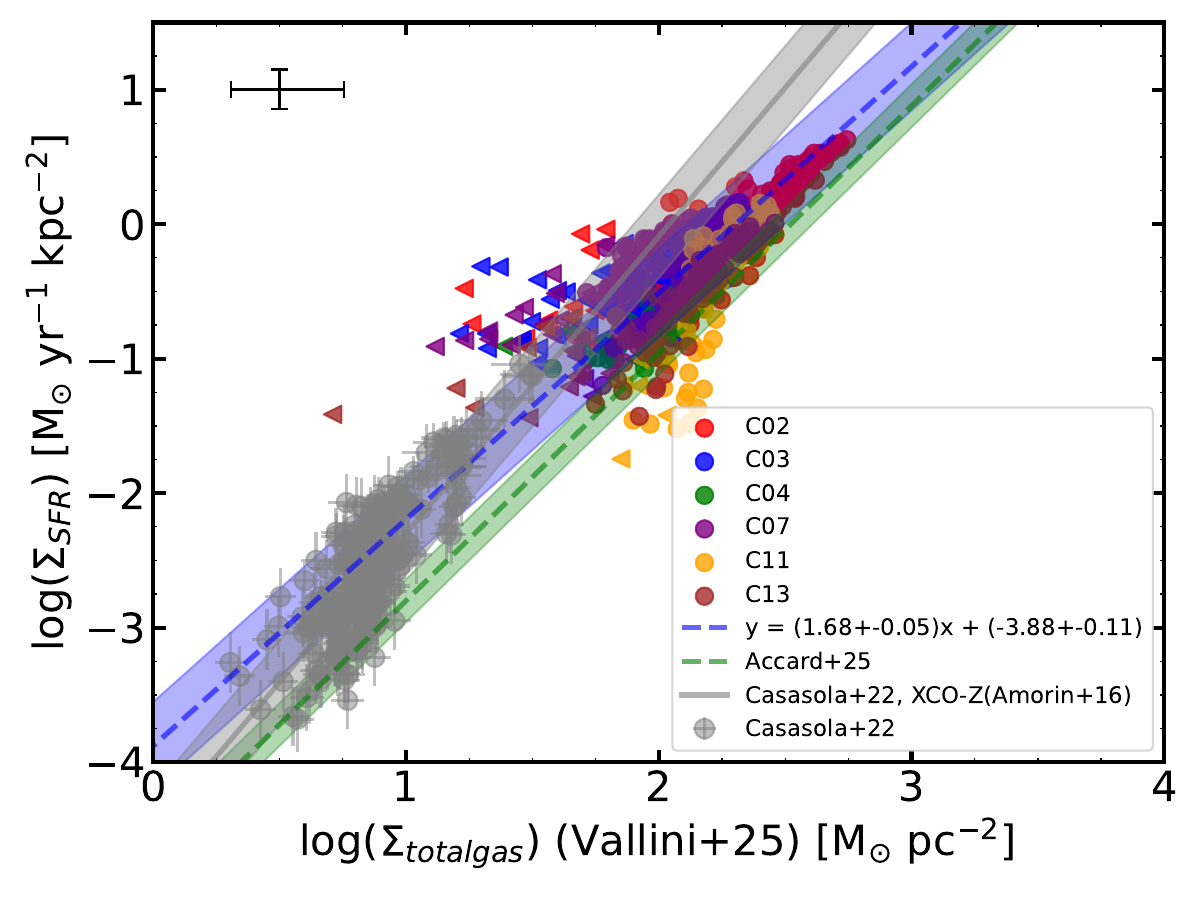}}
 \caption{$\Sigma_{SFR}-\Sigma_{totalgas}$ relation using the different ways of estimating the total gas mass surface density presented in Section\,\ref{sec:gascal}, the combination of atomic and molecular gas mass surface densities from \citet{Heintz_2021} and \citet{Vizgan_2022},  (left), and the total gas mass surface density obtained using the spatially resolved [C\,{\sc{ii}}] conversion factor from \citet{Vallini_25} (right). Each colour represents a different galaxy in our sample, while grey data points and continuous line show the observed values and the corresponding fit for local star-forming galaxies by \citet{Casasola}. The gas surface mass densities given in \citet{Casasola} were obtained using CO and H{\sc i} intensity maps from Dustpedia collaboration \citep{Casasola2017} (see Section\,\ref{sec:bestgasmass}). The blue dashed line corresponds to the orthogonal distance regression to the data points of our galaxy sample with the coloured area showing the standard deviation of the residuals. The green dashed line and coloured area shows the linear fit and corresponding dispersion from \citet{Accard2025}. In both figures, left triangles correspond to the pixels in the [C{\sc ii}] luminosity map with values below 3$\sigma$.}
  \label{fig:gasSFRcomparison}
\end{figure*}

In order to explore further the possible overestimation of the atomic  gas mass provided by \citet{Heintz_2021} (or the underestimation by \citet{Vizgan_2022} and \citet{Vallini_25} calibrations), we analyse how the total gas masses compare with the SFR obtained from the SED fitting done by \citet{Li_2024}, and with the results from nearby galaxies. In Figure\,\ref{fig:gasSFRcomparison} we show the relation between the SFR and the total gas mass surface density obtained with the combination of \citet{Heintz_2021} and  \citet{Vizgan_2022} calibrations (left panel), and the total gas mass surface density derived using \citet{Vallini_25} (right panel). We add the results from the spatially resolved studies for nearby galaxies from  \citet{Casasola} (grey data points). The gas surface mass densities reported in \citet{Casasola} were obtained using $ ^{12}\rm CO(1-0)$ and $^{12}\rm CO(2-1)$ emission line observations\footnote{ \citet{Casasola} assumed a metallicity-dependent $X_{\rm CO}$ to derive molecular gas surface mass densities. The uncertainties reported in \citet{Casasola} account for metallicity uncertainties and for the uncertainties on the calibration of the $X_{\rm CO}$ factor.} and H{\sc i} 21\,cm line intensity images from Dustpedia collaboration \citep{Casasola2017}.

We can see that for the case of the total gas mass surface density obtained using \citet{Heintz_2021} and \citet{Vizgan_2022} calibrations, the SFR-total gas mass relation for our galaxy sample is different from that observed in the nearby Universe, showing a flatter slope produced by the large total gas mass surface densities predicted when using the \citet{Heintz_2021} calibration. Vice versa, the SFR-total gas mass relation derived when the total gas masses are obtained using the  \citet{Vallini_25} calibration seems consistent with the one for galaxies in the nearby Universe. The linear fit to each dataset overlap within $\sim$\,2 orders of magnitude in $\Sigma_{totalgas}$. This suggests that the atomic gas masses derived using \citet{Heintz_2021} calibration are indeed overestimated and that a more robust quantification is provided by the calibration given in \citet{Vallini_25}. Indeed, it is suggested in \citet{Algera_Hz10_2025} that the calibration in \citet{Heintz_2021} provides an upper limit to the atomic gas mass associated to the [C\,{\sc{ii}}] emission for Hz10, a galaxy at $z=5.65$, and that a conversion factor more representative for the star-forming disk would give gas masses in agreement with the results from kinematic modelling. In the rest of this work, therefore, we will use gas masses  obtained by applying the spatially resolved [C\,{\sc{ii}}] conversion factor from \citet{Vallini_25}.

\begin{figure*}
    {\includegraphics[width=0.33\linewidth]{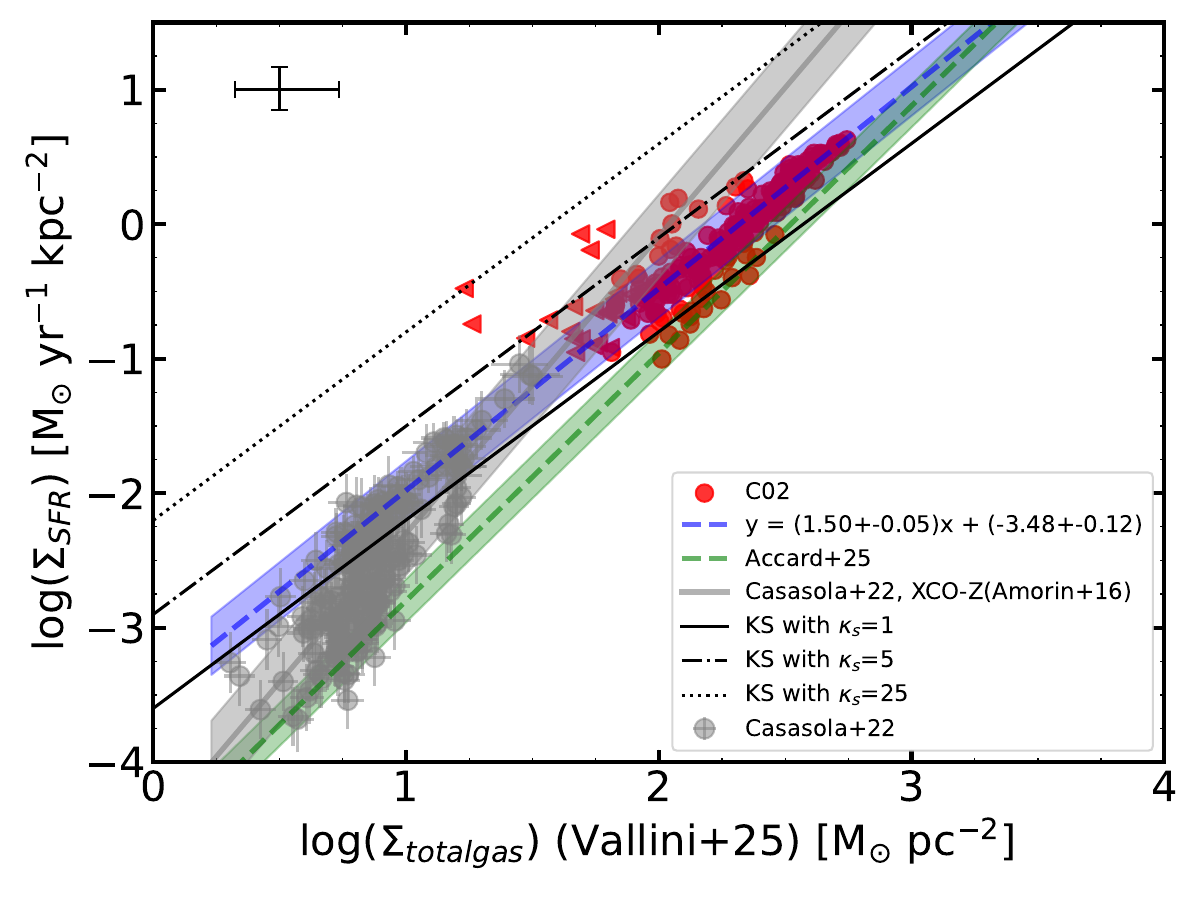}}
    {\includegraphics[width=0.33\linewidth]{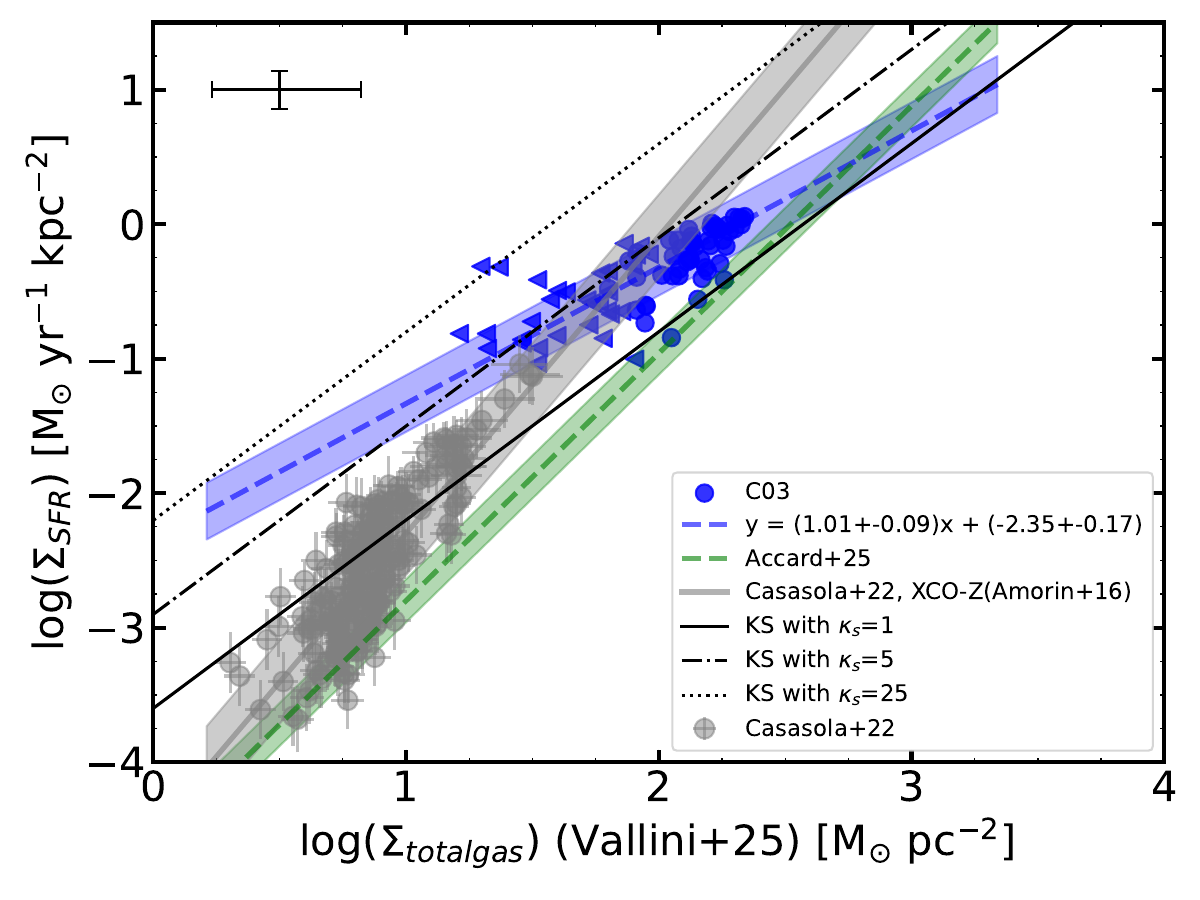}}
    {\includegraphics[width=0.33\linewidth]{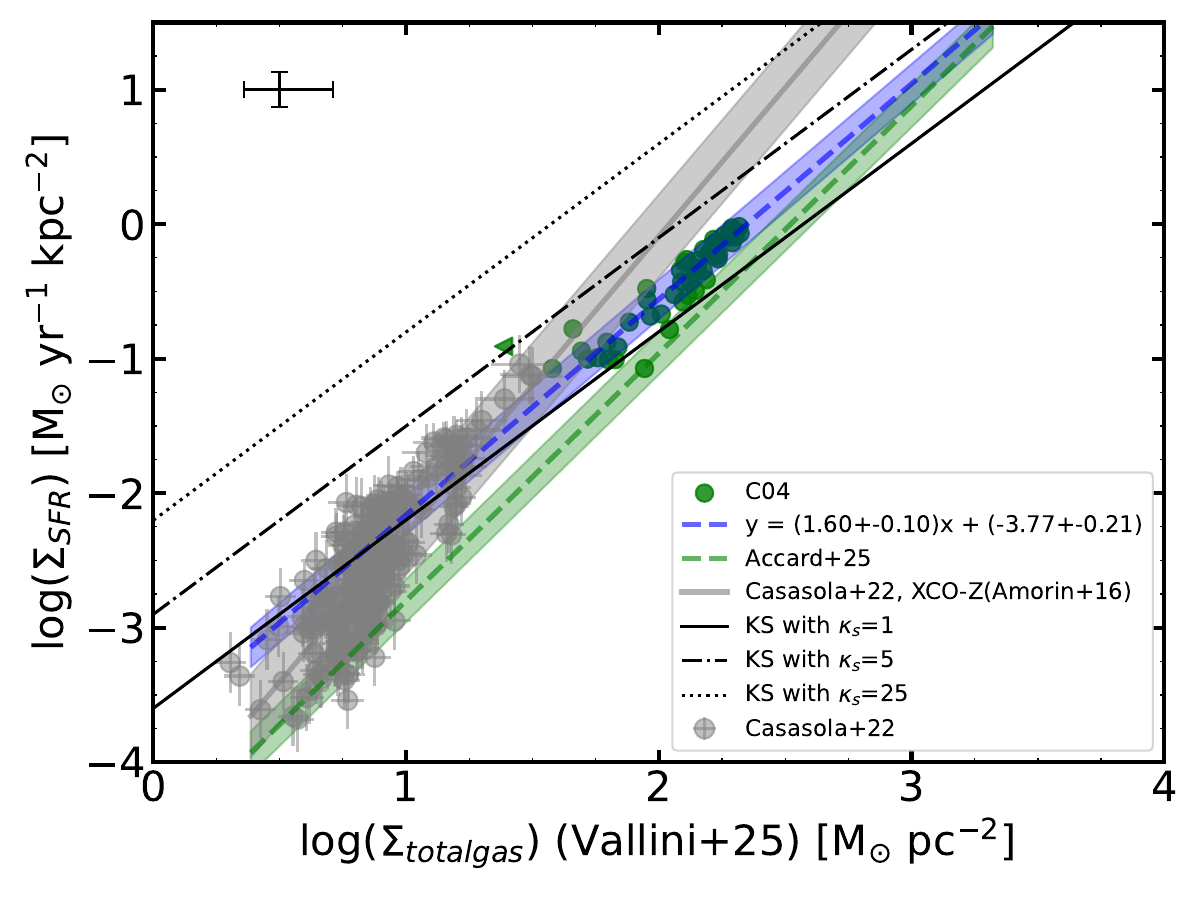}}
    {\includegraphics[width=0.33\linewidth]{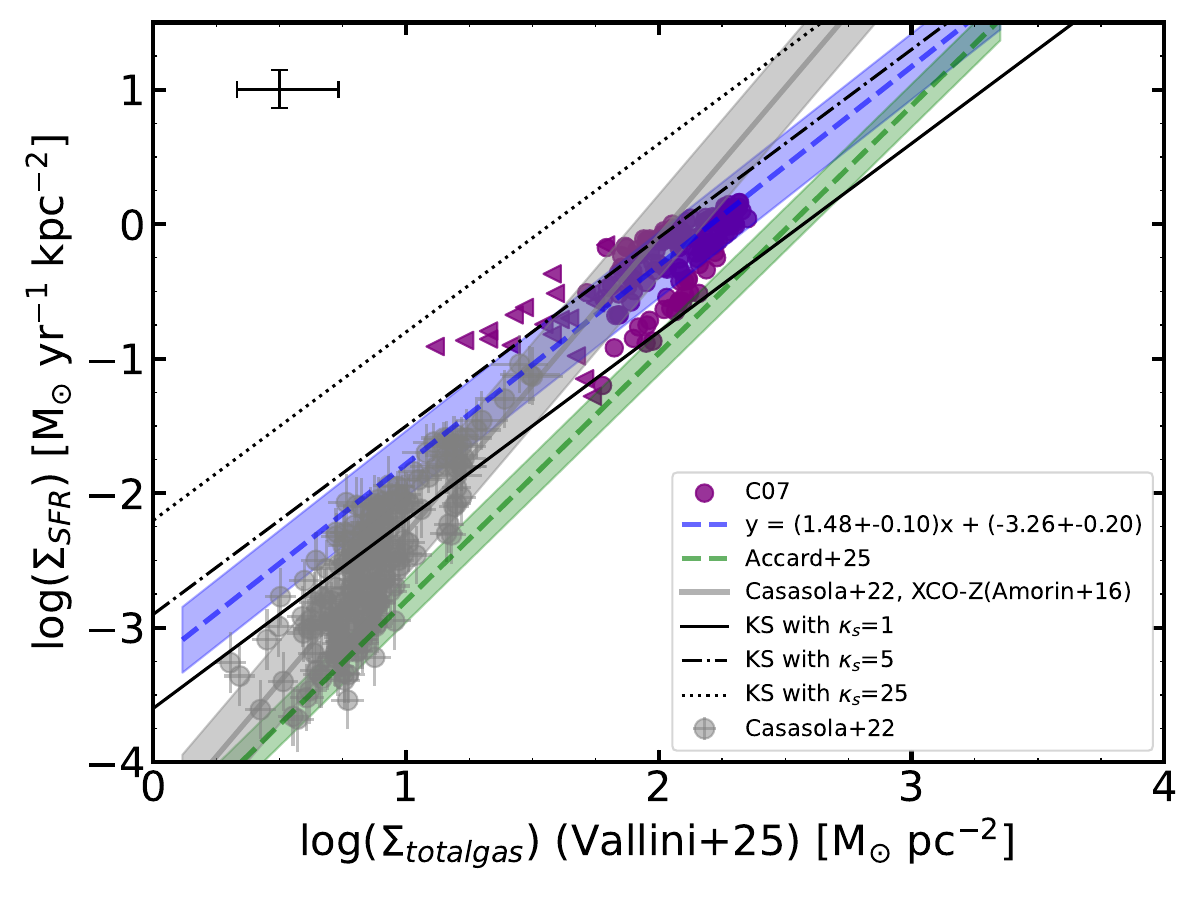}}\hfill
    {\includegraphics[width=0.33\linewidth]{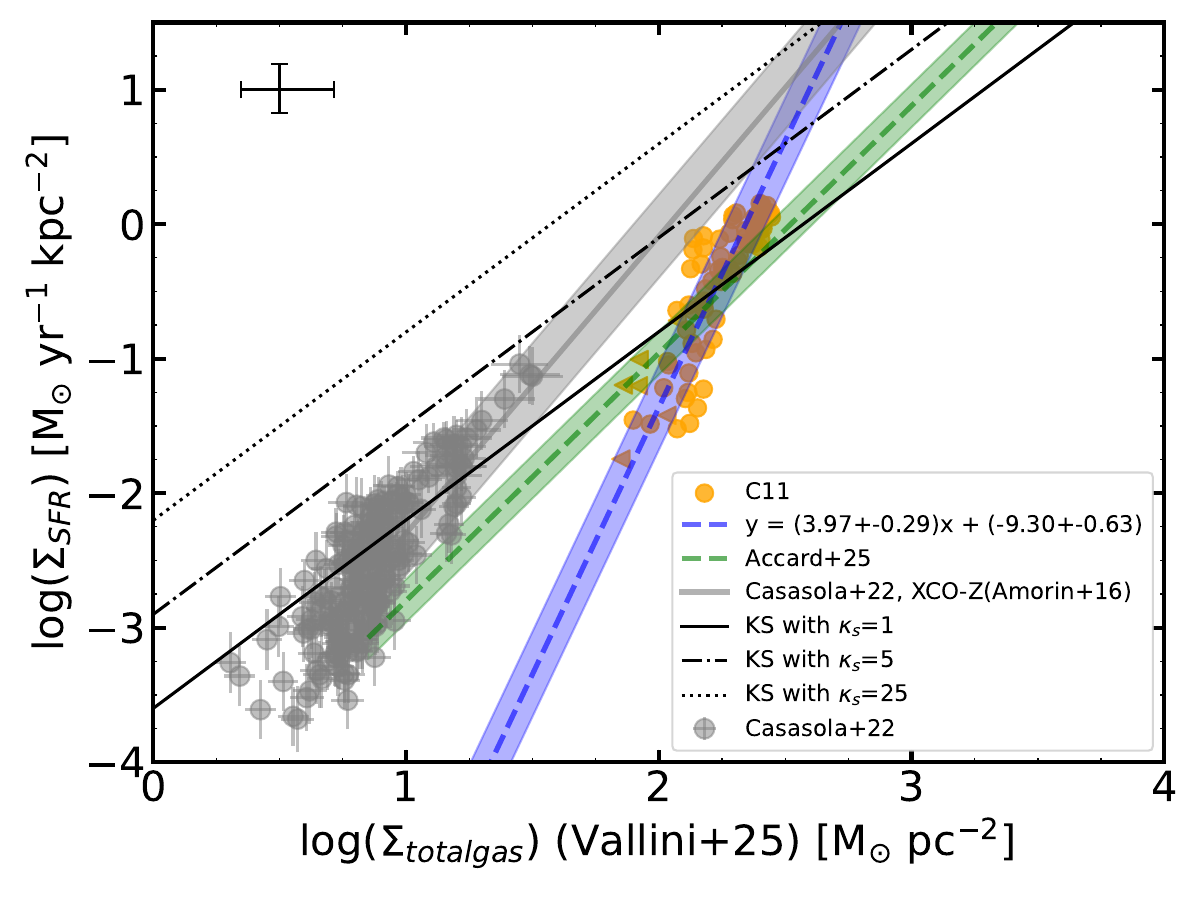}}\hfill
    {\includegraphics[width=0.33\linewidth]{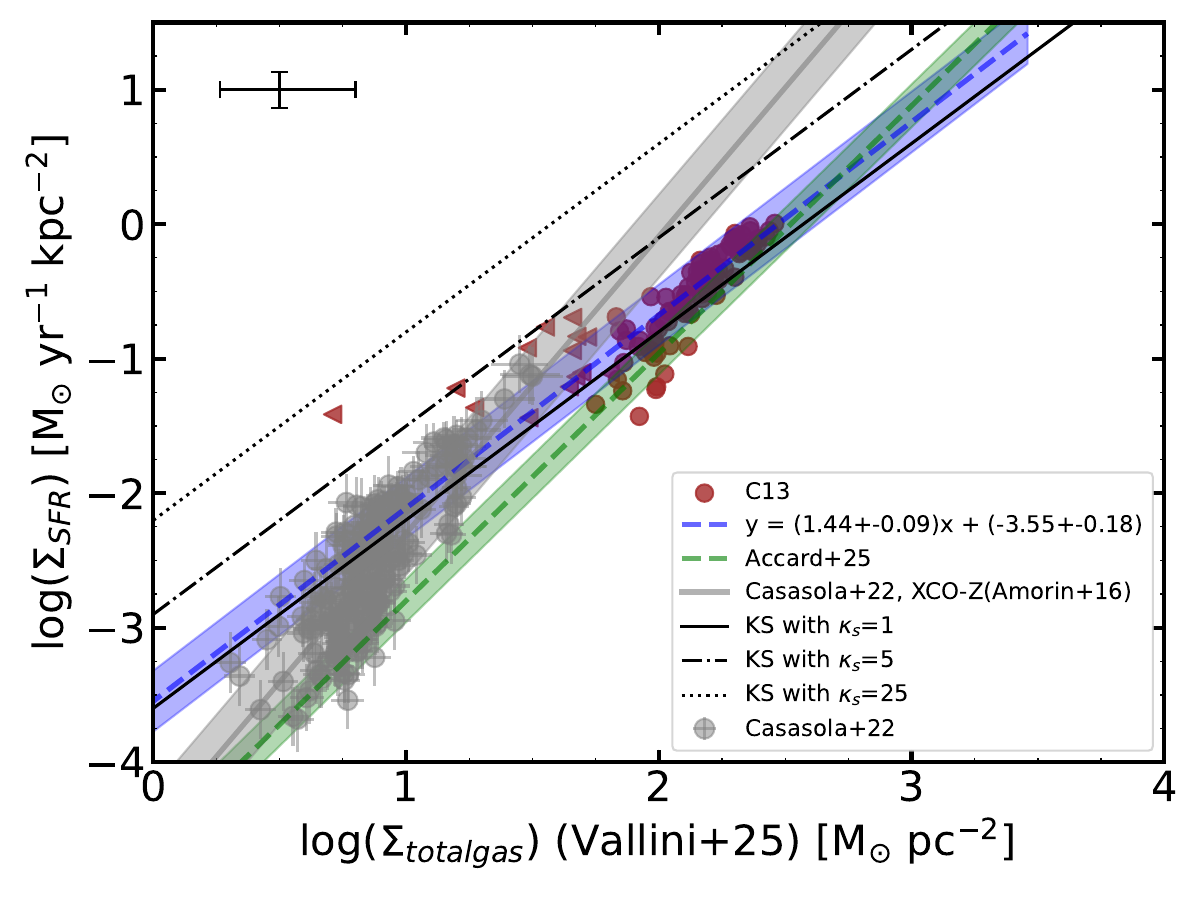}}
 \caption{$\Sigma_{SFR}-\Sigma_{totalgas}$ relation as in the left panel of Figure\,\ref{fig:gasSFRcomparison} but for each individual galaxy in our sample. In all the panels, left triangles correspond to the pixels in the [C{\sc ii}] luminosity map with values below 3$\sigma$.}
  \label{fig:gasSFRcomparison_individual}
\end{figure*}
Finally, in the right-hand side panel of Figure\,\ref{fig:gasSFRcomparison} we compare our results with those derived in \citet{Accard2025} for a larger sample of CRISTAL galaxies. The slopes for both relations are in agreement within the errors, but the total gas mass surface densities derived from \citet{Accard2025} seem to be shifted towards larger values. The differences might arise from the fact that \citet{Accard2025} applied a conversion factor from \citet{Vallini_25} that does not depend on the metallicity, and/or their SFRs are based on CIGALE \citep{Boquien2019}, while we use the SFR from MagPhys \citep{daCunha2008} derived in \citet{Li_2024}. 

In Figure\,\ref{fig:gasSFRcomparison_individual} we show the individual $\Sigma_{SFR}-\Sigma_{totalgas}$ relations with their corresponding slopes for each galaxy. We have also plot the Kennicutt-Schmidt (KS) relation for different burstiness ($\kappa_{\rm s}$) parameters. We see that DC-848185 (C02) and DC-873321 (C07ab) follow a relation with a slope close to the one in the KS relation,  but with a higher $\kappa_{\rm s}$. DC-536534 (C03) and DC-630594 (C11) present a shallower and steeper slope than the KS relation, respectively. The slopes derived for these galaxies could be biased by pixels located in the outer parts of the galaxy disks having low S/N [C\,{\sc{ii}}] luminosities, as it is the case for DC-536534 (C03). For DC-630594 (C11), however, we cannot rule out the possibility that the $\Sigma_{SFR}$ values derived in \citet{Li_2024} from SED fitting could be underestimated, as it is suggested in the corresponding panel for this galaxy in Figure\,11 of \citet{Li_2024}. VC-5100822662 (C04) and VC-51009947994 (C13), however, are galaxies with a slope close to the KS relation and with $\kappa_{\rm s}\sim1$. This seems to suggest that DC-848185 (C02) and DC-873321 (C07ab) are experiencing a strong burst of star formation compared to VC-5100822662 (C04) and VC-51009947994 (C13). We will see in Section\,\ref{sec:scalingdust} that the burst of star formation in these galaxies helps to understand the relative amount of dust in these galaxies.

\subsection{Scaling relations involving dust and metallicity}\label{sec:scalingdust}

\begin{figure*}
\centering
    {\includegraphics[width=0.5\linewidth]{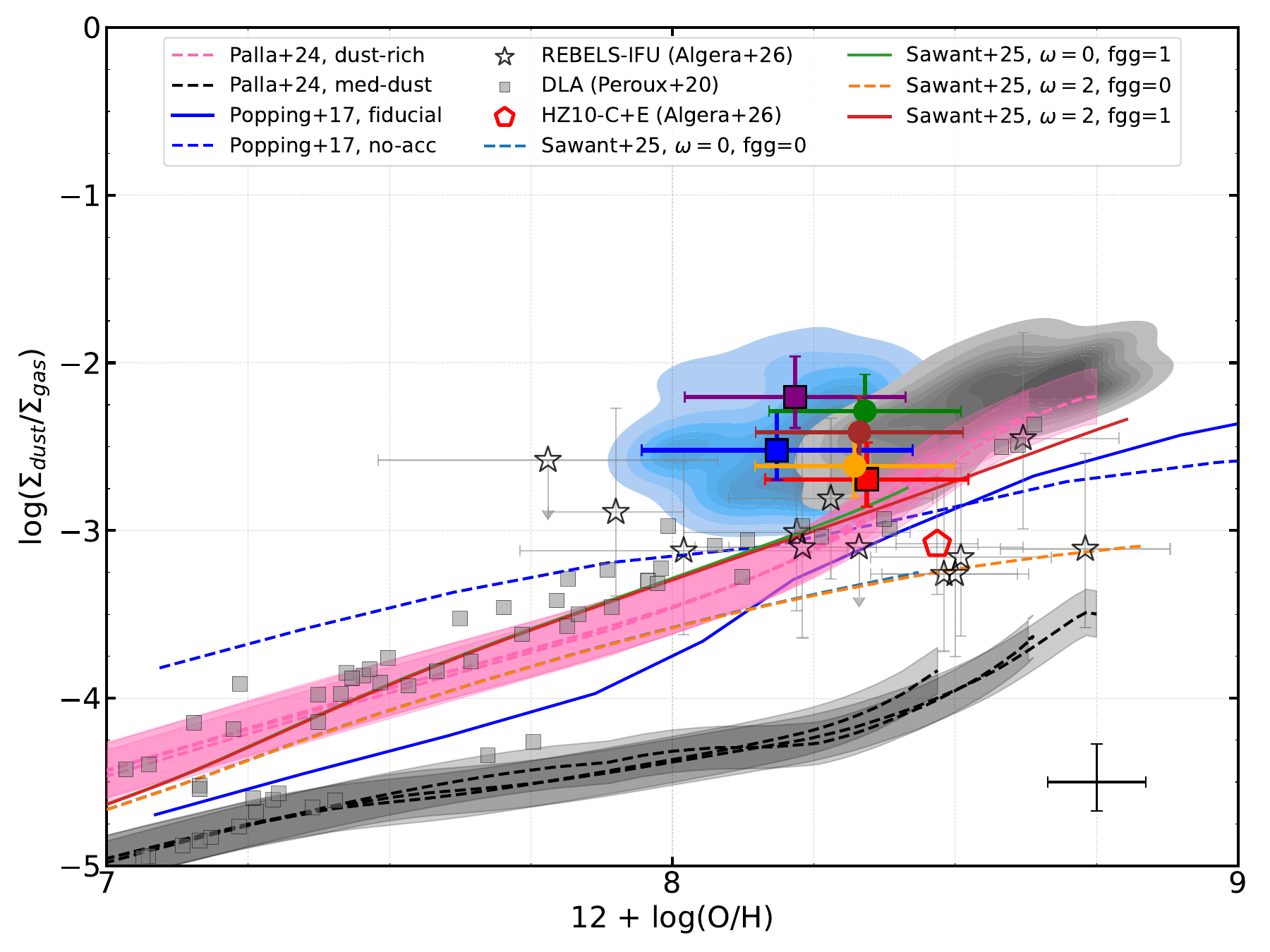}}
    {\includegraphics[width=0.5\linewidth]{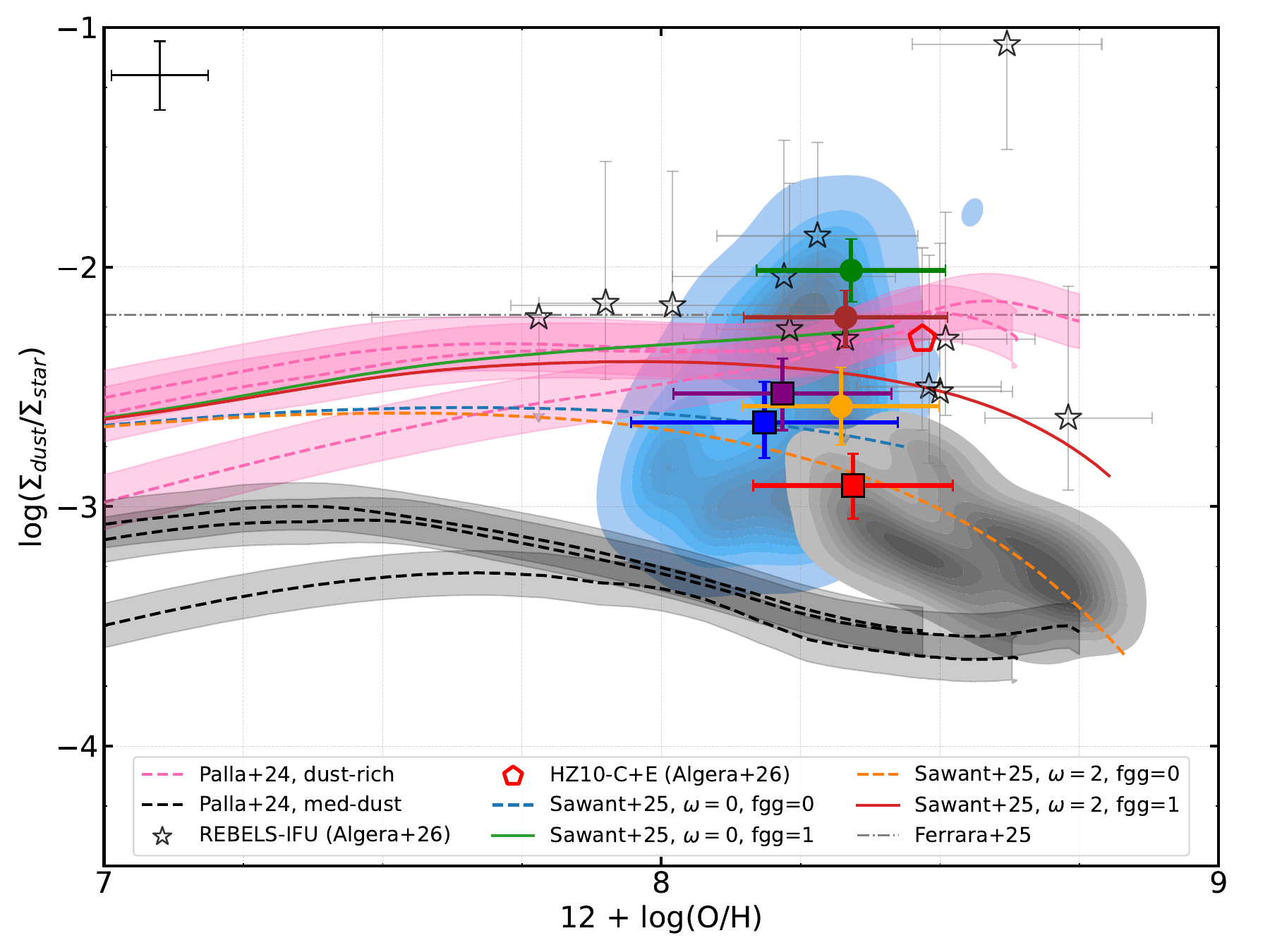}}
    {\includegraphics[width=0.5
    \linewidth]{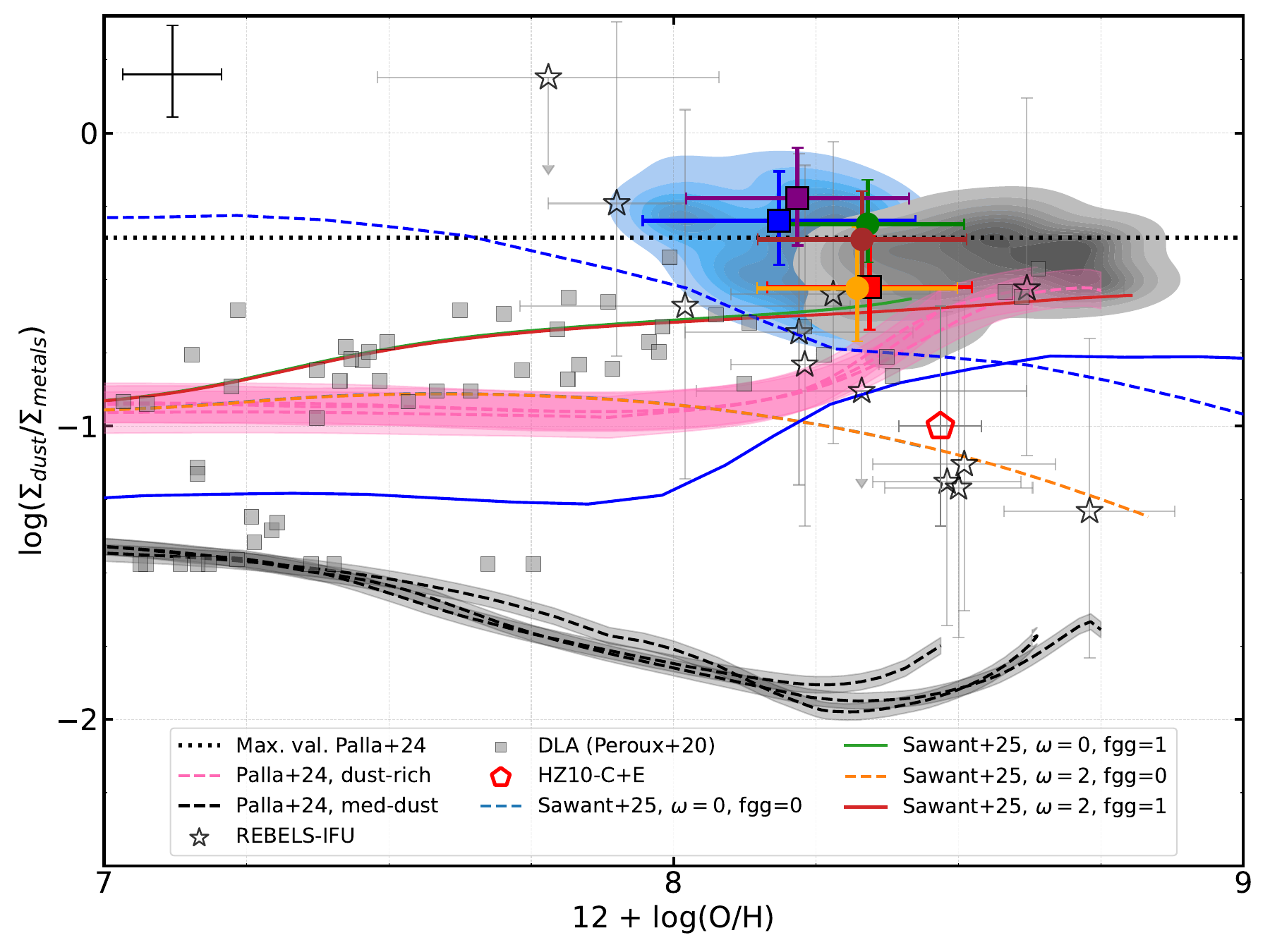}}
 \caption{Logarithm of dust-to-gas surface mass density ratio (top), dust-to-stellar surface mass density ratio (middle), and dust-to-metals surface mass density ratio (bottom) as a function of the oxygen abundance. The coloured circles, which follow the same colour scheme as in Figure\,\ref{fig:gasSFRcomparison}, correspond to the average values of the distribution data (blue density contours) within each galaxy, while the errors represent the dispersion of the data (see Table\,\ref{tab:properties}). Different set of models are presented: chemical evolution models from \citet{Palla2024} (pink and grey), \citet{Sawant2025} (coloured continuous lines) and the SAM from \citet{Popping_2017} (dashed and continuous blue line). The shaded regions around the model curves are defined using a fixed percentage of the difference between the maximum and minimum y-values. Error bars of the data points correspond to the standard deviation of the values found in each individual galaxy. Average errors obtained from uncertainties in the estimated magnitudes are located in the lower left corner of each plot. Gray density contours represent the distribution of local Universe galaxies given in \citet{Casasola} and empty stars correspond to REBELS galaxies \cite{Algera2026}.}
  \label{fig:dustandmet}
\end{figure*}

In the top panel of Figure\,\ref{fig:dustandmet} we present the spatially-resolved dust-to-gas surface mass density ratio as a function of metallicity of our galaxy sample (blue density contours) and compare with the results for nearby galaxies at $\sim$\,3\,kpc-scales from \citet{Casasola} (gray density contours). The dust-to-gas surface mass density ratios for our galaxies (coloured circles) are obtained as average values over each individual galaxy with errors representing the standard deviations of the dust-to-gas surface mass density ratios obtained for each galaxy. The dust-to-gas surface mass density ratios for the individual spaxels of each galaxy versus the metallicity are presented in Figure\,\ref{ap:fig:dustandmet}. In general, our galaxy sample shows a distribution similar to the galaxies in the local Universe with a more extended tail towards the low metallicity regime. The distribution, however, does not exhibit the decline towards low dust-to-gas surface mass density ratios found for the low metallicity regime that is typical of nearby galaxies \citep[grey countours representing data for local galaxies from][]{Casasola} and DLAs at $z> 3$ \citep{Peroux2020}. The black empty asterisks correspond to integrated dust-to-gas mass ratios derived for REBELS-IFU, a subsample of REBELS galaxies at $z\,\sim6.5-7.7$ with NIRSpec observations that allow estimation of their metallicities \citep{Algera2026}. The empty pentagon corresponds to the disturbed rotating disk Hz10-C+E within Hz10 proposed by \citet{Telikova2025} using [C\,{\sc{ii}}] kinematic analysis. We note here that we have derived surface mass density ratios and not total integrated mass ratios as it is the case of REBELS galaxies. Therefore, the comparison of our results with REBELS galaxies are not strictly equivalent. In Section\,\ref{ap:comp_ratios} of the Appendix we compare the surface mass density ratios with the total mass ratios derived from Table\,\ref{tab:gal} for our galaxy sample.

The average dust-to-gas surface mass density ratios for our galaxy sample show the same relatively flat distribution as that shown for REBELS-IFU galaxies, but with consistently higher ($\sim0.5-1$\,dex) dust-to-gas surface mass density ratios. Dust masses for REBELS galaxies were derived in \citet{Algera2026} assuming a fixed dust temperature of 45\,$\pm$\,15\,K. The dust temperature estimated for our ALPINE/CRISTAL galaxies are in general in agreement or slightly lower than the fixed value assumed by \citet{Algera2026} (Relaño et al., in prep). Regarding the gas masses, \citet{Algera2026} used the calibration from \citet{Zanella2018}, which gives a conversion factor ($\rm \alpha = 35\, M_{\odot}/L_{\odot}$) twice larger than that proposed in \citet{Vizgan_2022} ($\rm \alpha = 18\, M_{\odot}/L_{\odot}$). We note here that we are using the spatially resolved [C\,{\sc{ii}}] calibration from \citet{Vallini_25} to derive the gas masses. In general, there is a good agreement between the gas mass surface densities predicted from \citet{Vallini_25} and those using \citet{Vizgan_2022}. Differences are relevant ($\lesssim\,0.5$\,dex) at high gas mass surface densities (see right-hand side panel in Figure\,\ref{fig:gascomparison}). Taking into account these discrepancies, applying \citet{Zanella2018} would give at least a factor of 2 larger gas masses than those derived from \citet{Vallini_25}. This would give lower dust-to-gas mass ratios for the REBELS galaxies derived in \citet{Algera2026} than those if the conversion factor from \citet{Vallini_25} had been used. In this case, the predicted dust-to-gas mass ratios would be a factor of 2 larger, and the dust-to-gas mass ratios for REBELS and the surface mass density ratios for ALPINE/CRISTAL galaxies would be similar. As mentioned in \citet{Algera2026}, the REBELS galaxy sample shows gas mass fractions that are inconsistent with chemical evolution models \citep[see figure 3 in][]{Algera2026}, which points towards an overestimation of the gas masses, and therefore, an underestimation of the dust-to-gas mass ratios. 

Hz10-C+E, the rotating disk system of Hz10, falls within the same area in the dust-to-gas mass ratio versus metallicity diagram as REBELS galaxies, and below that one occupied by our galaxy sample. The dust mass for Hz10-C+E was estimated in \cite{Algera_Hz10_2025} using a well sampled FIR SED \citep[6 observed ALMA bands,][]{Faisst2020,Villanueva2024, Algera_Hz10_2025}. The gas mass, however, was derived obtaining the baryonic mass of the system from a rotation model applied to the [C\,{\sc{ii}}] observations \citep[see][]{Telikova2025}, and subtracting the stellar mass calculated from SED fitting. As in most REBELS galaxies, Hz10-C+E shows a gas mass fraction not compatible with chemical evolution models (see Figure\,\ref{fig:dustandgasfgas}), which points towards an overestimation of the gas mass for this system, and therefore, an underestimation of the dust-to-gas mass ratio.

The middle panel in Figure\,\ref{fig:dustandmet} shows the dust-to-stellar surface mass density ratio as a function of metallicity. The data distribution for our galaxy sample (blue distribution) exhibit a large range of dust-to-stellar surface mass density ratios that covers more than one order of magnitude. Two maxima at $\log(\Sigma_{\rm dust}/\Sigma_{\rm star})\sim-2$, corresponding to the dust-to-stellar surface mass density ratios from VC-5100822662 (C04a, green circle) and VC-5100994794 (C13, brown circle), and another one at $\log(\Sigma_{\rm dust}/\Sigma_{\rm star})\sim-3$, corresponding to dust-to-stellar surface mass density ratios from DC-848185 (C02, red square), DC-536534 (C03, blue square), DC-873321 (C07ab, purple square), and DC-630594 (C11, yellow circle). Interestingly, each galaxy shows quite homogeneous dust-to-stellar surface mass density ratios, which is visible by the small error bars of the data points. The distribution of the dust-to-stellar surface mass density ratios for each galaxy is shown in Figure\,\ref{ap:fig:dustandmet} of the Appendix. Most of the REBELS galaxies and Hz10-C+E occupy the top area  of the data distribution, having dust-to-stellar mass ratios similar to the surface mass density ratios of VC-5100822662 (C04a) and VC-5100994794 (C13). DC-848185 (C02), DC-536534 (C03), DC-873321 (C07ab), and DC-630594 (C11), however, are located in the lower part of the blue density contours and above the area of the distribution occupied by local galaxies (gray contours). 

The separation in the dust-to-stellar surface mass density ratios seems to be related to other physical properties of these galaxies. In Figure\,\ref{fig:dustandmet} we highlight those galaxies that have been classified as AGN candidates  exhibiting signatures of a broad line region with strong ($\sim$\,3000\,km\,s$^{-1}$) outflows \citep{Ren2025}. While DC-848185 (C02, red square) and DC-873321 (C07ab, purple squares) show signatures of broad components, which suggests the presence of a possible AGN, DC-536534 (C03, blue square) is robustly identified as a galaxy with AGN by \citet{Ren2025}. In particular, DC-848185 (C02) exhibits extremely strong outflows powered by a possible merger-induced starburst \citep{Davies2026}. All these galaxies have lower dust-to-stellar surface mass density ratios compared to the other galaxies in our sample (except for DC-630594 (C11), yellow circle)\footnote{The stellar masses for these sources were derived in \citet{Li_2024} without including AGN emission. \citet{Ren2025} quantified the AGN contribution to the stellar mass estimates and concluded that the  contamination is minimal.}. Their dust-to-gas surface mass density ratios are, however, similar to the rest of the galaxy sample. This would mean that either the stellar masses for these galaxies are overestimated or they have a low gas mass reservoir. Indeed, dust enrichment for these galaxies could not be as efficient as those presenting a higher dust-to-stellar surface mass density ratio. The strong outflows observed in these galaxies suggest that the second option is the most plausible explanation for having simultaneously similar dust-to-gas mass and lower dust-to-stellar surface mass density ratios than the other three galaxies of our sample. On the other hand, those galaxies without AGN or strong outflows, have high dust-to-stellar and dust-to-gas surface mass density ratios presenting a large boost of dust production compared with what is found in the local Universe (see gray density contours in the middle panel of Figure\,\ref{fig:dustandmet}). This agrees with previous findings suggesting a boost of dust enrichment in galaxies at $z\sim6$ \citep{Bakx2020,Pozzi2021,Sommovigo2022,Pozzi2021}. 

In the bottom panel of Figure\,\ref{fig:dustandmet} we show the dust-to-metals (the fraction of metals locked up in dust grains\footnote{ $\rm M_{\rm dust}/M_{metal}(gas + dust)$, see \cite{DeLooze2020}.})  surface mass density ratio as a function of metallicity. Our galaxy sample present dust-to-metals surface mass density ratios consistent with those of local galaxies (gray density contours). DC-848185 (C02) and DC-630594 (C11) show similar values as some of the REBELS galaxies, while the other four galaxies have dust-to-metals surface mass density ratio close to the maximum value estimated by \citet{Palla2024} (dotted-black line). We do not see a trend of declining dust-to-metals surface mass density ratio at higher metallicity, as it is observed for REBELS galaxies. However, the metallicity range covered by the average metallicity values is smaller compared with the metallicity range of the REBELS galaxies.

\subsection{Gas mass fraction}
\begin{figure}
    \centering
    \includegraphics[width=\linewidth]{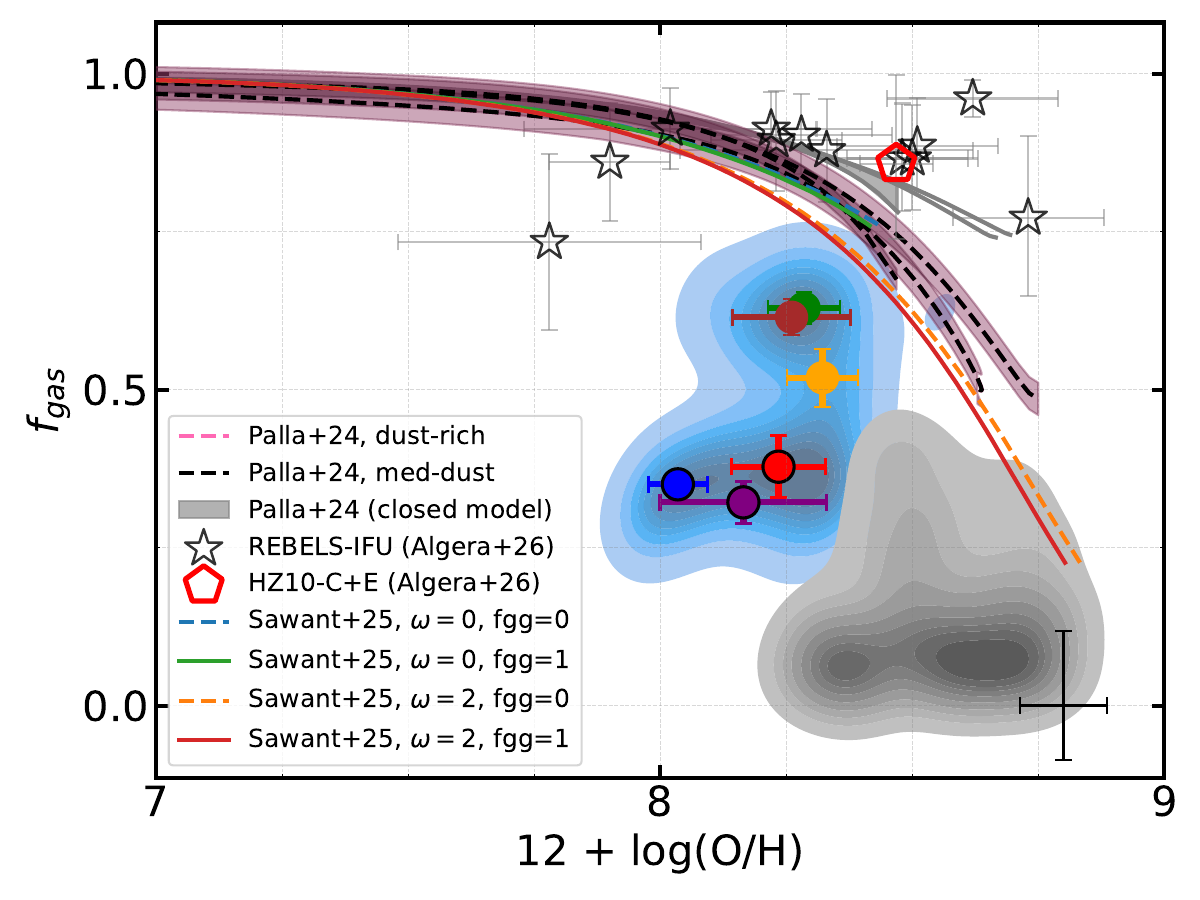}
    \caption {Logarithm of $f_{\text{gas}}$ as a function of 12 + log (O/H). Models are the same as in Figure\,\ref{fig:dustandmet}. Chemical models from \citet{Palla2024} assuming a closed-box approximation are shown in gray.}
    \label{fig:dustandgasfgas}
\end{figure}

In this section, we explore how the gas mass fraction, which is related to the amount of gas expelled from the galaxy and consumed by star formation, relates to other physical quantities. $f_{\text{gas}}$ is obtained using the expression:
\begin{equation}
    f_{gas} = \Sigma_{total gas}/(\Sigma_{total gas} + \Sigma_{star} ).
\end{equation}
In Figure\,\ref{fig:dustandgasfgas} we present the gas mass fraction as a function of the metallicity. Our galaxy sample have larger gas mass fractions than the galaxy sample in the local Universe, and lower than the REBELS galaxies and Hz10. The data distribution of the ALPINE/CRISTAL galaxies (blue contours) has two maxima, one related to VC-5100822662 (C04a) and VC-5100994794 (C13) at high $f_{\rm gas}$ and another at low $f_{\rm gas}$, corresponding to DC-848185 (C02), DC-536534 (C03) and DC-873321 (C07ab). These three galaxies are those with signatures of AGN and strong outflows in \citet{Ren2025}. The location of these objects in this diagram reinforces the suggestion that to explain simultaneously the high dust-to-gas mass and low dust-to-stellar surface mass density ratios for DC-848185 (C02), DC-536534 (C03) and DC-873321 (C07ab), gas masses should be lower than those in VC-5100822662 (C04a) and VC-5100994794 (C13). Indeed, VC-5100822662 (C04a) and VC-5100994794 (C13) are galaxies with a high gas mass fraction and similar dust-to-gas surface mass density ratios to the rest of our galaxy sample and to galaxies in the local Universe. This implies they have copious dust reservoirs, definitely larger than local Universe galaxies. 

\section{Comparison with chemical and dust models}\label{sec:models} 

\begin{table*}
\caption{Main parameters of the chemical and dust evolution models used here for comparison with our observations.}

\label{tab:models}      
\centering          
\begin{tabular}{lccc}    
\hline\hline
Parameter & \citet{Sawant2025} & \citet{Palla2024}  & \citet{Popping_2017} \\
\hline \\
 SFH & $\frac{t}{\tau_{\rm main}^2}e^{-t/\tau_{\rm main}}$  & REBELS 14, 25, 29  & merging of DM haloes \\
 \\
 Inflows & $\eta_{in}\times \rm SFR$, with  $\eta_{in}=0.6$ & $\propto e^{-t/\tau_{\rm inf}}$ & grow of DM haloes \\
 \\
 Outflows & $\eta_{out}\times \rm SFR$ with $\eta_{out}=0,2$ & $\omega\times \rm SFR$ with $\omega=3$ & stellar feedback \\
 \\
\hline
\\
SN dust & $\rm f_{cond}=1$ & SN reduction = $\times$3,  $\times$1   & $\rm f_{cond}=0.15$ \\
\\
Dust growth & key element & no key element & cycled through molecular clouds \\
\\
Prescription & \citet{Asano2013} & \citet{Mattsson2012}& \citet{Zhukovska2014} \\
 &  $f_{gg}=$0, 1 &   $\xi=1000,3000$ & $\tau_{\rm growth}= 15$\,Myr \\
 \\
 Dust Destruction & M$_{\rm clear}$ \citep{Asano2013}  &  M$_{\rm clear}$ \citep{Priestley2022} & M$_{\rm clear (carbon)}$, M$_{\rm clear (Sil)}$\\
\hline\hline 
\end{tabular}
\tablefoot{We refer the reader to the corresponding reference of each model for a further explanation. SN dust production is characterized by a factor, $\rm f_{cond}$, describing the SN dust condensation efficiency. Dust growth efficiency is parameterised by $f_{gg}$, dust growth efficiency, for the dust growth prescription in \citet{Asano2013}, while $\xi$ is a dust growth parameter that represents the dust growth efficiency in \citet{Mattsson2012}. In \citet{Zhukovska2014} dust growth is given by a dust growth time scale, $\tau_{\rm growth}$. Dust destruction is quantified by the mass cleared $\rm M_{clear}$ by the SN explosion and follow different prescriptions in each model. In \citet{Popping_2017} the amount of destroyed mass depends on the dust composition ($\rm M_{clear(carbon)}$ and $\rm M_{clear(Sil)}$ for the carbonaceous and silicate dust mass  destroyed), and in \citet{Sawant2025} an additional dust destruction efficiency ($\epsilon_{\rm SN}$) of 0.1 is assumed. $\tau_{\rm inf}$ is the gas mass infall time scale, and $\tau_{\rm main}$ is the e-folding time of the main stellar population in the galaxy (see section 3.1 of \citet{Sawant2025}). $\omega$ and $\eta_{out}$ refer to mass-loading factors, and  $\eta_{in}$ to the efficiency of the inflowing gas \citep{Sawant2025}.
}
\end{table*}

In this section, we compare our observational trends with chemical evolution models from the literature that have particularly been tuned for high-redshift galaxies. We choose three different models: i) Models from \citet{Palla2024}, tailored to describe three characteristic galaxies from the REBELS sample at $z\sim6-7$ \citep{Bouwens2022}, ii) models from \citet{Sawant2025}, with a star formation history (SFH) that reproduces the properties of ALPINE galaxies \citep{Burgarella2022}, and iii) semi-analytical models (SAM) from \citet{Popping_2017}, used to explore the variation of scaling relations at different redshifts. All these models take into account dust production from stellar sources and dust growth in the ISM, dust destruction from SNe, and astration as a form of dust sink. The main assumptions for these models are presented in Table\,\ref{tab:models}, for more specific details we refer the reader to the corresponding reference of each model. 

The models from \citet{Palla2024} follow the evolution of the gas, dust and stellar mass, as well as the metallicity of the galaxy, as a function of time, using three characteristic SFHs from \citet{Topping2022} and assuming a gas infall rate parameterised by a gas infall timescale and a total gas mass accreted onto the system. The outflow rate is proportional to the star formation rate with a fixed mass-loading factor of $\omega=3$, slightly higher than previously estimated mass-loading factors for ALPINE galaxies  \citep[$\omega\sim1-2$,][]{Ginolfi2020}. A certain amount of dust is expelled into the circumgalactic medium (CGM) as the outflowing material has the same dust-to-gas mass ratio as in the galaxy system. 
We choose the metal-rich scenario where the galaxy accretes an amount of gas tuned to obtain a galaxy with an --around or slightly less than solar-- metallicity \citep[see Figure\,2 in][]{Palla2024}. Within this scenario we select the {\it dust-rich} and the {\it intermediate-dust} models where the SN dust yields and the dust growth parameter (see Table\,\ref{tab:models}) are tuned to produce different dust reservoirs.

\citet{Sawant2025} produced chemical evolution models using the best fit SFH for ALPINE galaxies and allowing inflows of pristine gas and outflows of enriched gas with mass-loading factor as a free parameter. They followed the evolution of different dust species\footnote{Olivine, Pyroxene and carbonaceous dust grains, see Table\,2 in \citet{Sawant2025}.} incorporating the {{\it key-element approximation}}\footnote{The key-element is defined as the least abundant element needed to form the different dust species \citep[see][]{Ferrarotti2006,Zhukovska2008}}. We explore here models with SN condensation fraction equals 1 (no dust destruction by SNe) and dust growth efficiency values of 0 and 1. 

The SAM from \citet{Popping_2017} rely on a set of merging histories of dark matter haloes with a series of prescriptions that allow to track the size of the galactic disk and the molecular hydrogen abundance, as well as the star formation. Outflows due to SN and AGN feedback are included in the models assuming a fixed dust-to-gas mass ratio in the outflowing material, and therefore allowing part of the dust reservoir in the galaxy to be ejected into the CGM. A certain amount of dust is also accreted in the system depending on the cooling rate of the gas and the dust abundance of in the CGM. For the comparison with our data, we use the model predictions from \citet{Popping_2017} corresponding to $z\,=\,5$, suitable for the redshifts of our galaxy sample. We compare two types of models: i) their fiducial model (see Table\,\ref{tab:models}), and a model in which no dust accretion takes place and the SN and AGB dust condensation efficiencies are set to 1, maximizing the dust budget in the galaxy.

In the top panel of Figure\,\ref{fig:dustandmet} we compare the observed dust-to-gas surface mass density ratios for our galaxy sample  with the three chemical evolution models. Most of the observed values fall above the model trends in dust-to-gas versus metallicity relation. At high metallicity, the {\it dust-rich scenario} from \citet{Palla2024} and the models from \citet{Sawant2025} with $\rm f_{\rm gg}=1$ are able to explain the location of the galaxy with the lowest dust-to-gas surface mass density ratio (DC-848185, C02). The SAM models from  \citet{Popping_2017} predict lower dust-to-gas mass ratios than the observed ones. At low metallicity the models with {{\it no accretion}} from \citet{Popping_2017} and the  {\it intermediate-dust scenario} from \citet{Palla2024} describe a flat trend, similar to the observed one, albeit with lower values. The fact that we do not see an increase in the dust-to-gas surface mass density ratio above a certain critical metallicity for which dust accretion becomes efficient \citep[see][]{Asano2013,Galliano2018} would mean that either the galaxies have already built up their dust content via dust growth in the ISM, or dust growth is indeed not efficient in these galaxies. 
The discrepancies in the absolute values between the models and the observations could be explained if there is a significant amount of gas relative to the dust expelled outside the galaxy via AGN activity or strong outflows.

In the middle panel of Figure\,\ref{fig:dustandmet} we compare the observed dust-to-stellar surface mass density ratios and the predictions of the chemical evolution models. The {\it intermediate-dust} scenario from \citet{Palla2024} fall below the observed data distribution. The highest dust-to-stellar surface mass density ratios are, however, explained with the {\it dust-rich scenario} where dust growth is enhanced with high $\xi$ values, and dust destruction timescale is set to the minimum value (see Table\,\ref{tab:models}). The model from \citet{Sawant2025} with no outflows and $\rm f_{gg}=1$ explains the observed value for VC-5100994794 (C13). The dashed-dot line in the figure corresponds to the model prediction from \citet{Ferrara2025}, which does not account for dust growth\footnote{Completely inefficient dust growth models have their caveats. For instance, simulations including evolution of the grain size distribution find that an increase of small grains due to shattering can enhance significantly dust growth and explain the increase in dust content in galaxies in the early ($z= 12 \rightarrow6$) Universe \citep{Narayanan2025arXiv}}, maximize the SN dust destruction efficiency and does not include gas outflows. In these conditions of no outflows, no dust growth, and maximum dust destruction, the model from \citet{Ferrara2025} predicts compatible results with our galaxies having the largest dust-to-stellar surface mass density ratios. To explain lower dust-to-stellar surface mass density ratios, \citet{Ferrara2025} supports the attenuation-free scenario \citep{Ferrara2024a,Ferrara2024b} where dust is pushed through outflows into the halo of the galaxy.

The low dust-to-stellar surface mass density ratios of DC-848185 (C02), DC-536534 (C03), DC-873321 (C07ab), and DC-630594 (C11) are below the predictions from models in \citet{Ferrara2025} where outflows are not included. \citet{Sawant2025} models with $\omega=2.0$ show a decrease in the dust-to-stellar mass ratio at high metallicities, with the model having $\rm f_{gg}=1$ being above the average values of these galaxies, and the model with $\rm f_{gg}=0$ explaining all these galaxies except DC-630594 (C11). The model with no outflows and $\rm f_{gg}=0$, however, can reproduce DC-873321 (C07ab), and DC-630594 (C11). Finally, the {\it intermediate-dust} models from \citet{Palla2024} are below the observed dust-to-stellar surface mass density ratios. 

Three of the galaxies presenting the lowest dust-to-stellar surface mass density ratios (DC-848185 (C02), DC-536534 (C03), and DC-873321 (C07ab), square symbols in Figure\,\ref{fig:dustandmet}) are those with AGN signature and strong outflows in \citet{Ren2025}. These are galaxies having a $\Sigma_{SFR}-\Sigma{totalgas}$ relation with $\kappa_{\rm s}>1$ (see Figure\,\ref{fig:gasSFRcomparison_individual}). We propose that in these galaxies, dust could be removed from the system as: i) part of the material outflowing the galaxy and having a certain dust-to-gas mass ratio or, ii) due to radiation pressure affecting the dust grains \citep[][]{Ziparo2023,Ferrara2025,Lesniewska2025}. \citet{Ziparo2023} proposed a low limit for the burstiness parameter, $\kappa_{\rm s}\simeq3.3$, below which outflows cannot be launched. The $\kappa_{\rm s}$ values shown by these galaxies suggest that outflows are at least possible in these systems. 

Finally, in the bottom panel of Figure\,\ref{fig:dustandmet} we see that the dust-to-metals ratio of our galaxy sample are, in general, above the models. Only the dust models by \citet{Sawant2025} with $\rm f_{gg}=1$ are able to fall close to DC-848185 (C02) and DC-630594 (C11). Our galaxies have similar dust-to-metals surface mass density ratios to local Universe galaxies and close to the  maximum dust-to-stellar mass ratio estimated by \citet{Palla2024}.

Since our galaxies present higher dust-to-gas surface mass density ratios than the total dust-to-gas mass ratio for REBEL galaxies, we explore further whether the differences could be related to the estimation of the gas mass content in these galaxies. For that, we compare the gas mass fractions derived from observations with the results predicted by the models. In Figure\,\ref{fig:dustandgasfgas} we show the comparison of the gas mass fraction versus metallicity for our galaxies, the local galaxy sample from \citet{Casasola}, the REBELS sample, and the models tracks. All the models predict higher gas mass fractions than those observed in our galaxy sample, which points towards an overestimation of the gas masses for REBELS galaxies. In particular, we note here that most of the REBELS-IFU sample and HZ10-C+E fall above the model predictions, suggesting that the gas mass could be overestimated for these galaxies, as it was suggested in \citet{Algera2026}. Even closed box models from \citet{Palla2024} (gray lines in Figure\,\ref{fig:dustandgasfgas}) and \citet{Sawant2025} models with no outflows fall below the observations from REBELS-IFU and HZ10-C+E. An overestimation of the gas masses for REBELS and HZ10-C+E would cause an underestimation of the dust-to-gas mass ratios, as it is suspected when compared to our galaxy sample.

The dust-to-gas, dust-to-stars and dust-to-metals ratios, as well as the gas mass fractions observed in this small galaxy sample, show a wide range of values. From the independent study by \citet{Ren2025} we see that there is a relation between the different values and the possibility that these galaxies could have AGNs or strong outflows. Our comparison with three different evolution models suggests that outflows and dust growth play an important role in shaping the model tracks of the dust-to-gas, dust-to-stars and dust-to-metals ratios with the metallicity. We see here that to properly model the chemical and dust evolution of these galaxies, and probably galaxies at high-redshift in general, one needs to take into account the possible merging status, outflow strength and AGN contribution. 

\FloatBarrier
\section{Conclusions}\label{sec:conclusions}

We have analysed spatially resolved ($\sim$\,1\,kpc) rest-frame optical JWST/NIRSpec integral field observations of a sample of massive galaxies at $z\sim\,4-5$. We have derived fluxes of the most intense rest-frame optical emission lines, obtained metallicity maps, and compared them with SFR, gas mass, stellar mass, and dust mass surface density maps. We have also explored how current chemical and dust evolution models for high-redshift galaxies are able to reproduce the observational trends. Our main conclusions are: 

\begin{itemize}
    \item The average oxygen abundances derived from the maps show values slightly below Solar metallicity (12+log(O/H)$\sim$7.8-8.3), with variations up to 1\,dex within individual galaxies. The abundances derived here agree with those obtained in \cite{Faisst2026_survey,Faisst2025_massmetal} and \citet{Lee2026}. For some galaxies, the peak at H$\alpha$ emission, located at the centre of the main body of the galaxy, spatially correlates with areas of low (12+log(O/H)\,$\lesssim$\,8.0) metallicity, while in the outer parts the oxygen abundance tends to show larger values, in agreement with predictions from \citet{Lee2026}. 
    
    \item We have derived gas mass surface density maps using calibrations based on the [C{\sc{ii}}] emission line flux that have been presented in previous works. The atomic gas mass surface densities derived from the \citet{Heintz_2021} calibration are overestimated by 1-2 orders of magnitude. The total gas masses obtained from the spatially resolved [C{\sc{ii}}] conversion factor from \citet{Vallini_25} provides a better quantification of the gas mass surface densities to reproduce the KS relation. 

    \item Our galaxy sample shows similar dust-to-gas surface mass density ratios with a flatter relation with metallicity than that observed from local galaxies and from DLAs. Two galaxies of our sample have an order of magnitude higher dust-to-stellar surface mass density ratios than normal galaxies in the local Universe and in agreement with recently observed dusty galaxies at $z\sim7-8$. In contrast, four of our galaxies exhibit low dust-to-stellar  surface mass density ratios and therefore a low dust content. Three of the four show evidence of strong outflows, suggesting that dust could have been removed from these systems. 
        
    \item Chemical and dust evolution models favouring dust growth and minimizing dust destruction can explain the largest observed dust-to-stellar surface mass density ratios. In order to reproduce the lowest dust-to-stellar surface mass density ratios of our galaxy sample, models with outflows need to be taken into account.
    
\end{itemize}

There are recent works trying to explain the devoid of dust in luminous UV galaxies at high redshifts and the fast build up observed at $z\sim$\,6. We would like to note here that even at redshifts of $z\sim5-6$ we are able to find galaxies that have a high dust content as well as others presenting lower dust reservoirs compared to their stellar masses. Larger galaxy samples observed at spatially resolved scales will improve our knowledge of the different mechanisms in place that regulate the amount of dust, gas, stars and metals in galaxies at high redshifts. 

\begin{acknowledgements}
 We thank the referee for the comments that have helped to improve the previous version of this work. We also thank V. Casasola for sharing the data of the local Universe galaxies shown here. MR acknowledges support from project PID2023-150178NB-I00 financed by MCIU/AEI/10.13039/501100011033, and by FEDER, UE. 
IDL acknowledges funding from the European Research Council (ERC) under the European Union's Horizon 2020 research and innovation program DustOrigin (ERC-2019- StG-851622), from the Belgian Science Policy Office (BELSPO) through the PRODEX project “JWST/MIRI Science exploitation” (C4000142239) and from the Flemish Fund for Scientific Research (FWO-Vlaanderen) through the research project G0A1523N. 
MP acknowledges financial support from the project "LEGO – Reconstructing the building blocks of the Galaxy by chemical tagging" granted by the Italian MUR through contract PRIN2022LLP8TK-001. 
RA acknowledges financial support from projects PID2023-147386NB-I00 funded by MCIN/AEI/10.13039/501100011033 and by ERDF/EU and the Severo Ochoa grant CEX2021-001131-S to the IAA-CSIC. 
CA acknowledges that this work of the Interdisciplinary Thematic Institute IRMIA++, as part of the ITI 2021–2028 program of the University of Strasbourg, CNRS and Inserm, was supported by IdEx Unistra (ANR-10-IDEX-0002), and by the SFRI-STRAT’US project (ANR-20-SFRI-0012) under the framework of the French Investments for the Future Program.
MA is supported by FONDECYT grant number 1252054, and gratefully acknowledges support from ANID Basal Project FB210003,  ANID MILENIO NCN2024-112 and ANID + Vinculaci\'on Internacional + FOVI250261.
RJA was supported by FONDECYT grant number 1231718 and by the ANID BASAL project FB210003. 
MB acknowledges support from the ANID BASAL project FB210003. This work was supported by the French government through the France 2030 investment plan managed by the National Research Agency (ANR), as part of the Initiative of Excellence of Université Côte d’Azur under reference number ANR-15-IDEX-01. 
EdC acknowledges support from the Australian Research Council (project DP240100589).
EPM thanks the financial support from project Estallidos (PID2022-136598NB-C32, Spanish Ministerio de Ciencia e Innovaci\'on), the Severo Ochoa grant CEX2021-001131-S funded by MCIN/AEI/10.13039/501100011033, and the assistance from his guide dog Rocko, without whose daily help this work would have been much more difficult.
RLD is supported by the Australian Research Council through the Discovery Early Career Researcher Award (DECRA) Fellowship DE240100136 funded by the Australian Government.
AF is partly supported by the ERC Advanced Grant INTERSTELLAR H2020/740120, and by grant NSF PHY-2309135 to the Kavli Institute for Theoretical Physics. 
NGV acknowledges scholarship from ANID BECAS/Doctorado Nacional/2023-21231942. 
EI acknowledges support from ANID MILENIO NCN2024-112 and ANID FONDECYT Regular 1221846. 
JM gratefully acknowledges support from ANID MILENIO NCN2024-112. 
AN acknowledges support from the Narodowe Centrum Nauki (NCN), Poland, through the SONATA BIS grant UMO-2020/38/E/ST9/00077. 
PS was funded by National Science Centre, Poland 2023/50/E/ST9/00383 and UMO-2020/38/E/ST9/00077. 
LV acknowledges support from the INAF Minigrant "RISE: Resolving the ISM and Star formation in the Epoch of Reionization" (PI: Vallini, Ob. Fu. 1.05.24.07.01). 
VV acknowledges support from the Comité ESO Mixto 2024 and from the ANID BASAL project FB210003. 
SAvdG acknowledges support by the French National Research Agency under the contract REDEEMING (ANR-24-CE31-2530). 
\end{acknowledgements}

\bibliographystyle{aa}
\bibliography{aa59898}

\begin{appendix}

\section{Metallicity estimations: Comparison between different strong-line calibrations}

\begin{figure}[htbp]

    {\includegraphics[width=0.8\linewidth]{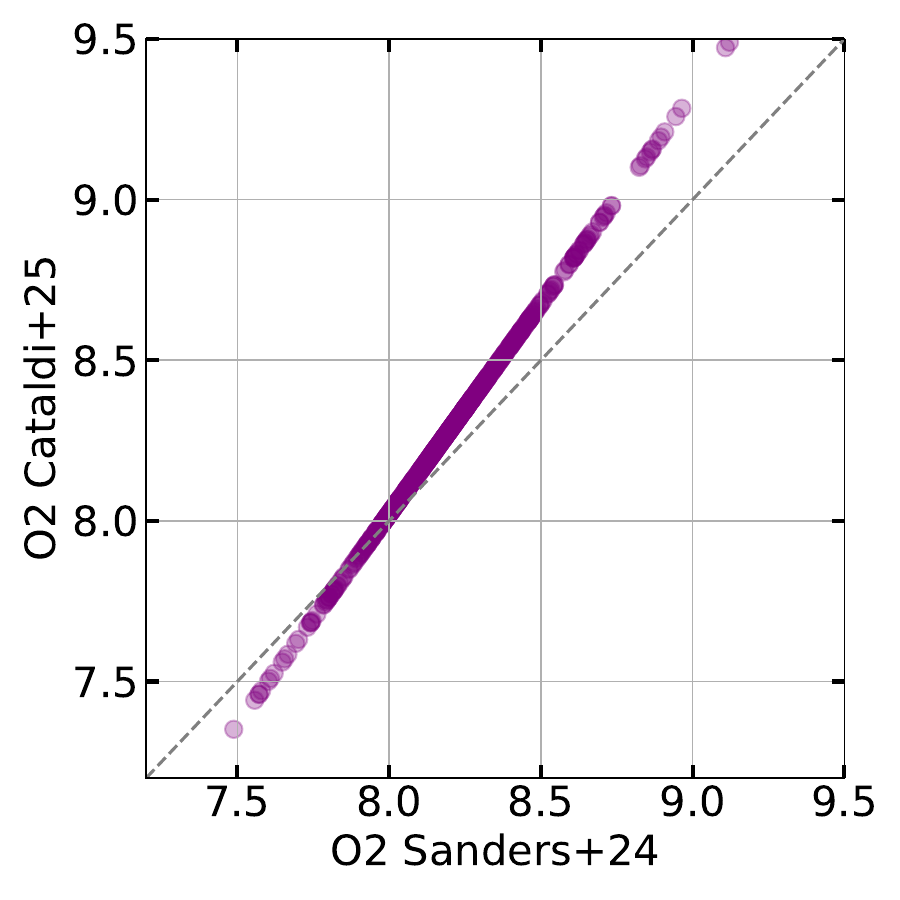}}
    {\includegraphics[width=0.8\linewidth]{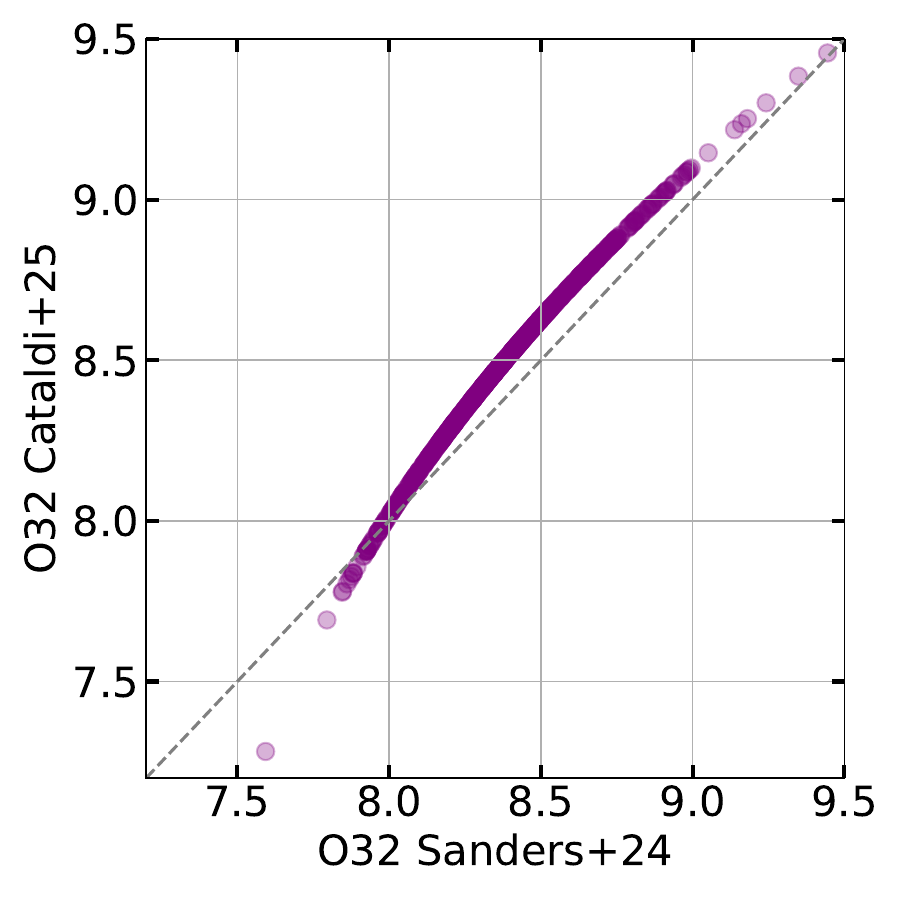}}
 \caption{Comparison of the metallicity values obtained from the different calibrations used in this paper. Top: Comparison between the O2 calibrations from \citet{Sanders} and \citet{Cataldi2025}. Bottom: the corresponding comparison for O32 calibration.}
  \label{fig:cataldi_sanders}
\end{figure}

In Table\,\ref{ap:tab:met_results} we show the median values, average uncertainties, and standard deviations of the oxygen abundances for the spaxel distribution of each galaxy using the strong-line calibrations from \citet{Cataldi2025}. In Figure\,\ref{fig:cataldi_sanders} we show the comparison between the O32 and O2 calibrations derived in \citet{Sanders} and \citet{Cataldi2025}. Note that the comparison is done for all the spaxels in the field of view of our galaxy sample, however the metallicity calibrations have in general a smaller metallicity range for which the different calibrations are valid. In the case of \citet{Cataldi2025}, the calibrations cover moderately high oxygen abundances (12+log(O/H)\,$\sim$\,8.0-8.4), while \citet{Sanders} calibrations are valid in a wider metallicity range (12+log(O/H)\,$\sim$\,7.4-8.3). 

\begin{table*} 
\caption{Oxygen abundances (12 + log (O/H)) using strong-line calibrations from \citet{Cataldi2025}.}
\label{ap:tab:met_results}      
\centering 
\begin{tabular}{cccccccccc}
\toprule
 & \multicolumn{9}{c}{\citet{Cataldi2025} } \\
\cmidrule(lr){2-10}
Diagnostic & \multicolumn{3}{c}{O2} & \multicolumn{3}{c}{O32} & \multicolumn{3}{c}{R23} \\
\midrule
Galaxy &  median  & av. err. & std &  median & av. err. & std &  median & av. err. & std  \\
\midrule
C02 & 8.26 & 0.11 & 0.27 & 8.33 & 0.04 & 0.14 & 8.36 & 0.09 & 0.22 \\
C03 & 8.18 & 0.09 & 0.24 & 8.33 & 0.05 & 0.23 & 8.18 & 0.12 & 0.22  \\
C04a & 8.35 & 0.10 & 0.19 & 8.36 & 0.05 & 0.15 & 8.36 & 0.10 & 0.11  \\
C07ab & 8.23 & 0.11 & 0.23 & 8.26 & 0.05 & 0.18 & 8.23 & 0.11 & 0.22  \\
C11 & 8.36 & 0.10 & 0.21 & 8.37 & 0.05 & 0.12 & 8.33 & 0.08 & 0.17\\
C13 & 8.25 & 0.12 & 0.21 & 8.39 & 0.05 & 0.17 & 8.46 & 0.08 & 0.19  \\
\bottomrule
\end{tabular}    
\tablefoot{We show the median values, average uncertainties, and standard deviations of the oxygen abundances for the spaxel distribution of each galaxy.
}             
\end{table*}

\section{Comparison with other gas mass estimates}\label{ap:comfgas}
\begin{figure}[h]
\begin{centering}
    {\includegraphics[width=\linewidth]{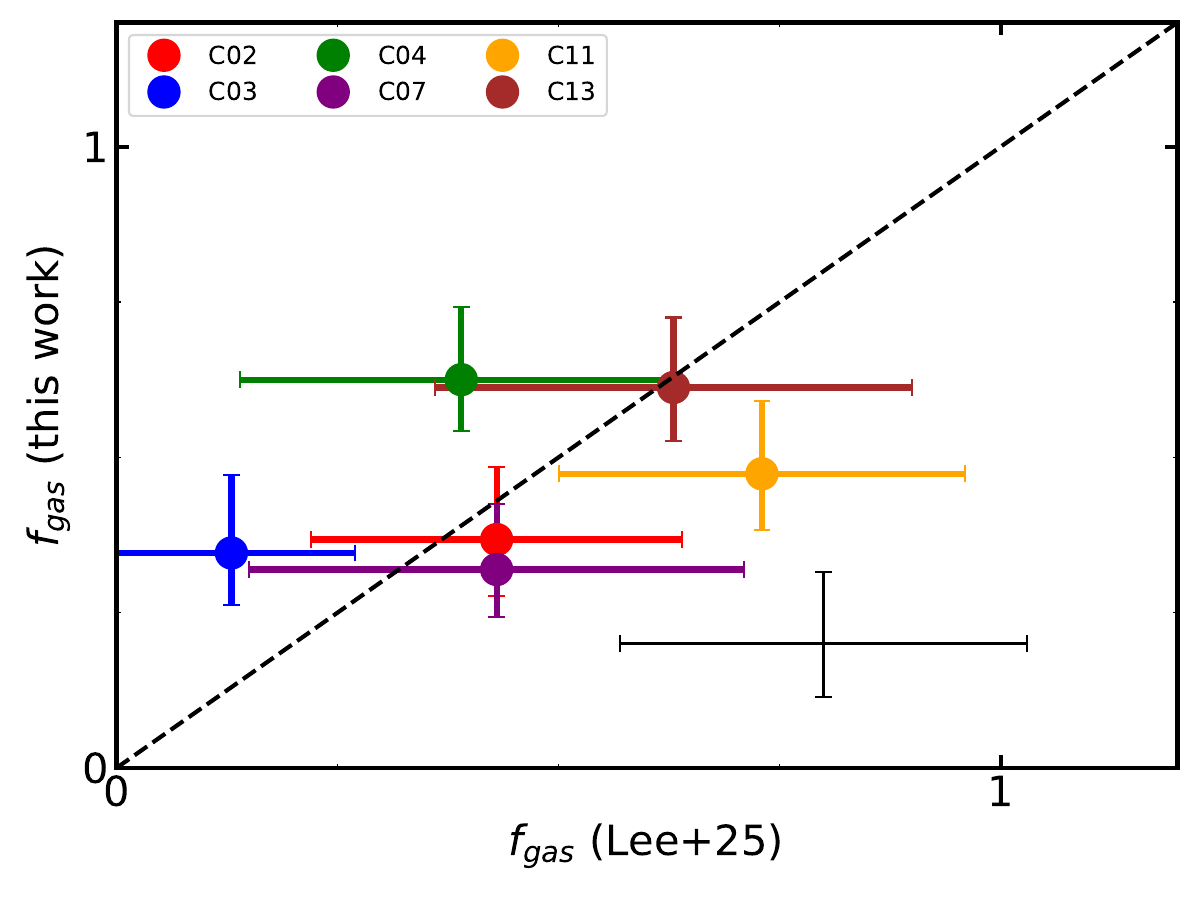}}
\end{centering}
 \caption{Comparison between the gas mass fractions derived in this work using \citet{Vallini_25} calibration and those derived in \citet{Lee2025} applying the calibration in \citet{Tacconi2020}. The dashed-line corresponds to one-to-one relation.}
  \label{ap:fig:compfgas}
\end{figure}

\citet{Lee2025} derived gas mass fraction with a different methodology based on the observed flux of the ALMA Band-7 dust continuum and following Eq.\,(3) from \citet{Tacconi2020} and assuming a typical dust temperature of $\sim$\,50\,K. We compare below the gas mass fractions we obtain with those derived in \citet{Lee2025}. In Figure\,\ref{ap:fig:compfgas} we show the comparison between both estimates of the gas mass fraction. In general there is a good agreement between the gas mass fractions estimated from both methods, only two galaxies (DC-536534 (C03) and  DC-630594 (C11)) depart from the one-to-one relation considering the uncertainties.   

\section{Scaling relations for individual galaxies}

\begin{figure}[h]
\begin{centering}
   
    {\includegraphics[width=\linewidth]{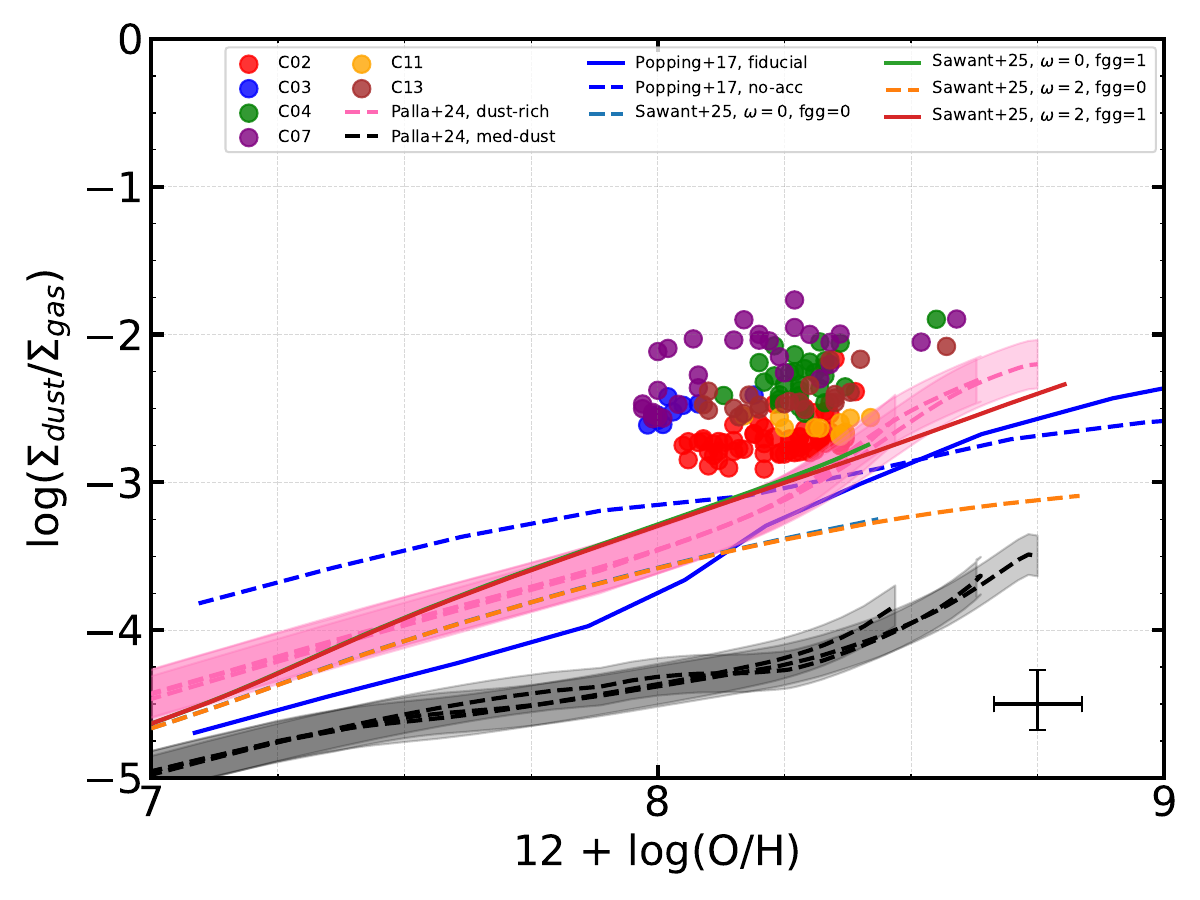}}
    
    {\includegraphics[width=\linewidth]{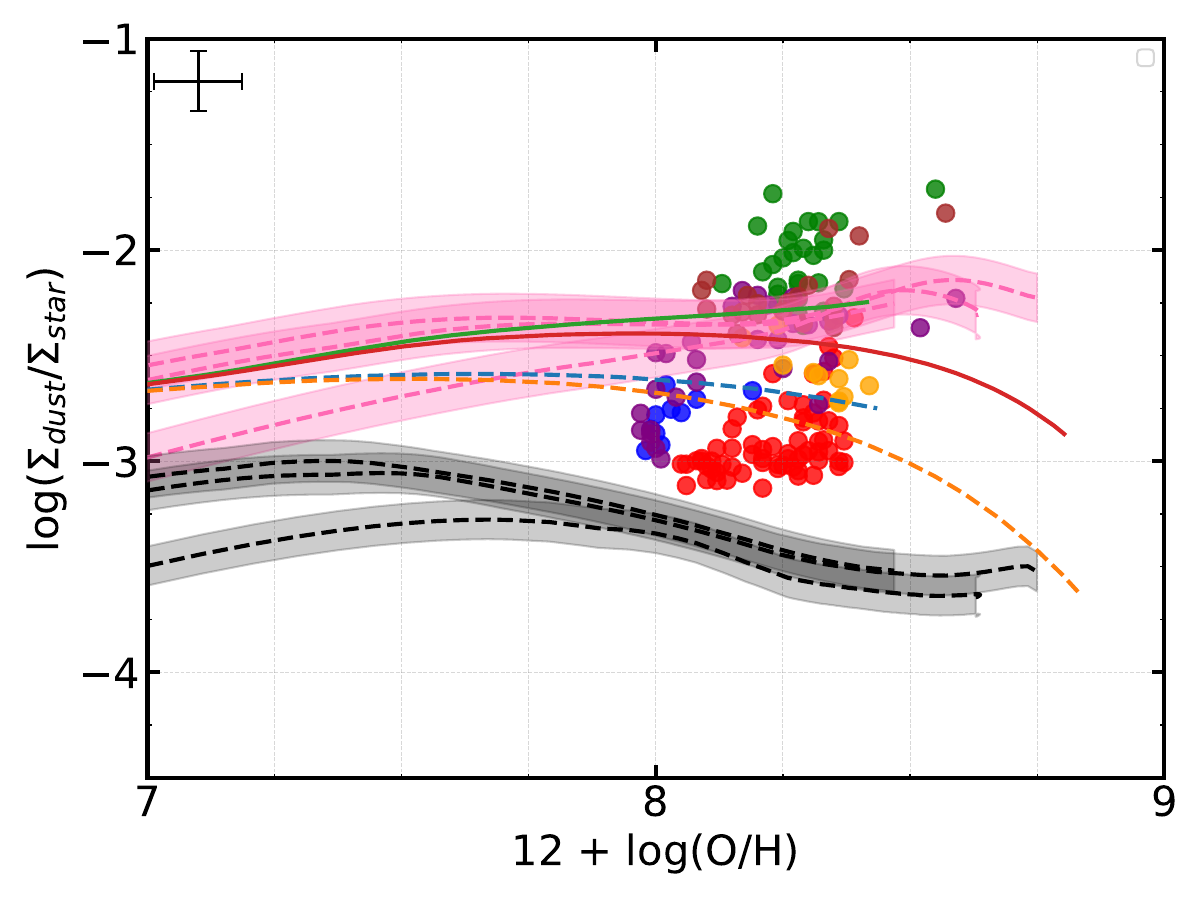}}

    {\includegraphics[width=\linewidth]{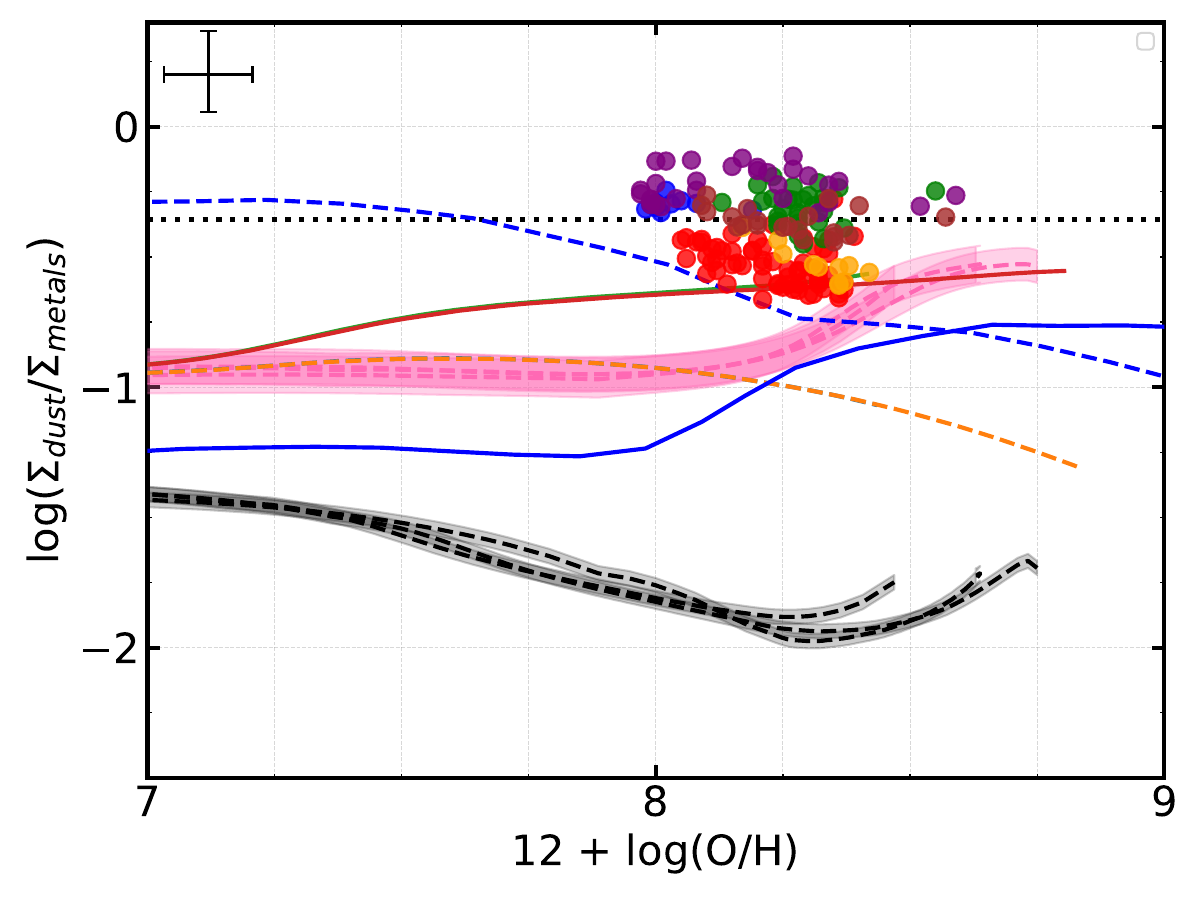}}
\end{centering}
 \caption{Logarithm of dust-to-gas surface mass density ratio (top panels), dust-to-stellar surface mass density ratio (middle panels), and dust-to-metals surface mass density ratio (bottom) as a function of the oxygen abundance. Two different set of models are presented, the chemical and dust evolution models from \citet{Palla2024}, and the SAM from \citet{Popping_2017} (see Section\,\ref{sec:models}). The shaded regions around the model curves are defined using a fixed percentage of the difference between the maximum and minimum y-values of each model.}
  \label{ap:fig:dustandmet}
\end{figure}

\begin{figure}
    \centering
    \includegraphics[width=\linewidth]{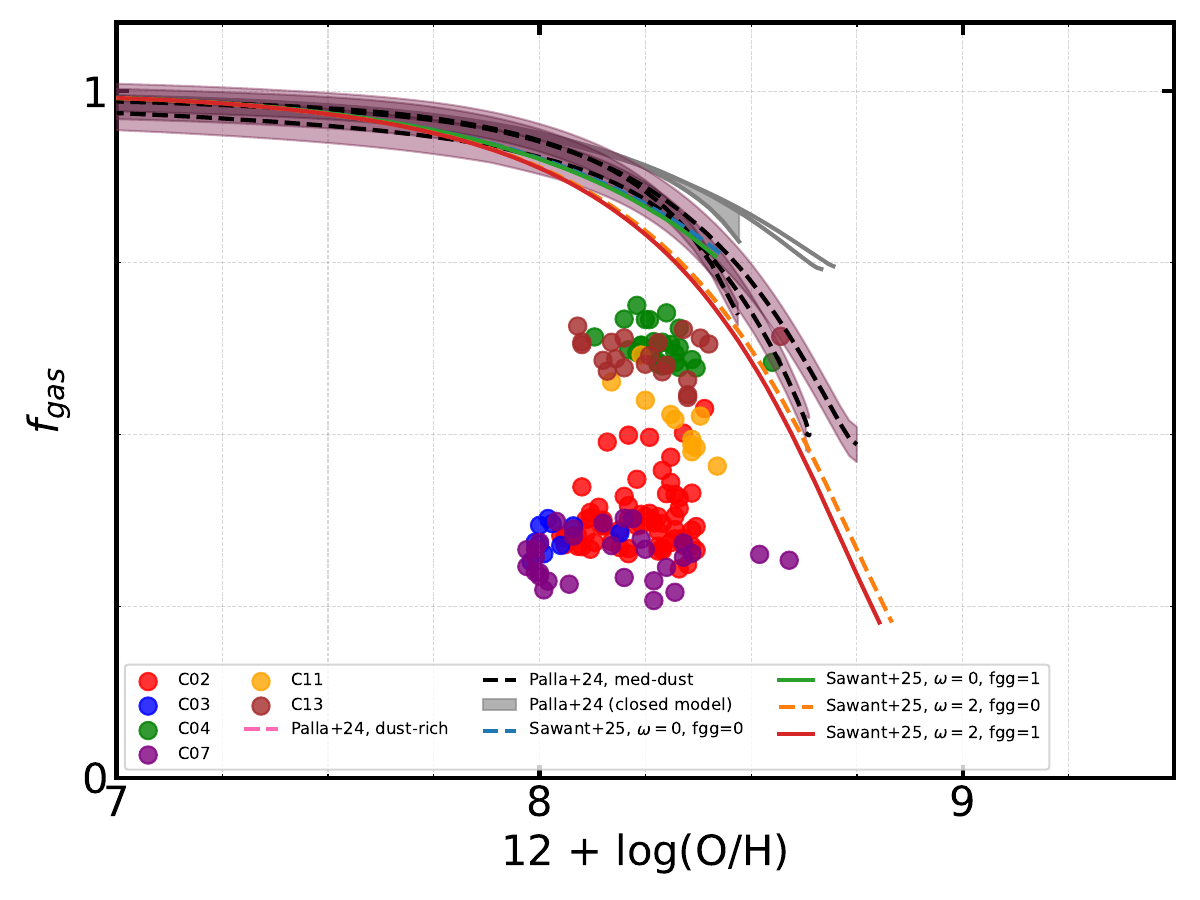}
    \caption {Logarithm of $f_{\text{gas}}$ as a function of 12 + log (O/H). Models are the same as in Figure\,\ref{fig:dustandmet}. Chemical models from \citet{Palla2024} assuming a closed-box approximation are depicted in grey.}
    \label{ap:fig:dustandgasfgas}
\end{figure}

In this Section we show the dust scaling relations and the gas mass fraction versus metallicity for the individual \-pixels\- of each galaxy. Each colour corresponds to one galaxy in our sample. The distribution of the data points shows how each galaxy follows these scaling relations. For instance, the data in DC-536534 (C03) are concentrated on a narrow metallicity and dust-to-gas surface mass density ratio range in the tope panel of Figure\,\ref{ap:fig:dustandmet}, while VC-5100994794 (C13) or DC-873321 (C07ab) cover a wider range of values for these quantities. In the bottom panel of Figure\,\ref{ap:fig:dustandmet} we see that while DC-848185 (C02) tends to show a decreasing trend of dust-to-metals ratio with metallicity the data distribution of the other galaxies follow a constant trend. Figure,\ref{ap:fig:dustandgasfgas} also shows that, while the individual pixels of some galaxies have gas mass fractions within a narrow range, those of DC-848185 (C02) and DC-630594 (C11) span a wider range. A detailed analysis of the data distribution of each galaxy in these scaling relations is beyond the scope of this paper.

\section{Surface mass density and integrated mass ratio comparison}\label{ap:comp_ratios}

In this Section we compare the dust-to-gas and dust-to-stellar surface mass density ratio with the dust-to-gas and dust-to-stellar mass ratios infer from integrated quantities as presented in Table\,\ref{tab:gal}. Except for C03 in the top panel, the rest of the galaxies have dust-to-gas ratios close to the one-to-one line. The larger value for the dust-to-gas mass ratio compared to the dust-to-gas surface mass density ratio for C03 is due to the high value of the total dust mass (see Table\,\ref{tab:gal}) for this galaxy. In particular the dust continuum radius inferred in \citet{Mitsuhashi} for this galaxy is a 1$\sigma$ upper limit. If the size for C03 would be indeed smaller than the value reported in table\,1 of \citet{Mitsuhashi}, then we would have recovered larger SFR surface density and the dust temperature estimate would have been larger\footnote{The dust temperature estimated in Relaño et al. (in prep) is T=23\,K, lower than the temperature estimates from the SFR surface density maps.}, with a corresponding lower dust mass estimates and a better agreement with the dust-to-gas surface mass density ratio.

Regarding the comparison with the dust-to-stellar ratios, C04 and C13 shows smaller dust-to-stellar mass ratios than dust-to-stellar surface mass density ratios. These galaxies are classified as {\it multiple} in \citet{Mitsuhashi} (see Table 1 in that paper). The classification is based on the HST/F160W images corresponding to UV stellar continuum, therefore it remains unclear whether dust continuum and UV emission are indeed co-spatial. This brings to an open question that is out of the scope of this paper. 

\begin{figure}[h]
\begin{centering}
    {\includegraphics[width=\linewidth]{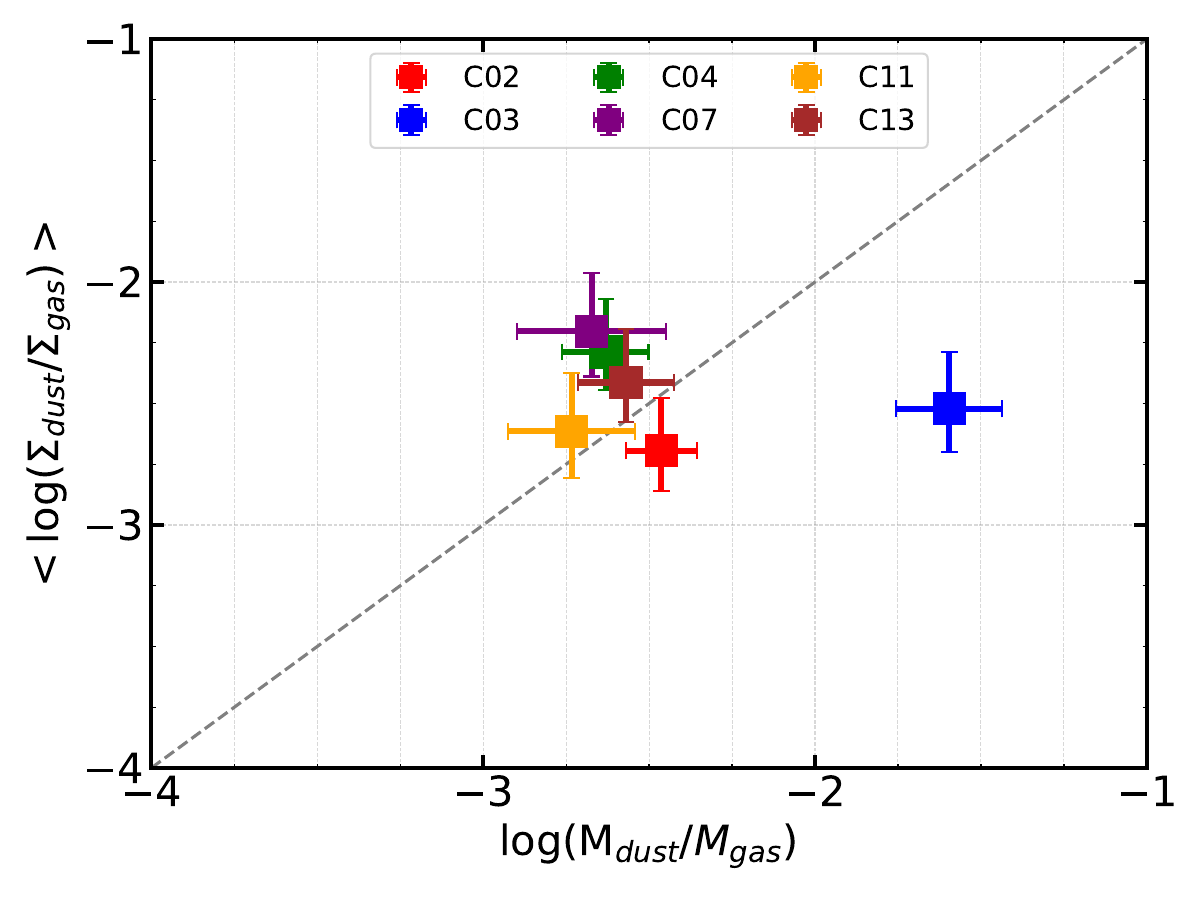}}
    {\includegraphics[width=\linewidth]{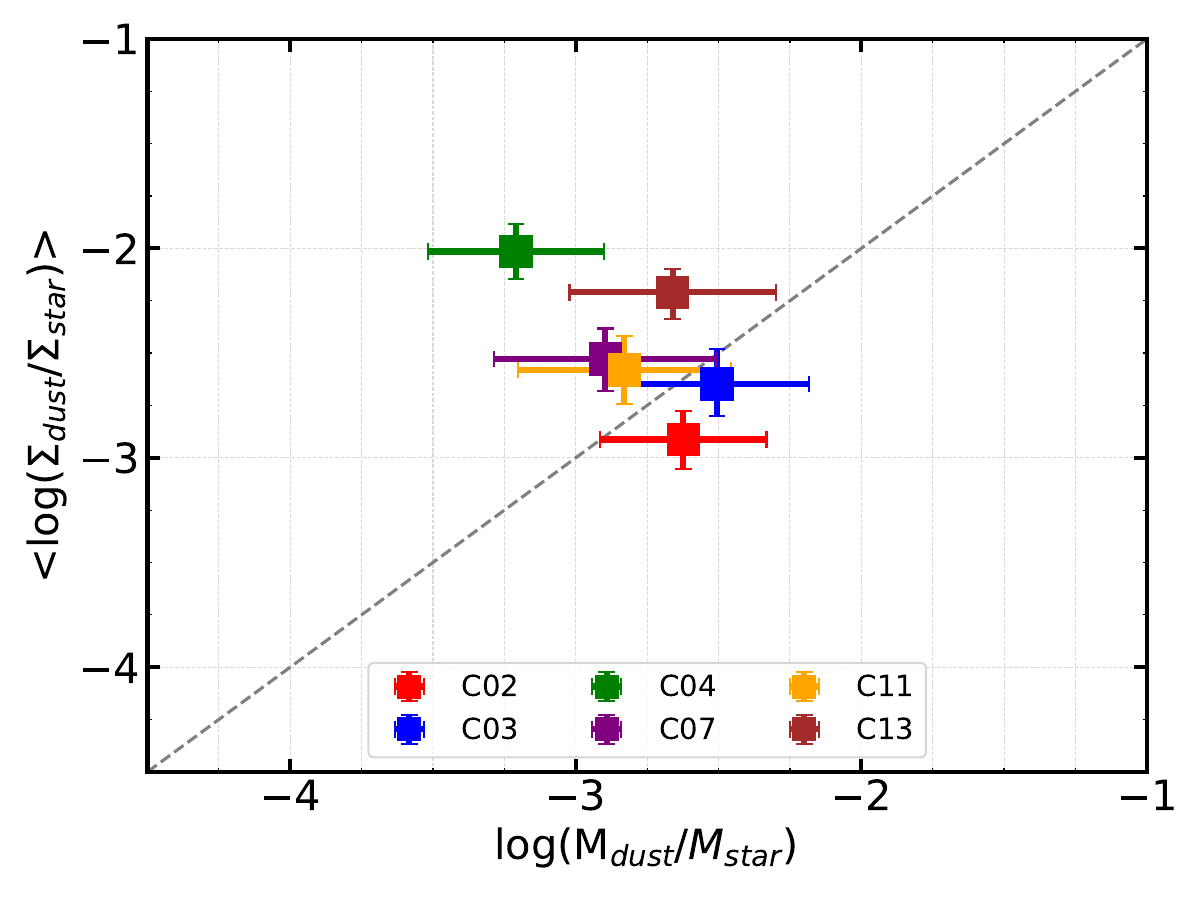}}  
    {\includegraphics[width=\linewidth]{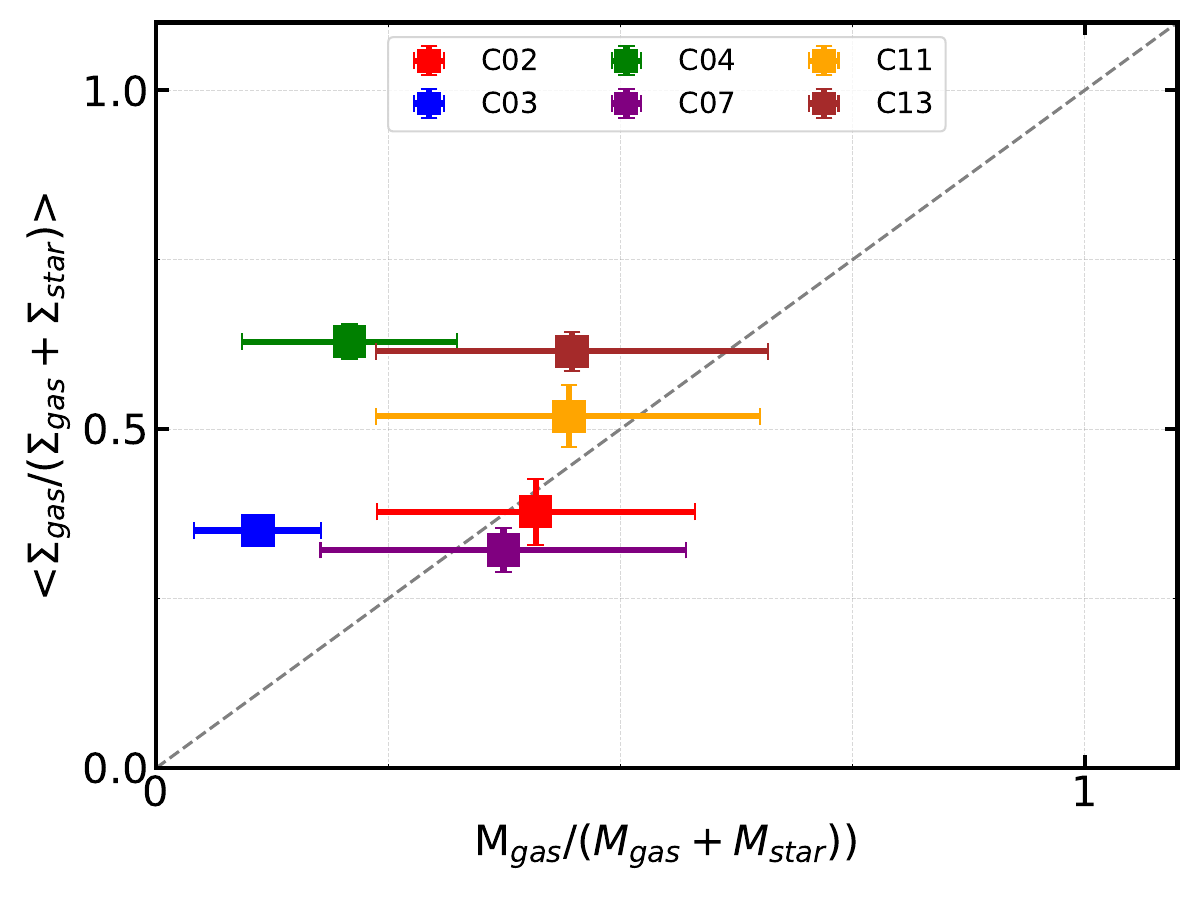}}
\end{centering}
 \caption{Comparison of the average values for the dust-to-gas (top) and dust-to-stellar (middle) surface mass density ratio found in this paper with the dust-to-gas and dust-to-stellar mass ratios inferred from integrated quantities. The bottom panel corresponds to the comparison for the gas mass fraction.}
  \label{ap:fig:comp_ratios}
\end{figure}

In the bottom panel of Figure\,\ref{ap:fig:comp_ratios} we show the comparison between the gas mass fraction obtained with surface mass densities and total integrated mass estimates. We see that, except C03 and C04, which present lower fractions when the integrated masses are used the other galaxies have consistent gas mass fractions. For both methodologies, gas mass fractions are below 0.6. For the metallicities of our galaxies, both methods give gas mass fractions that will not be larger than those estimated from chemical evolution models (see Figure\,\ref{fig:dustandgasfgas}).

\end{appendix}
\end{document}